\documentclass[opre,nonblindrev]{informs3arxiv}

\OneAndAHalfSpacedXI 

\usepackage{endnotes}
\let\footnote=\endnote

\usepackage[utf8]{inputenc}
\usepackage{graphicx,tikz}
\usepackage[tbtags]{mathtools}
\usepackage{amsmath,amsfonts,amssymb,bm}
\usepackage{color,dsfont}
\usepackage{url}
\usepackage{algorithm,algorithmic,setspace}
\usepackage{hhline}
\usepackage{enumerate}
\usepackage{array}
\usepackage{booktabs}
\usepackage{mathrsfs,bbm,bm}
\usepackage{tikz}
\usepackage{multirow}
\usepackage{hyperref,cleveref}
\usepackage{accents}
\usepackage[round]{natbib}
\usepackage{pgfplots}
\pgfplotsset{compat=1.9}
\usepackage{float}

\usetikzlibrary {positioning}
\usepackage{pgf}
\usepackage{pgfplots}
\usetikzlibrary{shapes.geometric, arrows}
\tikzstyle{node} = [ellipse, minimum width=1cm, minimum height=.5cm, text centered, text width=1cm, draw=black, fill=blue!40]
\tikzstyle{arrow} = [thick, ->, >=stealth]
\tikzstyle{arrow1} = [red, thin, ->, >=stealth]
\usepackage{rotating}
\usetikzlibrary{fillbetween}
\usetikzlibrary{intersections}
\pgfdeclarelayer{bg}
\pgfsetlayers{bg,main}
\usepgfplotslibrary{fillbetween}

\usepackage[caption=false]{subfig}

\usepackage{etoolbox}
\let\bbordermatrix\bordermatrix
\patchcmd{\bbordermatrix}{8.75}{4.75}{}{}
\patchcmd{\bbordermatrix}{\left(}{\left[}{}{}
\patchcmd{\bbordermatrix}{\right)}{\right]}{}{}

\newcounter{algo}
\AtBeginEnvironment{algorithm}{\OneAndAHalfSpacedXI}
\AtBeginEnvironment{tabular}{\OneAndAHalfSpacedXII}

\renewcommand{\o}{\omega}
\renewcommand{\O}{\Omega}
\newcommand{\A}{\mathcal{A}}
\newcommand{\Y}{\mathcal{Y}}
\newcommand{\X}{\mathcal{X}}

\newcommand{\R}{\mathbb{R}}

\newcommand{\N}{\mathbb{N}}

\DeclareMathOperator{\essinf}{ess\; inf}
\DeclareMathOperator{\bd}{bd}
\DeclareMathOperator{\cl}{cl}

\DeclareMathOperator{\Int}{int}

\DeclareMathOperator{\Leb}{Leb}
\DeclareMathOperator{\dom}{dom}

\newcommand{\var}{\text{V@R}}

\newcommand{\of}[1]{\ensuremath{\left( #1 \right)}}

\newcommand{\cb}[1]{\ensuremath{ \left\{ #1 \right\} }}

\newcommand{\sqb}[1]{\ensuremath{ \left[ #1 \right] }}

\newcommand{\ignore}[1]{}

\newcommand{\EN}{\text{EN}}

\usepackage{natbib}
\bibpunct[, ]{(}{)}{,}{a}{}{,}%
\def\bibfont{\small}%
\TheoremsNumberedThrough     
\ECRepeatTheorems

\EquationsNumberedThrough    

\begin{document}
	
	
	
	\RUNAUTHOR{AlAli and Ararat}
	
	\RUNTITLE{Systemic Values-at-Risk}
	
	\TITLE{Systemic Values-at-Risk and their Sample-Average Approximations}
	
	\ARTICLEAUTHORS{%
		\AUTHOR{Wissam AlAli}
		\AFF{Department of Industrial Engineering, University of Houston, Houston, TX,\\ \EMAIL{wialali@uh.edu}} 
	\AUTHOR{\c{C}a{\u{g}}{\i}n Ararat}
	\AFF{Department of Industrial Engineering, Bilkent University, Ankara, Turkey,\\ \EMAIL{cararat@bilkent.edu.tr}}
} 

\ABSTRACT{%
	This paper investigates the convergence properties of sample-average approximations (SAA) for set-valued systemic risk measures. We assume that the systemic risk measure is defined using a general aggregation function with some continuity properties and value-at-risk applied as a monetary risk measure. We focus on the theoretical convergence of its SAA under Wijsman and Hausdorff topologies for closed sets. After building the general theory, we provide an in-depth study of an important special case where the aggregation function is defined based on the Eisenberg-Noe network model. In this case, we provide mixed-integer programming formulations for calculating the SAA sets via their weighted-sum and norm-minimizing scalarizations. To demonstrate the applicability of our findings, we conduct a comprehensive sensitivity analysis by generating a financial network based on the preferential attachment model and modeling the economic disruptions via a Pareto distribution.
}%



\maketitle

%


\section{Introduction}

\subsection{Background and Literature Review}

Systemic risk refers to potential disturbances within the financial system that can spread across institutions, markets, and economies, causing adverse outcomes. This often triggers a cascading sequence leading to financial contagion. Hence, recognizing the potential repercussions underscores the importance of proactive measures and risk management strategies.

Risk management has gained substantial attention, particularly after the swift onset of the 2008 financial crisis. This crisis, marked by the rapid devaluation of mortgage-related securities, impacted both the United States and global financial systems, leading to the downfall of numerous prominent institutions \citep{kotz2009financial}. It underscored the profound interconnectedness of global financial systems, illustrating how minor shocks can generate significant disruptions in the economy \citep{silva2017systemic}.

Recent research has introduced innovative ideas on systemic risk measures and allocation methods, focusing on breaking down such measures into an aggregation function and a monetary risk measure \citep{iyengar,kromer2016systemic}. The aggregation function gauges societal impact, while capital allocations, representing potential bailout costs, are added post-aggregation. A noteworthy framework assigns capital allocations to institutions before aggregation, as discussed in \cite{birgit}, \cite{biagini2019unified}, \cite{armenti2018multivariate}, proving particularly useful for regulatory purposes and crisis prevention. These models yield multivariate risk measures based on the set of feasible capital allocations.

The intriguing aspect of systemic risk lies in its connection to complex interconnections among organizations, rather than individual entities. Traditional risk measures such as value-at-risk, conditional value-at-risk, and negative expectation are adequate when risk is solely tied to a single firm's performance, irrespective of its interconnections. However, systemic risk demands a broader view, requiring an analysis of the distribution of total profits and losses across all firms in the economy, as exemplified in \cite{acharya}, \cite{tarashev}, \cite{adrian}, \cite{amini2015systemic}, \cite{brunnermeier2019measuring}.


In this paper, we focus on financial networks where members are interconnected through contractual obligations, and interbank payments are determined by a clearing process based on uncertain operating cash flows. The model in \cite{enberg} depicts a financial system as a static directed network of banks with specified interbank liabilities. They propose two methods for calculating a clearing vector, representing payments to fulfill interbank obligations, assuming positive operating cash flows for each bank. The first method, the fictitious default algorithm, iteratively computes the clearing vector through a finite number of updates. The second approach formulates a concise continuous optimization problem based on liabilities, operating cash flows, and an increasing arbitrary objective function, yielding a linear programming problem when a linear objective function is selected.

The fictitious default algorithm is widely preferred by researchers working on financial network models. In \cite{suzuki2002valuing}, a similar method for evaluating clearing vectors is proposed, expanding on \cite{enberg} and considering cross-shareholdings.  In \cite{cifuentes2005liquidity}, systemic risk is examined in terms of institution liquidity and asset price instability. The model in \cite{elsinger2009financial} extends the Eisenberg-Noe model by integrating a cross-holdings structure, relaxing the assumption of positive cash flows and exploring the model under certain seniority assumptions. In \cite{rogers2013failure}, default costs are introduced to the Eisenberg-Noe model, and the necessity of bailing out delinquent institutions is investigated. A comprehensive network model is developed in \cite{weber2017joint} by incorporating cross-holdings, fire sales, and bankruptcy costs simultaneously. We refer the reader to \cite{kabanov2018clearing} for a detailed overview of these models and algorithms for calculating clearing vectors. Recently, in \cite{menj}, an extension of the Rogers-Veraart model is proposed in which the operating cash flows are unrestricted in sign. In contrast to these works, the focus in \cite{menj} is on the mathematical programming approach rather than fixed point algorithms, and it is shown that one can find a clearing vector for the Rogers-Veraart model by solving a mixed-integer linear programming problem.

The operating cash flows of the network members often face uncertainty due to correlated risk factors, represented as a random vector with possibly correlated components. The resulting clearing vector becomes a deterministic function of this random operating cash flow vector, determined by the clearing mechanism. Building on this concept, recent research initiated in \cite{iyengar} focuses on defining systemic risk measures capturing necessary capital allocations for network members to control certain (non-linear) averages across scenarios. By utilizing the clearing mechanism, a random aggregate quantity associated with the clearing vector—such as total debt paid or total equity generated—is defined. This quantity serves as a deterministic and scalar aggregation function of the operating cash flow vector. In \cite{iyengar}, a systemic risk measure is defined as a scalar functional of the operating cash flow vector, quantifying the risk of the random aggregate quantity using convex risk measures like negative expected value, average value-at-risk, or entropic risk measure.

The systemic risk measure in \cite{iyengar} denotes the total capital requirement for the system to maintain an acceptable risk level for the aggregate quantity. Since this total capital is only deployed after the shock is aggregated, allocating it back to system members poses an unresolved question, requiring a separate procedure. To address this, set-valued and scalar systemic risk measures that are sensitive to capital levels are proposed in \cite{birgit} and \cite{biagini2019unified}, respectively. These functionals aim to identify deterministic capital allocation vectors directly complementing the random operating cash flow vector. The augmented cash flow vector is aggregated, and the risk of the resulting random aggregate quantity is controlled using a convex risk measure, akin to \cite{iyengar}. The set-valued systemic risk measure in \cite{birgit} represents the set of all feasible capital allocation vectors, treating the measurement and allocation of systemic risk as a joint problem.

When the underlying aggregation function is suitably simple, the sensitive systemic risk measures studied in \cite{birgit} and \cite{biagini2019unified} exhibit advantageous theoretical properties. In \cite{ararat2020dual}, assuming a monotone and concave aggregation function and a convex base risk measure, it is shown that the set-valued sensitive systemic risk measure becomes a convex set-valued risk measure according to the framework in \cite{hamel2011set}, and dual representations are obtained in terms of the conjugate function of the aggregation function.

The computation of set-valued risk measures is usually handled by formulating an associated vector optimization problem whose upper image gives the risk measure; see \cite{sv-avar}, \cite{recursivealg}, \cite{superhedging}. In \cite{menj}, the mixed-integer programming formulations of clearing vectors for the Rogers-Veraart model are combined with the vector optimization approach to compute systemic risk measures based on polyhedral convex risk measures. Different from the vector optimization approach, a grid search algorithm is proposed in \cite{birgit} for computing systemic risk measures. Notably, all these works assume that the underlying probability space is finite to ensure that the corresponding vector optimization problems are finite-dimensional (or a finite number of clearing vectors are computed in the grid search). This is a significant drawback as, in practice, the operating cash flow vector may be better modeled via a continuous heavy-tailed distribution such as the Pareto distribution.

\subsection{Contribution}

In this paper, we study the convergence properties of sample-average approximations (SAA) for set-valued systemic risk measures within a general probability space, putting particular emphasis on the theoretical aspects of SAA convergence. We assume that the underlying monetary risk measure is value-at-risk, which enables us to formulate our optimization problems as chance-constrained programming problems. We name the resulting systemic risk measure as \emph{systemic value-at-risk}. By utilizing the SAA framework for chance-constrained programming proposed in \cite{Shapiro}, \cite{ahmed2008solving}, and \cite{chen2019convergence}, our primary contribution lies in extending the SAA framework to set-valued risk measures, a novel direction in the literature, and proving the almost sure convergence of these approximations under Wijsman and Hausdorff topologies for closed sets. Our convergence results encompass both sensitive and insensitive cases, and they allow for a general aggregation subject to some continuity conditions, demonstrating the robustness of our theoretical framework.

Specifically, our results reveal that for the sensitive case, where risk measures exhibit non-convexity and lack dual representations, the convergence of weighted-sum scalarizations and distance functions, often called norm-minimizing scalarizations in the current paper, is established through an intricate analysis of the underlying chance-constrained programming problems. On the technical level, the problems considered in \cite{Shapiro} are closely related to our scalarization problems. However, unlike the form considered there, our problems involve an almost sure selection constraint in addition to the usual chance constraint given by the value-at-risk. Hence, we start by proving a technical result (\Cref{prop:technical} in \Cref{sec:theory} below) that is well-suited for our purposes. With this tool, we prove convergence for a general scalarization, followed by the Wijsman and Hausdorff convergence of the sample-average approximations to the true systemic value-at-risk.

We apply our theory on the systemic value-at-risk in which the aggregation function is based on the Eisenberg-Noe model. We verify that this aggregation function satisfies the assumptions made for the general theory to work as long as the random operating cash flow vector has a continuous distribution with full support on the positive orthant. In addition to the convergence results, we also address the computational challenges inherent in calculating the SAA for this systemic value-at-risk. To that end, we propose mixed-binary linear and quadratic programming formulations for the weighted-sum and norm-minimizing scalarizations of the SAA, respectively. These formulations are embedded in the grid search algorithm in \cite{birgit} as well as a modified version of it proposed here.



In the last part of the paper, we undertake a comprehensive computational validation of our proposed methodologies. Drawing insights from \cite{bollobas2003directed} and \citet[Chapter 8]{hofstad_2016} on preferential attachment models, we employ this model detailed to generate network data. We model the random operating cash flow vectors via the Pareto distribution for its effective portrayal of heavy-tailed behavior, as indicated in \cite{cont2010network}. 

The rest of the main paper is organized as follows. In \Cref{sec:ENM}, we provide a general framework for systemic risk measures that is not particularly restricted to convex monetary risk measures. The core of the paper is \Cref{sec:theory}, where we prove all the convergence results for SAA. In \Cref{sec:systemicRM} we delve into the Eisenberg-Noe model, verify the assumptions of the general convergence results in this setting, and also provide mixed-integer programming formulations for computing the SAA sets. We present the computational results in \Cref{sec:comp}. Some additional theoretical results as well as all proofs are collected in the electronic companion at the end of the main text.

\section{Systemic Risk Measures} \label{sec:ENM}

We start by introducing some notation that will be used throughout the paper. Given real numbers $x,y\in\R$, we denote by $x \wedge y$ their minimum and by $x \vee y$ their maximum; we also define $x^{+} \coloneqq 0 \vee x$, and $x^{-} \coloneqq 0 \vee (-x)$. Let $d\in\N\coloneqq\{1,2,\dots\}$. We denote by $\R^d$ the $d$-dimensional Euclidean space whose elements $\bm{x}=(x_1,\ldots,x_d)^{\mathsf{T}}$ are expressed as column vectors. We define $\bm{1}_d$ as the vector of ones and $\bm{0}_d$ as the vector of zeros in $\mathbb{R}^d$; we define $\bm{I}_d$ as the $d\times d$ identity matrix. For each $i\in[d]$, we denote by $\bm{e}_{i,d}$ the $i^{\text{th}}$ standard unit vector in $\R^d$. Let $\bm{x}, \bm{y} \in \mathbb{R}^d$ and $\varepsilon\geq 0$. We write $|\bm{x}|_2$ for the $\ell^2$-norm of $\bm{x}$ and $\mathbb{B}(\bm{x},\varepsilon)$ denotes the closed $\ell^2$-ball centered at $\bm{x}$ with radius $\varepsilon$. We denote by $\bm{x} \wedge \bm{y}$ and $\bm{x} \vee \bm{y}$ the componentwise minimum and maximum of $\bm{x},\bm{y}$, respectively. We also write $\bm{x} \leq \bm{y}$ ($\bm{x} < \bm{y}$) whenever $x_i \leq y_i$ ($x_i < y_i$) for each $i\in[d]\coloneqq\{1,\ldots,d\}$. We also define $\mathbb{R}^{d}_+\coloneqq \{\bm{x}\in\R^d\mid \bm{0}_d\leq\bm{x}\}$ and $\mathbb{R}^{d}_{++}\coloneqq\{\bm{x}\in\R^d\mid \bm{0}_d<\bm{x}\}$. We say that a function $f \colon \R^d \to \R$ is increasing if $\bm{x}\leq \bm{y}$ implies $f(\bm{x})\leq f(\bm{y})$ for every $\bm{x},\bm{y}\in\R^d$; it is called strictly increasing if $\bm{x} \leq \bm{y}$ and $\bm{x} \neq \bm{y}$ imply $f(\bm{x}) < f(\bm{y})$ for every $\bm{x},\bm{y}\in\R^d$.

We denote by $\mathcal{B}(\R^d)$ the Borel $\sigma$-algebra on $\R^d$ and by $\Leb_d$ the Lebesgue measure on $(\R^d,\mathcal{B}(\R^d))$. Let $A,B\subseteq \R^d$ be sets. We define their Minkowski sum by $A+B\coloneqq\{\bm{x}+\bm{y}\mid \bm{x}\in A,\ \bm{y}\in B\}$, with the convention $A+\emptyset=\emptyset+A=\emptyset$, and write $\bm{x}+A\coloneqq \{\bm{x}\}+A$ for every $\bm{x}\in\R^d$. The interior, closure, boundary of $A$ are denoted by $\Int A, \cl A, \bd A$, respectively. We define the indicator function $1_A\colon \R^d\to \R$ of $A$ by $1_A(\bm{x})=1$ for $\bm{x}\in A$ and by $1_A(\bm{x})=0$ for $\bm{x}\in\R^d\setminus A$. Given a function $\varphi\colon\R^d\to\R\cup\{+\infty\}$ and parameters $\bm{w},\bm{v}\in\R^d$, we define
\[
\mathsf{u}(\varphi,A)\coloneqq \inf_{\bm{x}\in A}\varphi(\bm{x}),\quad \mathsf{s}(\bm{w},A)\coloneqq \inf_{\bm{x}\in A}\bm{w}^{\mathsf{T}}\bm{x},\quad \mathsf{d}(\bm{v},A)\coloneqq \inf_{\bm{x}\in A}|\bm{v}-\bm{x}|_2.
\]
The Hausdorff distance between $A$ and $B$ (with respect to the $\ell^2$-norm) is defined as $\mathsf{h}(A,B) \coloneqq  \bar{\mathsf{h}}(A,B)\vee \bar{\mathsf{h}}(B,A)$, where $\bar{\mathsf{h}}(A,B) \coloneqq \sup_{\bm{x} \in A}\mathsf{d}(\bm{x},B)$. Given a sequence $(A_N)_{N\in\N}$ of closed subsets of $\R^d$ and a nonempty closed set $A\subseteq \R^d$, we say that $(A_N)_{N\in\N}$ \emph{Wijsman-converges} to $A$ if $(\mathsf{d}(\bm{x},A_N))_{N\in\N}$ converges to $\mathsf{d}(\bm{x},A)$ for every $\bm{x}\in\R^d$; we say that $(A_N)_{N\in\N}$ \emph{Hausdorff-converges} to $A$ if $(\mathsf{h}(A_N,A))_{N\in \N}$ converges to $0$. It is well-known that Hausdorff convergence implies Wijsman convergence \citep[Corollary~3.65]{aliprantis06} and the converse also holds when $A,A_1,A_2,\ldots\subseteq\mathcal{Z}$ for some compact set $\mathcal{Z}\subseteq\R^d$ \citep[Theorem~2.5]{beer1987metric}.

Let us fix a probability space $(\Omega,\mathcal{F},\mathbb{P})$. We denote by $L^0(\R^d)$ the space of all $\mathcal{F}$-measurable random vectors $\bm{X}\colon \Omega\to\R^d$ that are identified up to $\mathbb{P}$-almost sure equality. Hence, (in)equalities between random variables are understood in the $\mathbb{P}$-almost sure sense. For each $\bm{X}\in L^0(\R^d)$, we define $\|\bm{X}\|_p:=(\mathbb{E}[|\bm{X}|_2^p])^{1/p}$ for $p\in[1,+\infty)$ and $\|\bm{X}\|_\infty:=\inf\{c>0\mid \mathbb{P}\{|\bm{X}|_2\leq c\}=1\}$. Then, for each $p\in[1,+\infty]$, the space $L^p(\R^d):=\{\bm{X}\in L^0(\R^d)\mid \|\bm{X}\|_p<+\infty\}$ is a Banach space with respect to the norm $\|\cdot \|_p$. A multifunction $F\colon \O\rightrightarrows\R^d$ is called $\mathcal{F}$-measurable if $\{\o\in\O\mid F(\o)\cap A\neq\emptyset\}\in\mathcal{F}$ for every closed set $A\subseteq \R^d$. A function $f\colon \R^d\times\O\to \R$ is called a \emph{normal integrand} if its epigraphical multifunction $\o\mapsto \{(\bm{x},\alpha)\in\R^d\times\R\mid f(\bm{x},\o)\leq\alpha\}$ is closed-valued and measurable.







In this section, we review the scalar-valued systemic risk measures introduced in \cite{iyengar} and the set-valued systemic risk measures introduced in \cite{birgit}. Since the expositions in these works and other related works, e.g., \citet{Biagini}, \citet{ararat2020dual}, \citet{menj}, assume either bounded random variables and/or finite-valued aggregation functions, we give a self-contained exposition tailored for the current paper. The proofs of all results are given in \Cref{app:ENM}.

\subsection{Aggregation Functions and Acceptance Sets}

Let us consider a financial network with $d\in\N$ institutions. The sample space $\O$ contains all possible scenarios that may affect the ability of the institutions to meet their obligations within the network. Hence, we model the operating cash flow vector of the network as a random vector $\bm{X}\in L^0(\R^d)$, where $X_i(\o)$ is understood as the future value of the external assets of institution $i\in[d]$ under scenario $\o\in\O$. For the definitions of the systemic risk measures, we fix a linear subspace $\X$ of $L^0(\R^d)$ that contains all deterministic random vectors, e.g., $\X=L^p(\R^d)$ with $p\in\{0\}\cup [1,+\infty]$, and consider only $\bm{X}\in \X$.

To define systemic risk measures, the realizations of the operating cash flow vector are aggregated into a scalar quantity by taking into account the structure of the network. This is achieved through an increasing function $\Lambda\colon \R^d \to \R\cup\{-\infty\}$, usually referred to as the \emph{aggregation function}. Given a realization $\bm{x}\in\R^d$ for the operating cash flow vector, the scalar $\Lambda(\bm{x})$ is a quantification of the impact of this cash flow profile on society. In this paper, most results will be shown by choosing $\Lambda$ based on an Eisenberg-Noe network, as we will define precisely in \Cref{sec:systemicRM}; in this case, $\Lambda$ will also be a concave function. Let us introduce effective domain of $\Lambda$ by
\[
\dom\Lambda \coloneqq \{\bm{x}\in\R^d\mid \Lambda(\bm{x})>-\infty\}
\]
and the collection of all random vectors that yield finite aggregate outcomes almost surely by
\[
\X_\Lambda\coloneqq\cb{\bm{X}\in \X\mid \mathbb{P}\{\bm{X}\in \dom \Lambda\}=1}.
\]
We work under the following assumption for $\Lambda$.

\begin{assumption}\label{asmp:dom}
	$\dom\Lambda$ is a nonempty closed subset of $\R^d$ and the restriction of $\Lambda$ on $\dom\Lambda$ is continuous and bounded.
\end{assumption}

Similar to $\X$, we fix a linear subspace $\Y\subseteq L^0(\R)$ that contains all deterministic random variables. Let $\rho\colon\Y\to\R$ be a monetary risk measure (see \citet[Section~4.1]{follmer}), that is, it satisfies the following two properties:
\begin{itemize}
	\item \textbf{Monotonicity}: $Y_1 \geq Y_2$ implies $\rho(Y_1) \leq \rho(Y_2)$ for every $Y_1,Y_2\in \Y$.
	\item \textbf{Translativity}: $\rho(Y+y) = \rho(Y)-y$ for every $Y\in \Y$, $y\in\R$.
\end{itemize}
The \emph{acceptance set} of $\rho$ is defined as
\[
\A:=\cb{Y\in \Y\mid \rho(Y)\leq 0}.
\]
Moreover, $\rho$ can be calculated using $\A$ via
\[
\rho(Y)=\inf\{y\in\R\mid Y+y\in \A\},\quad Y\in \Y.
\]
In this paper, most results will be shown by choosing $\rho$ as \emph{value-at-risk}, which is recalled in the next example.

\begin{example}\label{ex:var}
	Assume that $\Y=L^0(\R)$ and let $\alpha\in\R$, $\lambda\in(0,1)$. For each $Y\in \Y$, the value-at-risk of $Y$ at level $\lambda$ with threshold $\alpha$ is defined as
	\[
	\var_{\alpha,\lambda}(Y)\coloneqq\inf\{y\in\R\mid \mathbb{P}\{Y+y<\alpha\}\leq\lambda\}.
	\]
	It is easy to verify that the above infimum is attained, i.e., $\mathbb{P}\{Y+\var_{\alpha,\lambda}(Y)<\alpha\}\leq\lambda$. Clearly, $\var_{\alpha,\lambda}(Y)=\var_{0,\lambda}(Y)+\alpha$ and $\var_{0,\lambda}(Y)$ is called the value-at-risk of $Y$ at level $\lambda$ in the literature (see \citet[Section~4.4]{follmer}). We work with general $\alpha\in\R$ for convenience in the sequel. It is easy to verify that the corresponding acceptance set is given by
	\[
	\A_{\alpha,\lambda}\coloneqq \{Y\in \Y\mid \mathbb{P}\{Y<\alpha\}\leq \lambda\}.
	\]
\end{example}

We work under the following domain assumption that connects $\X_\Lambda$ and $\Y$:

\begin{assumption}\label{asmp:domain}
	For every $\bm{X}\in\X_\Lambda$, it holds $\Lambda(\bm{X})\in\Y$.
\end{assumption}

\subsection{Insensitive Systemic Risk Measures}

Using an aggregation function $\Lambda$ and an acceptance set $\A$ as primitives, in \cite{iyengar}, scalar-valued systemic risk measures of the form
\begin{equation}\label{eq:ins}
	r(\bm{X})\coloneqq\rho(\Lambda(\bm{X}))=\inf\{y\in\R\mid \Lambda(\bm{X})+y\in\A\},\quad \bm{X}\in\X_\Lambda,
\end{equation}
are studied. Here, $r(\bm{X})$ calculates the minimum capital requirement for the aggregate random variable $\Lambda(\bm{X})$ to be acceptable. Since the capital is allocated for the whole network after aggregation, such functionals are also called \emph{insensitive systemic risk measures} (see \cite{ararat2020dual}). The following property of $r$ is an immediate consequence of the monotonicity of $\Lambda$.

\begin{lemma}\label{lem:ins}
	Under \Cref{asmp:domain}, $\bm{X}^1 \geq \bm{X}^2$ implies $r(\bm{X}^1) \leq r(\bm{X}^2)$ for every $\bm{X}^1,\bm{X}^2\in\X_\Lambda$.
\end{lemma}

\subsection{Sensitive Systemic Risk Measures}

As a sensitive alternative to $r$ given by \eqref{eq:ins}, in \cite{birgit}, systemic risk measures are defined as set-valued functionals that assign to a random vector $\bm{X}\in \X$ all capital allocation vectors, assigned before aggregation, that yield acceptable aggregate outcomes. To keep the dimension of this set reasonably low compared to the size of the network, we formulate this systemic risk measure by using the concept of \emph{grouping} as discussed in \cite{birgit} and \cite{menj}. We assume that the financial institutions are partitioned into $g\in [n]$ nonempty groups, where the size of group $j\in [g]$ is denoted by $d_j$. In particular, $\sum_{j=1}^g d_j=d$. The institutions within each group will have an identical designated capital level. For this purpose, we introduce a binary matrix $\bm{B}\in \{0,1\}^{g\times d}$, called the \emph{grouping matrix}, where $B_{ji}=1$ if and only if institution $i\in[d]$ belongs to group $j\in[g]$. Then, for a capital allocation vector $\bm{z}\in\R^g$ for groups, the actual capital allocation vector for the network is calculated as $\bm{B}^{\mathsf{T}}\bm{z}\in\R^d$. 

\begin{example}
	The case of $g=2$ groups is suitable for core-periphery networks. Let us write $\mathcal{N}_{c}=[d_1]$ and $\mathcal{N}_p=\{d_1+1,\ldots,d\}$ for the sets of core and peripheral banks, respectively. Then, we may write the grouping matrix as
	\[
	\bm{B} = [\bm{B}_{c} \ \bm{B}_{p}],
	\]
	where $\bm{B}_{c}\coloneqq [\bm{1}_{d_1}\  \bm{0}_{d_1}]^{\mathsf{T}}\in\{0,1\}^{2\times d_1}$ and $\bm{B}_{p}\coloneqq [\bm{0}_{d-d_1}\ \bm{1}_{d-d_1}]^{\mathsf{T}}\in\{0,1\}^{2\times(d-d_1)}$. For instance, when $d_1=2$ and $d=6$, we have $\bm{B}^{\mathsf{T}}\bm{z}=(z_1,z_1,z_2,z_2,z_2,z_2)^{\mathsf{T}}$ for every $\bm{z}\in\R^2$. 
\end{example}

With the structure above, the set-valued systemic risk measure proposed in \cite{birgit} is defined as
\[
R(\bm{X})\coloneqq \{\bm{z}\in \R^g\mid \Lambda(\bm{X}+\bm{B}^{\mathsf{T}}\bm{z})\in\A,\ \bm{X}+\bm{B}^{\mathsf{T}}\bm{z} \in \X_\Lambda \},\quad \bm{X}\in\X.
\]
Thanks to the monotonicity of $\Lambda$, the set-valued functional $R$ has properties that are analogous to the properties of $\rho$ as a risk measure. These are given in the next lemma and they make $R$ an example of a set-valued risk measure in the sense of \cite{hh,hamel2011set}.

\begin{lemma}\label{lem:systrisk}
	Under \Cref{asmp:domain}, $R$ has the following properties:
	\begin{itemize}
		\item \textbf{Monotonicity}: $\bm{X}^1 \geq \bm{X}^2$ implies $R(\bm{X}^1) \supseteq R(\bm{X}^2)$ for every $\bm{X}^1,\bm{X}^2\in\X$.
		\item \textbf{Translativity}: $R(\bm{X}+\bm{B}^{\mathsf{T}}\bm{z}) = R(\bm{X})-\bm{z}$ for every $\bm{X}\in \X$, $\bm{z}\in\R^g$.
		\item \textbf{Upper set property:} $R(\bm{X})=R(\bm{X})+\R^g_+$ for every $\bm{X}\in\X$.
	\end{itemize}
\end{lemma}

\subsection{Scalarizations of Sensitive Systemic Risk Measures}

Given $\bm{X}\in\X$, the set $R(\bm{X})$ can be seen as the so-called \emph{upper image} of the vector optimization problem
\[
\text{minimize }\bm{z}\text{ with respect to }\R^g_+\text{ subject to }\Lambda(\bm{X}+\bm{B}^{\mathsf{T}}\bm{z})\in\A,\  \bm{X}+\bm{B}^{\mathsf{T}}\bm{z}\in\X_\Lambda,\ \bm{z}\in\R^g,
\]
and solving this problem (approximately) typically involves solving some scalar optimization problems, called scalarizations, associated to it. A general \emph{scalarization} problem can be formulated as
\[
\text{minimize }\varphi(\bm{z})\text{ subject to }\Lambda(\bm{X}+\bm{B}^{\mathsf{T}}\bm{z})\in\A,\  \bm{X}+\bm{B}^{\mathsf{T}}\bm{z}\in\X_\Lambda,\ \bm{z}\in\R^g,
\]
where $\varphi\colon \R^g\to \R\cup\{+\infty\}$ is a given function, its optimal value is given by
\[
\mathsf{u}(\varphi,R(\bm{X}))= \inf_{\bm{z}\in R(\bm{X})}\varphi(\bm{z})
\]
In this paper, we are interested in two types of scalarizations. The first is the well-known \emph{weighted sum scalarization}, which is formulated as
\[
\text{minimize }\bm{w}^{\mathsf{T}}\bm{z}\text{ subject to }\Lambda(\bm{X}+\bm{B}^{\mathsf{T}}\bm{z})\in\A,\  \bm{X}+\bm{B}^{\mathsf{T}}\bm{z}\in\X_\Lambda,\ \bm{z}\in\R^g
\]
for a given weight parameter $\bm{w}\in\R^g_+\setminus\{\bm{0}_g\}$, and its optimal value is given by
\[
\mathsf{s}(\bm{w},R(\bm{X}))= \inf_{\bm{z}\in R(\bm{X})}\bm{w}^{\mathsf{T}}\bm{z}.
\]
The second one is the \emph{norm-minimizing scalarization}, which is formulated as
\[
\text{minimize }|\bm{z}-\bm{v}|_2\text{ subject to }\Lambda(\bm{X}+\bm{B}^{\mathsf{T}}\bm{z})\in\A,\  \bm{X}+\bm{B}^{\mathsf{T}}\bm{z}\in\X_\Lambda,\ \bm{z}\in\R^g
\]
for a given reference point $\bm{v}\in \R^g$, and its optimal value is given by
\[
\mathsf{d}(\bm{v},R(\bm{X}))=\inf_{\bm{z}\in R(\bm{X})}|\bm{z}-\bm{v}|_2.
\]

\section{Systemic Values-at-Risk and Their Sample-Average Approximations}\label{sec:theory}

In this paper, we study systemic risk measures with $\rho$ taken as value-at-risk (see \Cref{ex:var}). For convenience, we rename them as systemic values-at-risk in the next definition. Throughout this subsection, we fix an aggregation function $\Lambda$ that satisfies \Cref{asmp:dom} and \Cref{asmp:domain}. We also fix $\alpha\in\R$ and $\lambda\in (0,1)$. The proofs of all results are given in \Cref{app:theory}.

\begin{definition}\label{Set-valued}
	(i) Let $\bm{X}\in\X_\Lambda$. The \textbf{insensitive systemic value-at-risk} of $\bm{X}$ at level $\lambda$ with threshold $\alpha$ is defined as
	\begin{align}
		r_{\alpha,\lambda}(\bm{X})\coloneqq & \inf\{y\in\R\mid \Lambda(\bm{X})+y\in\A_{\alpha,\lambda}\}\notag \\
		=&\inf\{y\in\R\mid \mathbb{P}\{\Lambda(\bm{X})+y<\alpha\}\leq \lambda\}.\notag 
	\end{align}
	(ii) Let $\bm{X}\in\X$. The \textbf{sensitive systemic value-at-risk} of $\bm{X}$ at level $\lambda$ with threshold $\alpha$ is defined as 
	\begin{align}
		R_{\alpha,\lambda}(\bm{X})\coloneqq& \{\bm{z}\in\R^g\mid \Lambda(\bm{X}+\bm{B}^{\mathsf{T}}\bm{z})\in \A_{\alpha,\lambda},\ \bm{X}+\bm{B}^{\mathsf{T}}\bm{z}\in\X_\Lambda\}
		\notag \\
		=&\{\bm{z}\in\R^g\mid \mathbb{P}\{\Lambda(\bm{X}+\bm{B}^{\mathsf{T}}\bm{z})<\alpha\}\leq\lambda,\ \mathbb{P}\{\bm{X}+\bm{B}^{\mathsf{T}}\bm{z}\in\dom \Lambda\}=1\}.\notag 
	\end{align}
\end{definition}

The calculations of $r_{\alpha,\lambda}(\bm{X})$ and $R_{\alpha,\lambda}(\bm{X})$ involve solving a chance-constrained optimization problem due to the probabilistic constraints in their definitions. When $\O$ is an infinite sample space, these constraints can be expressed in terms of multivariate integrals, which are difficult to deal with in an optimization problem, except for some special distribution classes for $\bm{X}$. In this section, we introduce sample-average approximations (SAA) for the sensitive value-at-risk and prove their almost sure convergence to its true value. Each SAA problem is essentially a special case of the true problem in which the random vector is defined on a finite probability space and, as we will do later, it can be reformulated as a mixed-integer programming problem using binary variables. We limit the scope of the main text to the sensitive case, which is more interesting from a capital allocation perspective and also more challenging mathematically due to the appearance of set-valued functions. In \Cref{app:ins}, we prove a convergence result for the insensitive systemic value-at-risk.

%

\Cref{prop:technical} below is a generalization of \citet[Proposition~2.1]{Shapiro} and it will be used repeatedly in the sequel as a technical tool for the convergence arguments. Before stating it, we recall the notion of epi-convergence for scalar-valued functions.

\begin{definition}\label{defn:syst}
\citep[Definition~7.1, Proposition~7.2]{rockafellar2009variational} Let $f,f^1,f^2,\ldots$ be extended real-valued functions on $\R^m$, where $m\in\N$. We say that the sequence $(f^N)_{N\in\N}$ \emph{epi-converges} to $f$ if the following conditions hold for every $\bm{y}\in\R^m$:
\begin{enumerate}[(i)]
\item For every sequence $(\bm{y^N})_{N \in \mathbb{N}}$ in $\R^m$ converging to $\bm{y}$, it holds $\liminf_{N \rightarrow \infty} f^N(\bm{y}^N) \geq f(\bm{y})$.
\item There exists a sequence $(\bm{y^N})_{N \in \mathbb{N}}$ in $\R^m$ converging to $\bm{y}$ such that $\limsup_{N \to \infty} f^N(\bm{y}^N) \leq f(\bm{y})$.
\end{enumerate}
\end{definition}

\begin{proposition}\label{prop:technical}
Let $\bm{X}\in \X$ with a sequence $(\bm{X}^N)_{N\in\N}$ of independent copies. Let $D\subseteq[-\infty,+\infty]$ ($D\subseteq \R^{d^\prime}$ with $d^{\prime}\in\N$) be an open set. Let $g\colon \R^m\times\R^d\to [-\infty,+\infty]$ ($g\colon \R^m\times\R^d\to \R^{d^\prime}$) be a Carath\'{e}odory function, where $m\in\N$. For each $N\in\N$, $\bm{x}^1,\ldots,\bm{x}^N\in\R^d$, and $\bm{y}\in\R^m$, let
\[
p^N(\bm{y};\bm{x}^1,\ldots,\bm{x}^N)\coloneqq \frac1N\sum_{n=1}^N 1_D(g(\bm{y},\bm{x}^n)),\quad p(\bm{y})\coloneqq \mathbb{P}\{g(\bm{y},\bm{X})\in D\}.
\]
Then, the following properties hold:
\begin{enumerate}[(i)]
\item For each $N\in\N$, the function $p^N$ is a normal integrand.
\item The function $p$ is lower semicontinuous.
\item The sequence $(p^N(\cdot;\bm{X}^1,\ldots,\bm{X}^N))_{N\in\N}$ epi-converges to $p(\cdot)$ almost surely.
\end{enumerate}
\end{proposition}

As an application of \Cref{prop:technical}, we verify the closed-valuedness of the sensitive systemic value-at-risk in the next lemma.

\begin{lemma}\label{lem:closed}
	For every $\bm{X}\in\X$, the set $R_{\alpha,\lambda}(\bm{X})$ is closed; in particular, it holds $R_{\alpha,\lambda}(\bm{X})=\cl(R_{\alpha,\lambda}(\bm{X})+\R^g_+)$.
\end{lemma}

\subsection{SAA for the Sensitive Systemic Value-at-Risk}\label{sec:SAA}

Let us fix a random vector $\bm{X}\in\X$. By passing to the countably infinite product space $(\O,\mathcal{F},\mathbb{P})^{\N}$ if necessary, we assume the existence of a sequence $(\bm{X}^N)_{N\in\N}$ of independent copies of $\bm{X}$. 

Let $N\in\N$. For each $\bm{x}^1,\ldots,\bm{x}^N\in\R^d$, we define
\[
R^N_{\alpha,\lambda}(\bm{x}^1,\ldots,\bm{x}^N)\coloneqq \cb{\bm{z}\in\R^g\mid \frac{1}{N}\sum_{n=1}^N 1_{[-\infty,\alpha)}(\Lambda(\bm{x}^n+\bm{B}^{\mathsf{T}}\bm{z}))\leq\lambda,\ \forall n\in[N]\colon \bm{x}^n+\bm{B}^{\mathsf{T}}\bm{z}\in \dom\Lambda}.
\]
Note that this is a special case of the sensitive systemic risk measure for a random vector whose support is $\{\bm{x}^1,\ldots,\bm{x}^N\}$. Unlike $r^N_{\alpha,\lambda}$ in the insensitive case, the calculation of $R^N_{\alpha,\lambda}$ is not straightforward and the difficulty of the calculation depends largely on the structure of the aggregation function. In \Cref{sec:comp}, we will discuss an algorithm for calculating $R^N_{\alpha,\lambda}$ when the aggregation function is based on the Eisenberg-Noe model.

We check the set-valued measurability of $R^N_{\alpha,\lambda}$ in the next lemma.

\begin{lemma}\label{lem:meas-sen}
Let $N\in\N$. The multifunction $R^N_{\alpha,\lambda}\colon \R^{dN}\rightrightarrows\R^g$ is closed-valued and measurable.
\end{lemma}

\subsubsection{Convergence of a General Scalarization}\label{sec:gen}

Thanks to \Cref{lem:meas-sen}, composing $R^N_{\alpha,\lambda}$ with the first $N$ independent copies of $\bm{X}$ yields the random set $R^N_{\alpha,\lambda}(\bm{X}^1,\ldots,\bm{X}^N)$. To prove the convergence of this random set to the true set $R_{\alpha,\lambda}(\bm{X})$ in various senses, we will first prove a convergence result for their scalarizations. To that end, let $\varphi\colon \R^g\to \R\cup\{+\infty\}$ be a function and $\mathcal{Z}\subseteq\R^g$ be a set. For each $\bm{x}^1,\ldots,\bm{x}^N\in\R^g$, we consider the value
\[
\mathsf{u}(\varphi,R^N_{\alpha,\lambda}(\bm{x}^1,\ldots,\bm{x}^N)\cap\mathcal{Z})=\inf\cb{\varphi(\bm{z})\mid \bm{z}\in R^N_{\alpha,\lambda}(\bm{x}^1,\ldots,\bm{x}^N)\cap \mathcal{Z}}.
\]

\begin{lemma}\label{lem:meas-sc}
Let $N\in\N$. Suppose that $\varphi$ is a continuous function and $\mathcal{Z}$ is a closed set. Then, the function $(\bm{x}^1,\ldots,\bm{x}^N) \mapsto \mathsf{u}(\varphi,R^N_{\alpha,\lambda}(\bm{x}^1,\ldots,\bm{x}^N)\cap\mathcal{Z})$ is  measurable.
\end{lemma}

Thanks to \Cref{lem:meas-sc}, $\mathsf{u}(\varphi,R^N_{\alpha,\lambda}(\bm{X}^1,\ldots,\bm{X}^N)\cap\mathcal{Z})$ is a random variable. We also consider the true value
\[
\mathsf{u}(\varphi,R_{\alpha,\lambda}(\bm{X})\cap\mathcal{Z})=\inf\cb{\varphi(\bm{z})\mid \bm{z}\in R_{\alpha,\lambda}(\bm{X})\cap\mathcal{Z}}.
\]
For the convergence results to follow, we work under the following assumption, which asserts that the above scalarization problem has an optimal solution that can be approximated by a sequence of feasible solutions that are strictly feasible in the value-at-risk constraint.

\begin{assumption}\label{asmp:slater}
There exist $\bar{\bm{z}}\in R_{\alpha,\lambda}(\bm{X})$ such that $\varphi(\bar{\bm{z}})=\mathsf{u}(\varphi,R_{\alpha,\lambda}(\bm{X})\cap\mathcal{Z})$, and a sequence $(\bm{z}^\ell)_{\ell\in\N}$ in $\mathcal{Z}$ that converges to $\bar{\bm{z}}$ and satisfies  $\mathbb{P}\{\Lambda(\bm{X}+\bm{B}^{\mathsf{T}}\bm{z}^\ell)<\alpha\}<\lambda$, $\mathbb{P}\{\bm{X}+\bm{B}^{\mathsf{T}}\bm{z}^\ell\in\dom\Lambda\}=1$ for every $\ell\in\N$.
\end{assumption}

\begin{theorem}\label{thm:generalconv}
Under \Cref{asmp:slater}, suppose that $\varphi$ is continuous, and $\mathcal{Z}$ is  nonempty and compact. Then, the sequence $(\mathsf{u}(\varphi,R^N_{\alpha,\lambda}(\bm{X}^1,\ldots,\bm{X}^N)\cap\mathcal{Z}))_{N\in\N}$ converges to $\mathsf{u}(\varphi,R_{\alpha,\lambda}(\bm{X})\cap\mathcal{Z})$ almost surely.
\end{theorem}


\subsubsection{Convergence of a Weighted-Sum Scalarization}\label{sec:scalar}

Weighted-sum scalarizations for systemic risk measures are useful when it is desirable to find the minimum total capital allocation required for the network or a subset of it. For instance, with $\bm{w}=(d_1,\ldots,d_g)^{\mathsf{T}}$, the value $\mathsf{s}(\bm{w},R_{\alpha,\lambda}(\bm{X}))$ gives the minimum total capital requirement of the entire network; given a group $j\in[g]$, the value $\mathsf{s}(\bm{e}_{j,g},R_{\alpha,\lambda}(\bm{X}))$ gives the minimum total capital requirement of each bank in group $j$. These scalarizations are also used in algorithms for calculating systemic risk measures. For instance, assuming the finiteness of the scalarizations, the vector $\bm{z}^{\text{ideal}}\coloneqq (\mathsf{s}(\bm{e}_{1,g},R_{\alpha,\lambda}(\bm{X})),\ldots, \mathsf{s}(\bm{e}_{g,g},R_{\alpha,\lambda}(\bm{X})))^{\mathsf{T}}$ is called the \emph{ideal point} of $R_{\alpha,\lambda}(\bm{X})$ and it can be used to construct an initial outer approximation for $R_{\alpha,\lambda}(\bm{X})$ since we have $R_{\alpha,\lambda}(\bm{X})\subseteq \bm{z}^{\text{ideal}}+\R^g_+$.

The result of this section states that the weighted-sum scalarizations of a restricted version of the SAA converges to that of the original systemic risk measure under the following assumption. This assumption will later be verified for the Eisenberg-Noe model in \Cref{sec:asmp}.

\begin{assumption}\label{asmp:scalar}
Given $\bm{w}\in\R^g_+\setminus\{\bm{0}_g\}$ and a compact set $\mathcal{Z}\subseteq\R^g$ such that $R_{\alpha,\lambda}(\bm{X})=R_{\alpha,\lambda}(\bm{X})\cap\mathcal{Z}+\R^g_+$, there exist $\bar{\bm{z}}\in R_{\alpha,\lambda}(\bm{X})$ such that $\bm{w}^{\mathsf{T}}\bar{\bm{z}}=\mathsf{s}(\bm{w},R_{\alpha,\lambda}(\bm{X}))$, and a sequence $(\bm{z}^\ell)_{\ell\in\N}$ in $\mathcal{Z}$ that converges to $\bar{\bm{z}}$ and satisfies  $\mathbb{P}\{\Lambda(\bm{X}+\bm{B}^{\mathsf{T}}\bm{z}^\ell)<\alpha\}<\lambda$, $\mathbb{P}\{\bm{X}+\bm{B}^{\mathsf{T}}\bm{z}^\ell\in\dom\Lambda\}=1$ for every $\ell\in\N$.
\end{assumption}

\begin{theorem}\label{thm:scalar}
Let $\bm{w}\in\R^g_+\setminus\{\bm{0}_g\}$. Assume that there exists a compact set $\mathcal{Z}\subseteq\R^g$ such that 
\[
R_{\alpha,\lambda}(\bm{X})=R_{\alpha,\lambda}(\bm{X})\cap\mathcal{Z}+\R^g_+
\]
and \Cref{asmp:scalar} holds for $\mathcal{Z}$. Then, the sequence $(\mathsf{s}(\bm{w},R^N_{\alpha,\lambda}(\bm{X}^1,\ldots,\bm{X}^N))\cap\mathcal{Z})_{N\in\N}$ converges to $\mathsf{s}(\bm{w},R_{\alpha,\lambda}(\bm{X}))$ almost surely.
\end{theorem}

\subsubsection{Wijsman Convergence of the Sensitive Value-at-Risk}\label{sec:wijsman}


Due to well-known lack of convexity for value-at-risk, the set $R_{\alpha,\lambda}(\bm{X})$ may fail to be convex in general. Hence, its weighted-sum scalarizations do not provide sufficient information to characterize the entire set. Hence, to study the convergence behavior of SAA as sets, we study the convergence of the distance functions of SAA in this subsection. This yields the Wijsman convergence of SAA.

For the results of this section, we will work under the following additional assumption on $\Lambda$.

\begin{assumption}\label{asmp:dom2}
$\dom\Lambda$ is bounded with respect to $\R^d_+$, i.e., there exists $\bm{x}^{\textup{LB}}\in\R^d$ such that $\dom\Lambda\subseteq \bm{x}^{\textup{LB}}+\R^d_+$.
\end{assumption}

The boundedness property in \Cref{asmp:dom2} is automatically transferred to the sensitive systemic value-at-risk as the next lemma shows.

\begin{lemma}\label{lem:bdd}
Suppose that \Cref{asmp:dom2} holds and $\bm{X}\in\X_\Lambda$. Then, $R_{\alpha,\lambda}(\bm{X})$ is bounded with respect to $\R^g_+$, i.e., there exists $\bm{z}^{\textup{LB}}\in\R^g$ such that $\bm{z}^{\textup{LB}}\notin R_{\alpha,\lambda}(\bm{X})$ and $R_{\alpha,\lambda}(\bm{X})\subseteq \bm{z}^{\textup{LB}}+\R^g_+$.
\end{lemma}

\begin{remark}
	As discussed in \Cref{sec:scalar}, the finiteness of the weighted-sum scalarizations along the standard unit vectors also guarantees that $R_{\alpha,\lambda}(\bm{X})$ is bounded with respect to $\R^g_+$.
	\end{remark}
	
\begin{remark}\label{rem:Motzkin}
	In general, the boundedness of a closed set $A\subseteq \R^g$ with respect to $\R^g_+$ is weaker than its Motzkin-decomposability, i.e., the existence of a compact set $\mathcal{Z}\subseteq\R^g$ such that $A=A\cap\mathcal{Z}+\R^g_+$. For instance, the set $A\coloneqq \{\bm{z}\in\R^g_{++}\mid y_1y_2\geq 1\}$ is bounded with respect to $\R^g_+$ but it is not Motzkin-decomposable; see \citet[Section~1]{lohneMotzkin}.
	\end{remark}

The following is the analogous version of \Cref{asmp:scalar} for distance functions.

\begin{assumption}\label{asmp:wijsman}
Given $\bm{z}^{\textup{LB}}\in\R^g$ such that $R_{\alpha,\lambda}(\bm{X})\subseteq \bm{z}^{\textup{LB}}+\R^g_+$ and $\bm{z}^{\textup{LB}}\notin R_{\alpha,\lambda}(\bm{X})$, let $\varepsilon\coloneqq \mathsf{d}(\bm{z}^{\textup{LB}},R_{\alpha,\lambda}(\bm{X}))>0$. For every $\bm{v}\in\bm{z}^{\textup{LB}}+\R^g_+$, there exist $\bm{z}^{\bm{v}}\in R_{\alpha,\lambda}(\bm{X})$ such that $|\bm{v}-\bm{z}^{\bm{v}}|_2=\mathsf{d}(\bm{v},R_{\alpha,\lambda}(\bm{X}))$, and a sequence $(\bm{z}^{\bm{v},\ell})_{\ell\in\N}$ in $\mathbb{B}(\bm{v},2\varepsilon)$ that converges to $\bm{z}^{\bm{v}}$ and satisfies  $\mathbb{P}\{\Lambda(\bm{X}+\bm{B}^{\mathsf{T}}\bm{z}^{\bm{v},\ell})<\alpha\}<\lambda$, $\mathbb{P}\{\bm{X}+\bm{B}^{\mathsf{T}}\bm{z}^{\bm{v},\ell}\in\dom\Lambda\}=1$ for every $\ell\in\N$.
\end{assumption}

The next theorem is one of the main results of the paper and it establishes the Wijsman convergence of SAA for a sensitive systemic value-at-risk that is bounded with respect to $\R^g_+$.

\begin{theorem}\label{thm:wijsman}
Assume that $\bm{X}\in\X_{\Lambda}$. Suppose that \Cref{asmp:dom2} and \Cref{asmp:wijsman} hold for $\bm{z}^{\textup{LB}}\in\R^g$ given by \Cref{lem:bdd}. Then, the sequence $(R_{\alpha,\lambda}^N(\bm{X}^1,\ldots,\bm{X}^N))_{N\in\N}$ Wijsman-converges to $R_{\alpha,\lambda}(\bm{X})$ almost surely.
\end{theorem}

\subsubsection{Hausdorff Convergence of the Sensitive Systemic Value-at-Risk}

For the final general result of the paper, we study the Hausdorff convergence of SAA to the sensitive systemic value-at-risk. This convergence type is often preferred for being naturally generated by the Hausdorff distance $\mathsf{h}$, which is a metric with values in $[0,+\infty]$.

As a preparation for the convergence result, we state some useful properties of distance functions in the next two lemmata.

\begin{lemma}\label{lem:dopt}
Assume that $R_{\alpha,\lambda}(\bm{X})\neq\emptyset$ and let $\bm{v}\in \R^g$. Then, there exists $\bm{z}^\ast\in R_{\alpha,\lambda}(\bm{X})$ such that
\[
|\bm{v}-\bm{z}^\ast|_2=\mathsf{d}(\bm{v},R_{\alpha,\lambda}(\bm{X})).
\]
Moreover, such $\bm{z}^\ast$ satisfies $\bm{v}\leq \bm{z}^\ast$.
\end{lemma}

\begin{lemma}\label{lem:d}
Assume that $R_{\alpha,\lambda}(\bm{X})\neq\emptyset$ and there exist $\bm{z}^{\textup{LB}},\bm{z}^{\textup{UB}}\in\R^g$ with $\bm{z}^{\textup{LB}}\leq \bm{z}^{\textup{UB}}$ such that 
\begin{equation}\label{eq:cpc}
R_{\alpha,\lambda}(\bm{X})=R_{\alpha,\lambda}(\bm{X})\cap\mathcal{Z}+\R^g_+,
\end{equation}
where $\mathcal{Z}\coloneqq [\bm{z}^{\textup{LB}},\bm{z}^{\textup{UB}}]$. Let $\bm{v}\in\mathcal{Z}$. Then, it holds
\[
\mathsf{d}(\bm{v},R_{\alpha,\lambda}(\bm{X}))=\mathsf{d}(\bm{v},R_{\alpha,\lambda}(\bm{X})\cap \mathcal{Z}).
\]
\end{lemma}

As noted in \Cref{sec:ENM}, Hausdorff convergence is stronger than Wijsman convergence, and the two convergence types coincide for closed subsets of a compact set. To obtain Hausdorff convergence, we will work under the Motzkin-decomposability assumption as in \Cref{sec:scalar}; see \Cref{rem:Motzkin} as well. The next assumption is a modification of \Cref{asmp:wijsman} for our current purpose.

\begin{assumption}\label{asmp:hausdorff}
Given a compact set $\mathcal{Z}\subseteq\R^g$ such that $R_{\alpha,\lambda}(\bm{X})=R_{\alpha,\lambda}(\bm{X})\cap\mathcal{Z}+\R^g_+$, for every $\bm{v}\in\mathcal{Z}$, there exist $\bar{\bm{z}}\in R_{\alpha,\lambda}(\bm{X})$ such that $|\bm{v}-\bar{\bm{z}}|_2=\mathsf{d}(\bm{v},R_{\alpha,\lambda}(\bm{X}))$, and a sequence $(\bm{z}^\ell)_{\ell\in\N}$ in $\mathcal{Z}$ that converges to $\bar{\bm{z}}$ and satisfies  $\mathbb{P}\{\Lambda(\bm{X}+\bm{B}^{\mathsf{T}}\bm{z}^\ell)<\alpha\}<\lambda$, $\mathbb{P}\{\bm{X}+\bm{B}^{\mathsf{T}}\bm{z}^\ell\in\dom\Lambda\}=1$ for every $\ell\in\N$.
\end{assumption}

\begin{remark}
	It is easy to check that \Cref{asmp:hausdorff} implies \Cref{asmp:wijsman}.
	\end{remark}

\begin{theorem}\label{thm:hausdorff}
Assume that there exist $\bm{z}^{\textup{LB}},\bm{z}^{\textup{UB}}\in\R^g$ with $\bm{z}^{\textup{LB}}\leq \bm{z}^{\textup{UB}}$ such that 
\[
R_{\alpha,\lambda}(\bm{X})=R_{\alpha,\lambda}(\bm{X})\cap\mathcal{Z}+\R^g_+
\]
and
\[
R^N_{\alpha,\lambda}(\bm{X}^1,\ldots,\bm{X}^N)=R^N_{\alpha,\lambda}(\bm{X}^1,\ldots,\bm{X}^N)\cap\mathcal{Z}+\R^g_+
\]
almost surely for every $N\in\N$, where $\mathcal{Z}\coloneqq [\bm{z}^{\textup{LB}},\bm{z}^{\textup{UB}}]$. Suppose that \Cref{asmp:hausdorff} holds for this choice of $\mathcal{Z}$. Then, the sequence $(R^N_{\alpha,\lambda}(\bm{X}^1,\ldots,\bm{X}^N))_{N\in\N}$ Wijsman- and Hausdorff-converges to $R_{\alpha,\lambda}(\bm{X})$ almost surely.
\end{theorem}

\section{Sensitive Systemic Value-at-Risk for Eisenberg-Noe Model}\label{sec:systemicRM}

This section serves as a case study of the general theory established in \Cref{sec:theory}. We consider the Eisenberg-Noe network model, which introduces a mechanism for clearing the interbank liabilities within the network. Based on this mechanism, we define an aggregation function, and study the corresponding sensitive systemic value-at-risk as well as its SAA. One of the crucial tasks is to verify the assumptions on the existence of strictly feasible approximating sequences for optimizers, e.g., \Cref{asmp:scalar} and \Cref{asmp:hausdorff}, for this model. It turns out that this is a highly nontrivial point and it can be addressed by a deeper analysis of the aggregation function as will be done in the proof of \Cref{thm:ENscalar}.

Unless stated otherwise, the proofs of all results are given in \Cref{app:systemicRM}.

\subsection{Eisenberg-Noe Model}

In this subsection, we recall the network model in \cite{enberg} and introduce its aggregation function to be used in the definition of the sensitive systemic value-at-risk. We consider a financial network with $d$ nodes representing the member institutions.

\begin{definition} \citep[Section 2.2]{enberg}
A triplet $(\bm{\pi}, \bar{\bm{p}}, \bm{x})$ is called an Eisenberg-Noe network with $d$ institutions if $\bm{\pi}=(\pi_{ij})_{i,j\in[d]}\in \R^{d\times d}$ is a right stochastic matrix with $\pi_{ii}=0$ for every $i\in[d]$, $\bar{\bm{p}}\in\R^d_{++}$, and $\bm{x}\in\R^d_+$.
\end{definition}

In the above definition,
\begin{itemize}
\item $\bar{p}_i$ denotes the total amount of liabilities of node $i\in [d]$ and we refer to $\bar{\bm{p}}$ as the total obligation vector;
\item $\pi_{ij}$ denotes the fraction of total liabilities of node $i\in[d]$ owed to node $j\in[d]$ and we refer to $\bm{\pi}$ as the relative liability matrix; note that $\sum_{j=1}^{d} \pi_{ji} < d$ holds for each $i \in [d]$, i.e., not all claims are owed a single node;
\item $x_i$ denotes the total value of the assets of node $i\in[d]$ and we refer to $\bm{x}$ as the operating cash flow vector.
\end{itemize}


Let us fix an Eisenberg-Noe network $(\bm{\pi}, \bar{\bm{p}}, \bm{x})$. Next, we recall the notion of a clearing (payment) vector, which adheres to the principles of limited liability and absolute priority.

\begin{definition} \citep[Definition~1]{enberg}
A vector $\bm{p} \in [\bm{0}_d, \bar{\bm{p}}]$ is called a \textbf{clearing vector} for $(\bm{\pi}, \bar{\bm{p}}, \bm{x})$ if it satisfies the following properties:
\begin{itemize}
\item \textbf{Limited liability:} For each $i\in[d]$, it holds $p_i \leq \sum_{j=1}^{d}\pi_{ji}p_j + x_i$.
\item \textbf{Absolute priority:} For each $i \in [d]$, either $p_i = \bar{p_i}$ or $p_i = \sum_{j=1}^{d}\pi_{ji}p_j + x_i$ holds.
\end{itemize}
\end{definition}

Limited liability indicates that a node cannot pay more than it has and absolute priority means that each node either meets its obligations in full or it defaults by paying as much as it has. It is easy to see that a clearing vector $\bm{p} \in [\bm{0}_d, \bar{\bm{p}}]$ is equivalently characterized as a solution of the fixed point equation $\bm{p}=\bm{\Phi}(\bm{p})$, where $\bm{\Phi}\colon [\bm{0}_d, \bar{\bm{p}}]\to  [\bm{0}_d, \bar{\bm{p}}]$ is defined by
\begin{equation}\label{eq:fixed}
\bm{\Phi}(\bm{p}) \coloneqq  (\bm{\pi}^{\mathsf{T}}\bm{p} + \bm{x}) \wedge \bar{\bm{p}},\quad \bm{p}\in [\bm{0}_d,\bar{\bm{p}}].
\end{equation}
According to \citet[Theorem~1]{enberg}, with respect to the componentwise order $\leq$, a maximal and minimal clearing vector always exist. Moreover, the maximal clearing vector can be found by solving an optimization problem as recalled below.


\begin{lemma} \citep[Lemma 6]{enberg} \label{lem:mp}
Let $f\colon \R^d\to\R$ be a strictly increasing function and consider the optimization problem
\begin{equation} \label{LP}
\textup{maximize}\quad f(\bm{p})\quad \textup{subject to} \quad \bm{p} \leq \bm{\pi}^{\mathsf{T}}\bm{p} + \bm{x},\quad	\bm{p} \in [\bm{0}_d,\bar{\bm{p}}].	
\end{equation}
If $\bm{p}^\ast$ is an optimal solution to \eqref{LP}, then it is a clearing vector for $(\bm{\pi}, \bar{\bm{p}}, \bm{x})$.
\end{lemma}

In \Cref{app:all}, we prove a novel result that nicely complements \Cref{lem:mp}: we show that the set of \textbf{all} clearing vectors of an Eisenberg-Noe network can be seen as a mixed-integer linear \emph{projection problem}.

In \Cref{lem:mp}, when we take $f(\bm{p}) = \bm{1}_d^{\mathsf{T}}\bm{p}$ for each $\bm{p}\in\R^d$, the optimal value of \eqref{LP} yields the total payment made at clearing. We define this quantity as an aggregation function of the Eisenberg-Noe model. More precisely, let us define $\Lambda^{\text{EN}} \colon \mathbb{R}^d \to \mathbb{R} \cup \{-\infty\}$ by
\begin{equation}\label{primal}
\Lambda^{\EN}(\bm{x}):=
\begin{cases}
\sup \{ \bm{1}_d^{\mathsf{T}}\bm{p} \mid \bm{p} \leq \bm{\pi}^{\mathsf{T}}\bm{p} + \bm{x},\  \bm{p} \in [\bm{0}_d,\bar{\bm{p}}] \}, & \text{if }\bm{x} \in \mathbb{R}^{d}_{+}, \\
-\infty, & \text{if }\bm{x} \notin \mathbb{R}^{d}_{+}. \\
\end{cases}
\end{equation}
The next lemma shows that $\Lambda^{\text{EN}}$ is an aggregation function that satisfies the assumptions posed in \Cref{sec:ENM} and \Cref{sec:theory}.

\begin{lemma} \label{lem:enmon}
$\Lambda^{\textup{EN}}$ is a concave and increasing function that satisfies \Cref{asmp:dom} and \Cref{asmp:dom2}.
\end{lemma}

%

\subsection{Sensitive Systemic Value-at-Risk and its SAA}\label{sec:asmp}

Let us fix a pair $(\bm{\pi},\bar{\bm{p}})$, where $\bar{\bm{p}}\in\R^d_{++}$ and $\bm{\pi}\in\R^{d\times d}$ is a right stochastic matrix with $\pi_{ii}=0$ for every $i\in[d]$. Then, for every $\bm{X}\in L^0(\R^d_+)$, the triplet $(\bm{\pi},\bar{\bm{p}},\bm{X}(\o))$ is an Eisenberg-Noe network with $d$ institutions for almost every $\o\in\O$.

Within the framework of \Cref{sec:systemicRM}, let us take $\X=L^0(\R^d)$ and $\Y=L^\infty(\R)$, $\Lambda=\Lambda^{\text{EN}}$, and $\Y=L^0(\R)$. By \Cref{lem:enmon}, \Cref{asmp:dom} and \Cref{asmp:dom2} hold. Since $\dom\Lambda^{\text{EN}}=\R^d_+$, we have $\X_{\Lambda^{\text{EN}}}=L^0(\R^d_+)$. Moreover, since $\Lambda^{\text{EN}}$ is bounded on $\dom\Lambda^{\text{EN}}$, we have $\Lambda^{\text{EN}}(\bm{X})\in L^\infty(\R)$ for every $\bm{X}\in L^0(\R^d_+)$ so that \Cref{asmp:domain} also holds.

Let us fix $\alpha> 0$ and $\lambda\in(0,1)$. Let us also fix $\bm{X}\in L^0(\R^d_+)$ as a random operating cash flow vector. The sensitive systemic value-at-risk corresponding to the Eisenberg-Noe model evaluated at $\bm{X}$ is given by
\[
R^{\text{EN}}_{\alpha,\lambda}(\bm{X})\coloneqq\cb{\bm{z}\in\R^g\mid \mathbb{P}\{\Lambda^{\text{EN}}(\bm{X}+\bm{B}^{\mathsf{T}}\bm{z})<\alpha\}\leq\lambda,\ \mathbb{P}\{\bm{X}+\bm{B}^{\mathsf{T}}\bm{z}\geq 0\}=1}.
\]
We discuss the nonemptiness of $R^{\text{EN}}_{\alpha,\lambda}(\bm{X})$ in the next proposition.

\begin{proposition} \label{prop:nonempty}
It holds $R^{\textup{EN}}_{\alpha,\lambda}(\bm{X}) \neq \emptyset$ if and only if $\alpha \leq \bm{1}_d^{\mathsf{T}}\bar{\bm{p}}$.
\end{proposition}

Since \Cref{asmp:domain} holds, $R^{\text{EN}}_{\alpha,\lambda}(\bm{X})$ is bounded with respect to $\R^g_+$ by \Cref{lem:bdd}; indeed, as a lower bound vector we may take $\bm{z}^{\text{LB}}\in\R^g$ defined by
\[
z_j^{\text{LB}}\coloneqq -\min_{i\in[n]\colon B_{ji}=1}\essinf X_i,\quad j\in[g], 
\]
since the requirement $\mathbb{P}\{\bm{X}+\bm{B}^{\mathsf{T}}\bm{z}\geq 0\}=1$ is equivalent to $\bm{z}\geq \bm{z}^{\text{LB}}$. Let us also define $\bm{z}^{\textup{UB}}\in\R^g$ by
\[
z^{\textup{UB}}_j\coloneqq \max_{i\in[n]\colon B_{ji}=1} \bar{p}_i,\quad j\in[g],
\]
and set
\begin{equation}\label{eq:Z}
\mathcal{Z}\coloneqq [\bm{z}^{\textup{LB}},\bm{z}^{\textup{UB}}].
\end{equation}

\begin{remark}\label{rem:zub}
	The definition of $\bm{z}^{\textup{UB}}$ guarantees that $\bar{\bm{p}}$ is feasible for the problem of calculating $\Lambda^{\textup{EN}}(\bm{B}^{\mathsf{T}}\bm{z}^{\textup{UB}})$. In particular, we have $\Lambda^{\textup{EN}}(\bm{B}^{\mathsf{T}}\bm{z}^{\textup{UB}})=\bm{1}_d^{\mathsf{T}}\bar{\bm{p}}$.
\end{remark}

In order for the SAA convergence results of \Cref{sec:SAA}, we verify the assumptions of these results for the Eisenberg-Noe model. We first show that the ``interesting" part of $R^{\text{EN}}_{\alpha,\lambda}(\bm{X})$ can be bounded by a compact set as assumed in \Cref{thm:scalar} and \Cref{thm:hausdorff}.

\begin{proposition}\label{prop:compact}
With $\mathcal{Z}$ defined in \eqref{eq:Z}, it holds
\[
R^{\textup{EN}}_{\alpha,\lambda}(\bm{X}) = R^{\textup{EN}}_{\alpha,\lambda}(\bm{X})\cap \mathcal{Z} + \R^g_+.
\]
\end{proposition}

Next, we verify \Cref{asmp:scalar} for the Eisenberg-Noe model.

\begin{theorem}\label{thm:ENscalar}
	Let $\mathcal{Z}$ be defined by \eqref{eq:Z} and let $\bm{w}\in\R^g_+\setminus\{\bm{0}_g\}$. Suppose that $\alpha< \bm{1}_d^{\mathsf{T}}\bar{\bm{p}}$. 
	\begin{enumerate}[(i)]
		\item There exists $\bar{\bm{z}}\in R^{\textup{EN}}_{\alpha,\lambda}(\bm{X})\cap\mathcal{Z}$ such that $\bm{w}^{\mathsf{T}}\bar{\bm{z}}=\mathsf{s}(\bm{w},R^{\textup{EN}}_{\alpha,\lambda}(\bm{X}))$.
		\item Suppose further that $\bm{X}$ has a continuous distribution with full support on $\R^d_+$, i.e., the measure $\mathbb{P}\circ \bm{X}^{-1}$ is equivalent to $\Leb_d$ restricted to $\R^d_+$. Then, there exists a sequence $(\bm{z}^\ell)_{\ell\in\N}$ in $\mathcal{Z}$ that converges to $\bar{\bm{z}}$ and satisfies  $\mathbb{P}\{\Lambda^{\textup{EN}}(\bm{X}+\bm{B}^{\mathsf{T}}\bm{z}^\ell)<\alpha\}<\lambda$, $\mathbb{P}\{\bm{X}+\bm{B}^{\mathsf{T}}\bm{z}^\ell\geq 0\}=1$ for every $\ell\in\N$.
	\end{enumerate}
\end{theorem}

Finally, we verify \Cref{asmp:hausdorff} for the Eisenberg-Noe model.

\begin{theorem}\label{thm:ENscalar2}
	Let $\mathcal{Z}$ be defined by \eqref{eq:Z} and let $\bm{v}\in\mathcal{Z}$. Suppose that $\alpha< \bm{1}_d^{\mathsf{T}}\bar{\bm{p}}$. 
	\begin{enumerate}[(i)]
		\item There exists $\bar{\bm{z}}\in R^{\textup{EN}}_{\alpha,\lambda}(\bm{X})\cap\mathcal{Z}$ such that $|\bm{v}-\bar{\bm{z}}|_2=\mathsf{d}(\bm{v},R^{\textup{EN}}_{\alpha,\lambda}(\bm{X}))$.
		\item Suppose further that $\bm{X}$ has a continuous distribution with full support on $\R^d_+$. Then, there exists a sequence $(\bm{z}^\ell)_{\ell\in\N}$ in $\mathcal{Z}$ that converges to $\bar{\bm{z}}$ and satisfies  $\mathbb{P}\{\Lambda^{\textup{EN}}(\bm{X}+\bm{B}^{\mathsf{T}}\bm{z}^\ell)<\alpha\}<\lambda$, $\mathbb{P}\{\bm{X}+\bm{B}^{\mathsf{T}}\bm{z}^\ell\geq 0\}=1$ for every $\ell\in\N$.
	\end{enumerate}
\end{theorem}

Thanks to the above verifications, the convergence results of \Cref{sec:SAA} are applicable for $R^{\text{EN}}_{\alpha,\lambda}$. We summarize them in the next corollary.

\begin{corollary}\label{cor:EN}
	Let $\mathcal{Z}$ be defined by \eqref{eq:Z}. Suppose that $\alpha< \bm{1}_d^{\mathsf{T}}\bar{\bm{p}}$ and $\bm{X}$ has a continuous distribution with full support on $\R^d_+$.
	\begin{enumerate}[(i)]
		\item For every $\bm{w}\in\R^g_+\setminus\{\bm{0}_g\}$, the sequence $(\mathsf{s}(\bm{w},R^{\textup{EN},N}_{\alpha,\lambda}(\bm{X}^1,\ldots,\bm{X}^N)\cap\mathcal{Z}))_{N\in\N}$ converges to $\mathsf{s}(\bm{w},R^{\textup{EN}}_{\alpha,\lambda}(\bm{X}))$ almost surely.
		\item The sequence $(R^{\textup{EN},N}_{\alpha,\lambda}(\bm{X}^1,\ldots,\bm{X}^N))_{N\in\N}$ Wijsman- and Hausdorff-converges to $R^{\textup{EN}}_{\alpha,\lambda}(\bm{X})$ almost surely.
		\end{enumerate}
	\end{corollary}

\subsection{Mixed-integer programming formulations for SAA}\label{sec:MILP}

As in the previous subsection, we fix $\alpha> 0$, $\lambda\in(0,1)$, and a random operating cash flow vector $\bm{X}\in L^0(\R^d_+)$. Let $(\bm{X}^N)_{N\in\N}$ be a sequence of independent copies of $\bm{X}$. Let $N\in\N$. The set-valued SAA function defined in \Cref{sec:SAA} is given by
\[
R^{\text{EN},N}_{\alpha,\lambda}(\bm{x}^1,\ldots,\bm{x}^N)\coloneqq \cb{\bm{z}\in\R^g\mid \frac{1}{N}\sum_{n=1}^N 1_{[-\infty,\alpha)}(\Lambda^{\text{EN}}(\bm{x}^n+\bm{B}^{\mathsf{T}}\bm{z}))\leq\lambda,\ \forall n\in[N]\colon \bm{x}^n+\bm{B}^{\mathsf{T}}\bm{z}\geq 0}.
\]
for each $\bm{x}^1,\ldots,\bm{x}^N\in\R^d$. Then, $R^{\text{EN},N}_{\alpha,\lambda}(\bm{X}^1,\ldots,\bm{X}^N)$ is the SAA for $R^{\text{EN}}_{\alpha,\lambda}(\bm{X})$ corresponding to a random sample of size $N$. In practice, once a scenario $\o\in\O$ is observed, then the SAA is realized as $R^{\text{EN},N}_{\alpha,\lambda}(\bm{X}^1(\o),\ldots,\bm{X}^N(\o))$. Our goal is to argue that this set can be calculated by solving a finite-dimensional mixed-integer linear vector programming problem. To that end, we first reformulate the values of $R^{\text{EN},N}_{\alpha,\lambda}$ using binary variables in the next theorem.

\begin{theorem}\label{thm:milvp}
For each $\bm{x}^1,\ldots,\bm{x}^N\in\R^d$, it holds
\begin{align}
R^{\textup{EN},N}_{\alpha,\lambda}(\bm{x}^1,\ldots,\bm{x}^N)= \Bigg\{\bm{z}\in\R^g\mid & \sum_{n=1}^N y^n\geq N(1-\lambda),\ \forall n\in[N]\colon \bm{1}_d^{\mathsf{T}}\bm{p}^n\geq \alpha y^n,\ \bm{x}^n+\bm{B}^{\mathsf{T}}\bm{z}\geq 0, \notag \\
& \bm{p}^n\leq \bm{\pi}^{\mathsf{T}}\bm{p}^n+\bm{x}^n+\bm{B}^{\mathsf{T}}\bm{z},\ \bm{p}^n\in [\bm{0}_d,\bar{\bm{p}}],\ y^n\in\{0,1\}\Bigg\}.\label{eq:milvp}
\end{align}
\end{theorem}

As a consequence of \Cref{thm:milvp}, we may view $R^{\text{EN},N}_{\alpha,\lambda}(\bm{x}^1,\ldots,\bm{x}^N)$ as the \emph{upper image} of the finite-dimensional mixed-integer linear vector programming problem
\begin{align}
&\text{minimize}\quad \bm{z}\quad \text{with respect to}\quad \R^g_+\notag \\
&\text{subject to}\quad \sum_{n=1}^N y^n\geq N(1-\lambda),\notag \\
&\quad\quad\quad\quad\quad\; \bm{1}_d^{\mathsf{T}}\bm{p}^n\geq \alpha y^n,\quad n\in[N],\notag \\ 
&\quad\quad\quad\quad\quad\;  \bm{x}^n+\bm{B}^{\mathsf{T}}\bm{z}\geq 0,\quad n\in[N],\notag \\
&\quad\quad\quad\quad\quad\;  \bm{p}^n\leq \bm{\pi}^{\mathsf{T}}\bm{p}^n+\bm{x}^n+\bm{B}^{\mathsf{T}}\bm{z},\quad n\in[N], \notag \\
&\quad\quad\quad\quad\quad\; \bm{p}^n\in [\bm{0}_d,\bar{\bm{p}}],\ y^n\in\{0,1\},\quad n\in[N],\notag \\
&\quad\quad\quad\quad\quad\; \bm{z}\in\R^g.\notag 
\end{align}
Moreover, the constraint $\bm{x}^n+\bm{B}^{\mathsf{T}}\bm{z}\geq 0$ is equivalent to $\bm{z}\geq \bm{z}^{\text{LB},N}$, where $\bm{z}^{\text{LB},N}\in\R^g$ is defined by
\[
z^{\text{LB},N}_j\coloneqq -\min_{i\in[d]\colon B_{ji}=1}x^n_j,\quad j\in[g].
\]
Hence, $R^{\text{EN},N}_{\alpha,\lambda}(\bm{x}^1,\ldots,\bm{x}^N)$ is naturally bounded with respect to $\R^g_+$ since we have 
\[
R^{\text{EN},N}_{\alpha,\lambda}(\bm{x}^1,\ldots,\bm{x}^N)\subseteq \bm{z}^{\text{LB},N}+\R^g_+.
\]
Hence, $R^{\text{EN},N}_{\alpha,\lambda}(\bm{x}^1,\ldots,\bm{x}^N)$ can be calculated using the available algorithms for solving bounded nonconvex vector optimization problems, such as the one in \cite{nobakhtian2017benson}. As immediate corollaries of \Cref{thm:milvp}, we provide a mixed-integer linear programming (MILP) formulation of the weighted-sum scalarization and a mixed-integer quadratic programming (MIQP) formulation of the norm-minimizing scalarization in \Cref{app:milvp}; both of which will be used in the modified grid search algorithm described in \Cref{sec:comp}.


\section{Computational Results and Analysis}\label{sec:comp}

\subsection{Algorithms for Calculating SAA} \label{alg section}

In this section, we explore techniques to approximate the SAA for the sensitive systemic value-at-risk of the Eisenberg-Noe model. In addition to the direct implementation of the grid search algorithm in \cite{birgit} (\Cref{alg:acceptability}), we also leverage a modified version of it using the norm-minimizing scalarizations (\Cref{alg:acceptabilitynm}).

Let us fix $\bm{\pi},\bar{\bm{p}},\bm{X},\alpha,\lambda$ as in \Cref{sec:systemicRM}. Let $\bm{x}^1,\ldots,\bm{x}^N\in\R^d_+$ with $N\in\N$. Our goal is to calculate $R^{\text{EN},N}_{\alpha,\lambda}(\bm{x}^1,\ldots,\bm{x}^N)$. Let $\mathcal{Z}=[\bm{z}^{\text{LB}},\bm{z}^{\text{UB}}]$. To implement the grid algorithm more effectively, we first replace $\bm{z}^{\text{LB}}$ with a tighter lower bound. For this purpose, we calculate the ideal point $\bm{z}^{\text{ideal},N}$ of $R^{\text{EN},N}_{\alpha,\lambda}(\bm{x}^1,\ldots,\bm{x}^N)$ by solving the weighted-sum scalarizations along the standard unit vectors $\bm{e}_{1,g},\ldots,\bm{e}_{g,g}$ using the MILP formulation (see \Cref{cor:ws}). In particular, 
\[
\bm{z}^{\text{ideal},N}_j=\mathsf{s}(\bm{e}_{j,g},R^{\text{EN},N}_{\alpha,\lambda}(\bm{x}^1,\ldots,\bm{x}^N)),\quad j\in[g].
\]
Given a predefined approximation error $\epsilon>0$, we take a finite subset $\texttt{G}\subseteq [\bm{z}^{\text{ideal}},\bm{z}^{\text{UB}}]$, called the \emph{grid}, such that, for every $\bm{v}\in [\bm{z}^{\text{ideal}},\bm{z}^{\text{UB}}]$, there exists $\bm{z}\in\texttt{G}$ with $\bm{v}\leq \bm{z}\leq \bm{v}+\frac{\epsilon}{\sqrt{g}}\bm{1}_g$.
The goal of the algorithms is to detect the intersection $\texttt{AS}\coloneqq \texttt{G}\cap R^{\text{EN},N}_{\alpha,\lambda}(\bm{x}^1,\ldots,\bm{x}^N)$, called the \emph{acceptable set}. Then, $\hat{R}^{\text{EN},N}_{\alpha,\lambda}(\bm{x}^1,\ldots,\bm{x}^N)\coloneqq \texttt{AS}+\R^g_+$ is returned as an inner approximation of $R^{\text{EN},N}_{\alpha,\lambda}(\bm{x}^1,\ldots,\bm{x}^N)$.

The individual details of the algorithms are described below.
\begin{enumerate}[(i)]
	\item \textbf{SAA Computation With Clearing Vectors:} At each iteration, a point $\bm{z}\in\texttt{G}$ is chosen from the current grid and the aggregate values $\Lambda^{\text{EN}}(\bm{x}^n+\bm{B}^{\mathsf{T}}\bm{z})$ are calculated for each $n\in[N]$ using either a fixed point algorithm for the characterization in \Cref{eq:fixed} or by solving a linear programming problem based on \Cref{lem:mp}. Then, the condition $\frac{1}{N}\sum_{n=1}^N 1_{(-\infty,\alpha)}(\Lambda^{\text{EN}}(\bm{x}^n+\bm{B}^{\mathsf{T}}\bm{z}))\leq \lambda$ is checked. If satisfied, then $\bm{z}\in R^{\text{EN},N}_{\alpha,\lambda}(\bm{x}^1,\ldots,\bm{x}^N)$ so that all grid points in $\bm{z}+\R^g_+$ are labeled as \emph{acceptable}. Otherwise, all points in $\bm{z}-\R^g_+$ are labeled as \emph{not acceptable} so that they are removed from the grid and eliminated from future consideration. The algorithm proceeds to subsequent grid points along the diagonal trajectory until all points are classified as \emph{acceptable} and \emph{not acceptable}. The pseudocode is given in \Cref{alg:acceptability}.
	\item \textbf{SAA Computation Using Norm-Minimizing Scalarizations:} At each iteration, a  point $\bm{z}\in\texttt{G}$ is chosen from the current grid and the distance $\gamma\coloneqq \mathsf{d}(\bm{z},R^{\text{EN},N}_{\alpha,\lambda}(\bm{x}^1,\ldots,\bm{x}^N))$ is calculated by using the MIQP formulation of the norm-minimizing scalarization problem (see \Cref{cor:nm}), along with an optimizer $\bar{\bm{z}}$. All grid points in $\bar{\bm{z}} + \mathbb{R}_{+}^g$ are labeled as \emph{acceptable}. Moreover, all grid points that are within strictly less than $\gamma$-distance from the current grid point $\bm{z}$ are labeled as \emph{not acceptable} as these points do not belong to $R^{\text{EN},N}_{\alpha,\lambda}(\bm{x}^1,\ldots,\bm{x}^N)$ due to the structure of the norm-minimizing scalarization. The algorithm systematically progresses to subsequent grid points in a diagonal direction until all points are categorized as \emph{acceptable} and \emph{not acceptable}. The pseudocode is given in \Cref{alg:acceptabilitynm}.
\end{enumerate} 

\begin{minipage}[t]{0.48\textwidth}
	\begin{algorithm}[H]
		\centering
		\caption{SAA Computation Using Clearing Vectors}\label{alg:acceptability}
		\footnotesize 
		\begin{algorithmic}[1]
			\REQUIRE $\bm{x}^1,\ldots,\bm{x}^N\in\R^d_+$; \texttt{G} \COMMENT{Grid points}
			\STATE Initialize $\texttt{counter} \gets 0$
			\STATE Initialize $\texttt{AS} \gets \emptyset$ \COMMENT{Set of grid points in $R^{\text{EN}, N}_{\alpha,\lambda}(\bm{x}^1,\ldots,\bm{x}^N)$}
			\STATE Initialize $\texttt{NS} \gets \emptyset$ \COMMENT{Set of grid points that are not in $R^{\text{EN}, N}_{\alpha,\lambda}(\bm{x}^1,\ldots,\bm{x}^N)$}
			\WHILE{\texttt{G}$\neq\emptyset$}
			\STATE Choose $\bm{z}\in$ \texttt{G}
			\FOR{$n$ in $[N]$}
			\STATE Compute $\Lambda^{\text{EN}}(\bm{x}^n + \bm{B}^{\mathsf{T}}\bm{z})$ 
			\IF{$\Lambda^{\text{EN}}(\bm{x}^n + \bm{B}^{\mathsf{T}}\bm{z}) < \alpha$}
			\STATE $\texttt{counter} \gets \texttt{counter} + \frac{1}{N}$
			\ENDIF
			\ENDFOR
			\IF{$\texttt{counter} \leq \lambda$}
			\STATE Add all grid points in $\bm{z} + \mathbb{R}^g_{+}$ to \texttt{AS}
			\STATE Remove all grid points in $\bm{z} + \mathbb{R}^g_{+}$ from \texttt{G}
			\ELSE
			\STATE Add all grid points in $\bm{z} - \mathbb{R}^{g}_{+}$ to \texttt{NS}
			\STATE Remove all grid points in $\bm{z} - \mathbb{R}^{g}_{+}$ from \texttt{G}
			\ENDIF
			\STATE $\texttt{counter}$ $\gets$ $0$
			\ENDWHILE
			\RETURN $\hat{R}^{\text{EN}, N}_{\alpha,\lambda}(\bm{x}^1,\ldots,\bm{x}^N) \coloneqq \texttt{AS} + \mathbb{R}_+^g$
		\end{algorithmic}
	\end{algorithm}
\end{minipage}%
			\vspace{1em}
\setcounter{algo}{2}
\hspace{0.02\textwidth}
\begin{minipage}[t]{0.48\textwidth}
	\begin{algorithm}[H]
		\centering 
		\caption{SAA Computation Using Norm-Minimizing Scalarizations}\label{alg:acceptabilitynm}
		\footnotesize
		\begin{algorithmic}[1]
			\REQUIRE $\bm{x}^1,\ldots,\bm{x}^N\in\R^d_+$; \texttt{G} \COMMENT{Grid points}
			\STATE Initialize $\texttt{counter} \gets 0$
			\STATE Initialize $\texttt{AS} \gets \emptyset$ \COMMENT{Set of grid points in $R^{\text{EN}, N}_{\alpha,\lambda}(\bm{x}^1,\ldots,\bm{x}^N)$}
			\STATE Initialize $\texttt{NS} \gets \emptyset$ \COMMENT{Set of grid points that are not in $R^{\text{EN}, N}_{\alpha,\lambda}(\bm{x}^1,\ldots,\bm{x}^N)$}
			\WHILE{\texttt{G}$\neq\emptyset$}
			\STATE Choose $\bm{z}\in$ \texttt{G}
			\STATE Compute $\bar{\bm{z}}\in R^{\text{EN},N}_{\alpha,\lambda}(\bm{x}^1,\ldots,\bm{x}^N)\cap [\bm{z}^{\text{ideal},N},\bm{z}^{\text{UB}}]$ such that $|\bm{z}-\bar{\bm{z}}|_2=\mathsf{d}(\bm{z},R^{\text{EN},N}_{\alpha,\lambda}(\bm{x}^1,\ldots,\bm{x}^N))$
			\STATE Add all grid points in $\bar{\bm{z}} + \mathbb{R}^{g}_{+}$ to \texttt{AS}
			\STATE Remove all grid points in $\bar{\bm{z}}+ \mathbb{R}^{g}_{+}$ from \texttt{G}
			\STATE Add all grid points in $\Int \mathbb{B}(\bm{z},|\bm{z}-\bar{\bm{z}}|_2)$ to \texttt{NS}
			\STATE Remove all grid points in $\Int \mathbb{B}(\bm{z},|\bm{z}-\bar{\bm{z}}|_2)$ from \texttt{G}
			\ENDWHILE
			\RETURN $\hat{R}^{\text{EN}, N}_{\alpha,\lambda}(\bm{x}^1,\ldots,\bm{x}^N) \coloneqq \texttt{AS} + \mathbb{R}_+^g$
		\end{algorithmic}
	\end{algorithm}
\end{minipage}

The following theorem states the finiteness and correctness of these algorithms. Its proof is given in \Cref{app:alg}.

\begin{theorem} \label{thm:correctness}
	Let an approximation error $\epsilon> 0$ and a finite grid $\texttt{G}\subseteq [\bm{z}^{\text{ideal}},\bm{z}^{\text{UB}}]$ be given such that, for every $\bm{v}\in [\bm{z}^{\text{ideal}},\bm{z}^{\text{UB}}]$, there exists $\bm{z}\in\texttt{G}$ with $\bm{v}\leq \bm{z}\leq \bm{v}+\frac{\epsilon}{\sqrt{g}}\bm{1}_g$. Both \Cref{alg:acceptability} and \Cref{alg:acceptabilitynm} terminate in finitely many iterations and the returned approximation satisfies $\hat{R}^{\textup{EN},N}_{\alpha,\lambda}(\bm{x}^1,\ldots,\bm{x}^N)\subseteq R^{\textup{EN},N}_{\alpha,\lambda}(\bm{x}^1,\ldots,\bm{x}^N)$
	and $\mathsf{h}(\hat{R}^{\textup{EN},N}_{\alpha,\lambda}(\bm{x}^1,\ldots,\bm{x}^N),R^{\textup{EN},N}_{\alpha,\lambda}(\bm{x}^1,\ldots,\bm{x}^N)) \leq \epsilon$.
\end{theorem}

\subsection{Data Generation} \label{DG}

Our analysis focuses on computing the systemic set-valued risk measure within a core-periphery framework, examining a financial system with institutions divided into core ($\mathcal{C}$) and peripheral ($\mathcal{P}$) groups. The computational approach employs Matlab with CVX and Gurobi on an Intel(R) Core(TM) i7-6500U CPU @2.50GHz 2.59GHz processor with 8.00 GB RAM.

We outline data generation for an Eisenberg-Noe network with $d=20$ banks. The network is constructed using the directed preferential attachment model in \cite{bollobas2003directed}, capturing a scale-free growth \citep[Chapter 8]{hofstad_2016}. As a result, it is observed that the vertices with the longest existence tend to have the highest degrees, a phenomenon commonly known as the ``rich-get-richer" effect; see \cite{deijfen2009preferential, barabasi1999emergence, bollobas2003directed, bollobas2003mathematical, cooper2003general}. Before enumerating the specific rules of the model, we clarify the notation and parameters: $d_{\text{in}}(v)$ and $d_{\text{out}}(v)$ represent the in-degree and out-degree of vertex $v$ at time $t$, respectively, while $n(t)$ denotes the number of vertices and $\mathcal{G}(t)$ denotes the vertex set of the graph. The parameters $\delta_{\text{in}}$ and $\delta_{\text{out}}$ are smoothing constants that ensure well-defined probabilities even when degrees are zero, influencing the initial attractiveness of vertices. The probabilities $\theta$, $\eta$, and $\zeta$, constrained by $\theta + \eta + \zeta = 1$, govern the likelihood of different types of events in the network's evolution. Starting with an initial graph with vertex set $\mathcal{G}(0)$, the model evolves over time on some probability space $(\tilde{\Omega},\tilde{\mathcal{F}},\tilde{\mathbb{P}})$, adhering to the following specified rules:
\begin{enumerate}[(i)]
	\item With probability $\theta$, a new vertex $v$ is added along with an edge from $v$ to an existing vertex $w$, where $w$ is chosen with respect to the distribution,
	\[
	\tilde{\mathbb{P}}\{w = w_i\} = \frac{d_{\text{in}}(w_i) + \delta_{\text{in}}}{t + \delta_{\text{in}} n(t)} ,\, w_i \in \mathcal{G}(t).
	\]
	
	\item With probability $\eta$, an edge is added from an existing vertex $v$ to an existing vertex $w$. The selection of $v$ and $w$ is independent and follows the distribution,
	\[
		\tilde{\mathbb{P}}\{v = v_i\} = \frac{d_{\text{out}}(v_i) + \delta_{\text{out}}}{t + \delta_{\text{out}} n(t)} ,\, v_i \in \mathcal{G}(t),\quad 
	\tilde{\mathbb{P}}\{w = w_i\} = \frac{d_{\text{in}}(w_i) + \delta_{\text{in}}}{t + \delta_{\text{in}} n(t)},\, w_i \in \mathcal{G}(t).
	\]
	
	\item With probability $\zeta$, a new vertex $w$ is added along with an edge from an existing vertex $v$ to $w$, where $v$ is chosen with respect to the distribution,
	\[
	\tilde{\mathbb{P}}\{v = v_i\} = \frac{d_{\text{out}}(v_i) + \delta_{\text{out}}}{t + \delta_{\text{out}} n(t)} ,\, v_i \in \mathcal{G}(t).
	\]
\end{enumerate}

It is worth noting that the model allows for loops and multiple edges, although their presence does not significantly affect the conclusions drawn. The paper showed that the in-degree and out-degree distribution follow a power law distribution  \cite[Theorem 3.1]{bollobas2003directed}.

The matrix $\pmb{\pi}$ and vector $\boldmath{\bar{p}}$ were generated based on the adjacency matrix of the aforementioned network. The core banks set $\mathcal{C}$ comprises the top four banks based on node degree in $\bm{A}$, while $\mathcal{P}$ includes the remaining 16 banks. Intergroup liability matrix $\bm{M}$ distinguishes core-to-core (CC), core-to-periphery (CP), periphery-to-core (PC), and periphery-to-periphery (PP) linkages as follow
\[
\bm{M} = \begin{pmatrix}
	CC & CP \\
	PC & PP
\end{pmatrix}.
\]

The interbank nominal liability matrix $\bm{l}$ is constructed from $\bm{A}$ and $\bm{M}$. For core banks $i$ and $j$, $l_{ij}$ equals $CC$ if $A_{ij}=1$ and zero if $A_{ij}=0$. Similar rules apply for other bank classifications and connections. After computing $\bm{l}$, the total liability vector $\boldmath{\bar{p}}$ sums its rows, and the relative liability matrix $\pmb{\pi}$ is derived by normalizing $\bm{l}$.

Subsequently, the operating cash flow vector $\bm{X}$ in $L^0(\R^d_+)$ is a multivariate random vector, generated through Pareto-distributed shocks assigned to each bank. The correlation coefficient $\rho$ measures the dependence among banks. The groups use a shared shape parameter $\nu$ and distinct scale parameters $\beta_{\mathcal{C}}$ and $\beta_{\mathcal{P}}$ for core and peripheral banks, respectively.

\subsection{Runtime Performance Analysis} \label{bench}

In this subsection, we compare the runtime performance of \Cref{alg:acceptabilitynm} and \Cref{alg:acceptability}. For network construction, we use the Bollobás algorithm (\Cref{DG}) with the following parameters: $\theta = 0.2$, $\eta = 0.6$, $\zeta = 0.2$, $\delta_{\text{in}} = 0.5$, and $\delta_{\text{out}} = 0.5$.

Recapping, the Eisenberg-Noe network has $d=20$ banks ($4$ core and $16$ periphery) generated with parameters $N= 50$, $\rho = 0.3$, $\beta = [100, 50]$, and $\nu = 3$. The intergroup liability matrix $\bm{M}$ has values
\[
\bm{M} = \begin{pmatrix}
	400 & 200 \\
	300 & 150
\end{pmatrix}.
\]
The value-at-risk parameters are taken as $\alpha = (0.8)\times\bm{1}^{\mathsf{T}}_d\bar{\bm{p}}$ and $\lambda = 0.2$. The grid algorithm uses six $\epsilon$ values for different degrees of approximation. Performance is assessed in \Cref{AlgPerfomance} for $\epsilon$ ranging from 50 to 1, with their corresponding grid points.

\begin{table}
	\centering
	\resizebox{0.6\textwidth}{!}{
		\begin{tabular}{|c|c|c|c|}
		\hline
		\textbf{$\epsilon$} & \textbf{Grid Points \#} & \textbf{Algorithm 1 Runtime (s)} & \textbf{Algorithm 2 Runtime (s)} \\ \hline
		\textbf{50} & 570 & 24.72 & 50.04 \\ \hline
		\textbf{25} & 2175 & 45.68 & 99.96\\ \hline
		\textbf{10} & 13464 & 114.19 & 505.83 \\ \hline
		\textbf{1} & 1328148 & 13846.72 & 163242.83\\ \hline
	\end{tabular}%
	}
	\vspace{5pt}
	\caption{Runtime performance of the two algorithms for varying $\epsilon$.}
	\label{AlgPerfomance}
\end{table}

\begin{figure}[htbp]
	\centering
		\includegraphics[width=1.0\textwidth]{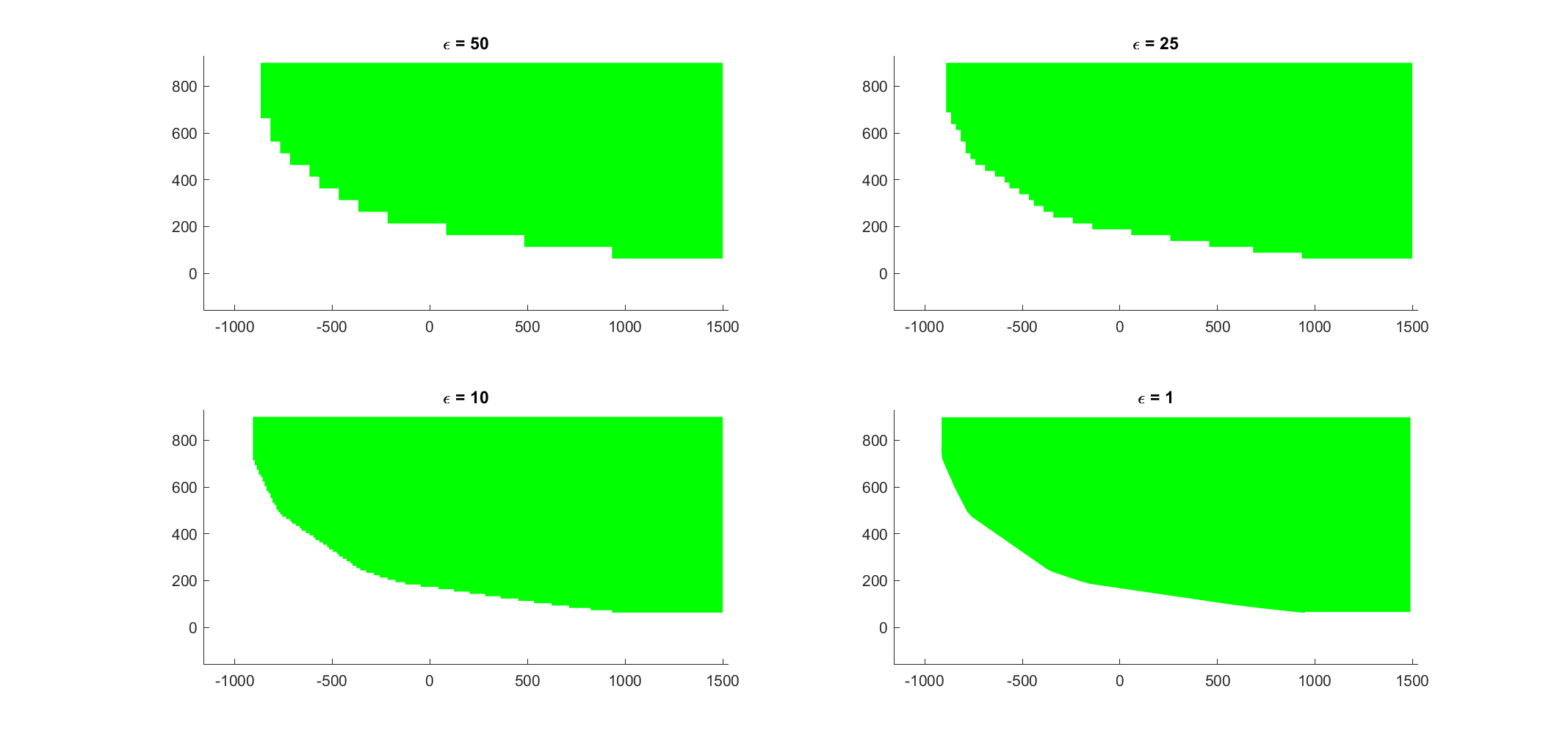}\label{epsilon}
		\caption{Approximations for the SAA of sensitive systemic value-at-risk for varying $\epsilon$.}
	\label{combinedPerformanceRiskMeasure}
\end{figure}

In \Cref{combinedPerformanceRiskMeasure}, the $x$-axis and $y$-axis of each subplot represent capital injection to core and peripheral banks, respectively. The visualization of approximation sets in \Cref{combinedPerformanceRiskMeasure} reveals that a smaller $\epsilon$ value leads to a more precise approximation of the set-valued systemic risk measure. Additionally, results in \Cref{AlgPerfomance} show that \Cref{alg:acceptability} outperforms, evaluating grid point acceptability using the optimization formulation in \Cref{sec:ENM}. This likely stems from the need for binary variables in the norm-minimizing scalarization problem used by \Cref{alg:acceptabilitynm}.

\subsection{Sensitivity of the Risk Measure}	

This subsection aims to assess the sensitivity of the set-valued systemic risk measure to the Bollobás network under constant parameters. Despite identical parameter values, the stochastic nature of the network results in varied network configurations across multiple iterations. Consistency is maintained with the network, operating cash flow, and set-valued risk measure parameters from \Cref{bench}. This ensures isolation of the network structure's influence on the set-valued systemic risk measure. A comprehensive analysis involves 10 independently simulated networks, capturing the range of configurations possible with identical Bollobás parameters.

\begin{figure}[htbp]
	\centering
	\subfloat[Sensitivity of Set-Valued Risk Measure.]{\includegraphics[width=0.50\textwidth]{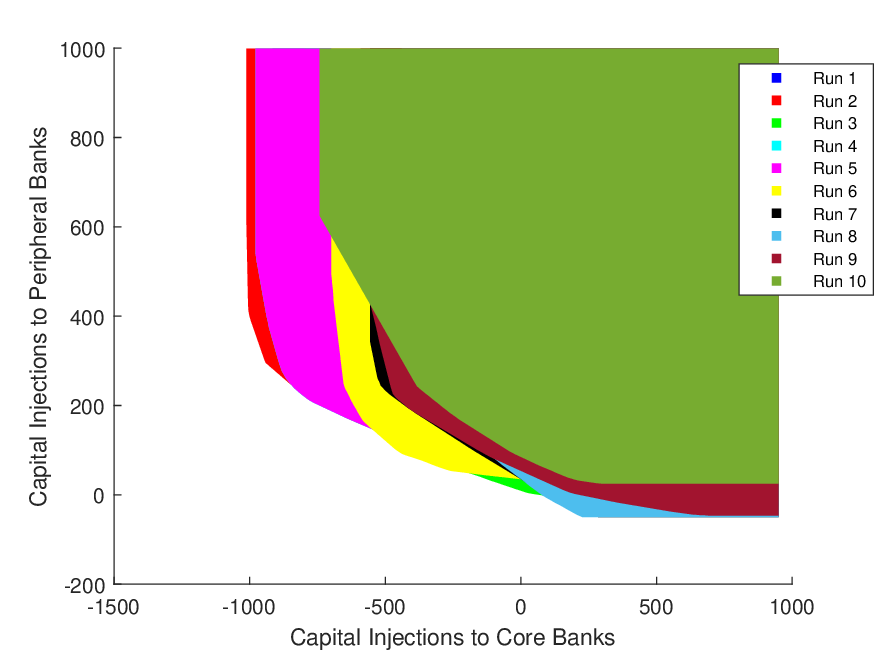}\label{SensitivityofRM}}
	\hfill
	\subfloat[Average of Set-Valued Risk Measure.]{\includegraphics[width=0.50\textwidth]{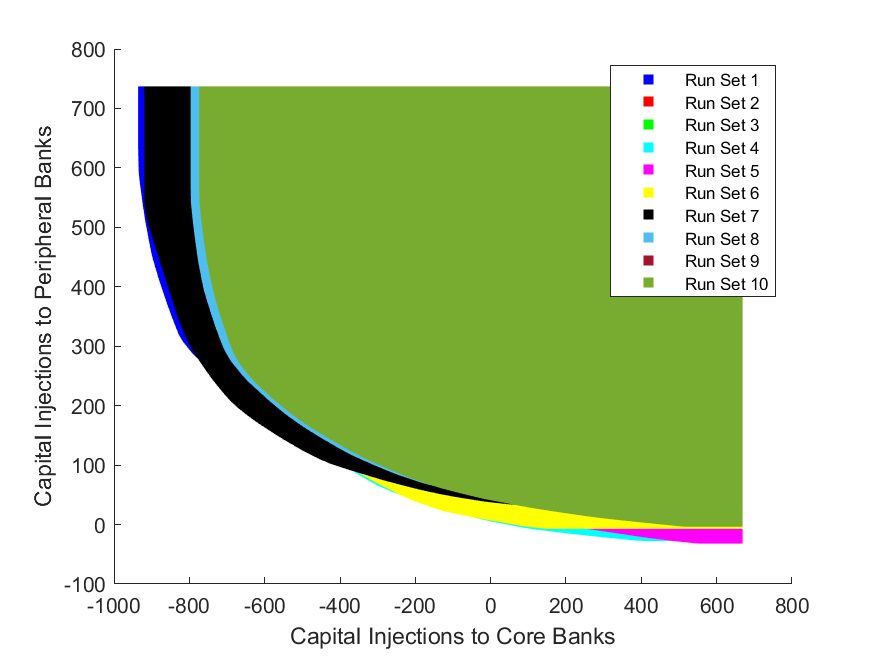}\label{AverageofRM}}
	\caption{Comparison of Sensitivity and Averaging of Set-Valued Risk Measures.}
	\label{CombinedFigures}
\end{figure}

\Cref{SensitivityofRM} vividly demonstrates our set-valued risk measure's sensitivity to network shape, notably with $\epsilon = 1$. Even minor changes in the network's structure significantly impact the computed systemic risk measure.

Continuing, we conduct a thorough analysis by evaluating systemic risk across 100 independently simulated networks with identical Bollobás parameters. The approach involves computing the mean value of 10 unique sets within the grid area, where each set is derived from the average of vertices obtained in 10 independent runs. This results in 10 distinct sets, each representing the mean value from 10 separate runs illustrated in \Cref{AverageofRM}.

\subsection{Impact of the Sample Size}

In this section, we examine the impact of varying the sample size (or number of stress scenarios), denoted as $N$, on both computational performance and systemic risk measures. All parameters mentioned in \Cref{bench} are held constant, with the exception of the scenarios count $N$, enabling us to conduct a sensitivity analysis using this particular parameter. This analysis offers valuable insights into the convergence behavior observed with an increasing number of scenarios.

We examine a set of values for the variable $N$, which are defined as follows: $N$ takes on the values of 5, 10, 20, 50, 100, 120, 140, 160, 180, and 200. Although the network structure remains consistent, it is anticipated that the systemic risk measures will display slight fluctuations. Nevertheless, it is expected that the computation time will increase as the number of scenarios increases, primarily because each scenario involves $d$ continuous variables. 

\begin{table}
\centering
\resizebox{0.35\textwidth}{!}{
	\begin{tabular}[b]{|c|c|c|}
	\hline
	\textbf{$N$} & \textbf{Grid Points \# ($\times 10^{5}$)} & \textbf{Time (s)} \\ \hline
	\textbf{5} & 6.21 & 2908.67 \\ \hline
	\textbf{10} & 5.13 & 2020.09 \\ \hline
	\textbf{20} & 5.68 & 2515.88 \\ \hline
	\textbf{50} & 5.08 & 2244.23 \\ \hline
	\textbf{100} & 5.12 & 2677.92 \\ \hline
	\textbf{120} & 5.63 & 3403.30 \\ \hline
	\textbf{140} & 5.84 & 3553.30 \\ \hline
	\textbf{160} & 6.30 & 4314.21  \\ \hline
	\textbf{180} & 5.13 & 3361.87  \\ \hline
	\textbf{200} & 5.40  & 3821.29 \\ \hline
\end{tabular}%
}
\vspace{5pt}
\caption{Runtime performance of \Cref{alg:acceptability} with respect to sample size.}
\label{ScenarioPerfomance}
\end{table}

\begin{figure}[htbp]
	\centering
		\includegraphics[width=0.55\textwidth]{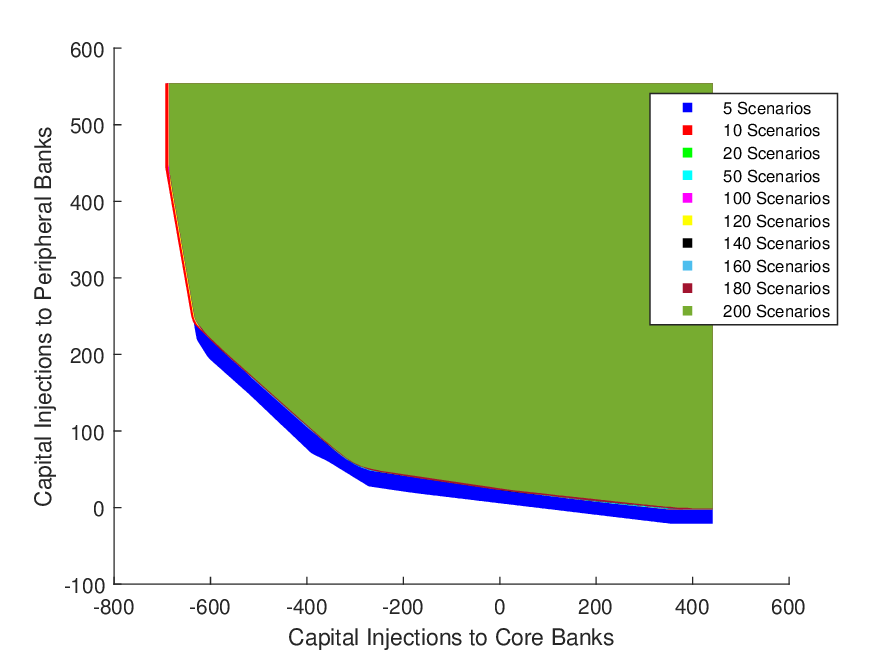}
	\caption{Comparison of SAA sets (overlapping regions) for different sample sizes.}
	\label{overlapping_regions_convergence}
\end{figure}

The approximations of set-valued systemic risk measures, specifically for $\epsilon = 1$, are shown in \Cref{overlapping_regions_convergence}. This figure highlights the relative stability of these approximations across different $N$ values, aligning with our initial hypotheses. It supports the expected connection between set-valued risk measures and the number of scenarios. In contrast, \Cref{ScenarioPerfomance} displays computational performance and grid points for the \Cref{alg:acceptability}, maintaining consistent scenario counts and approximation error. Despite irregular fluctuations in grid points, the overall algorithm performance time shows inconsistent variations as the number of scenarios increases. This can be attributed to the algorithm's sensitivity to the quantity of grid points.

\subsection{Network Parameters}

\begin{figure}[htbp]
	\centering
	\hspace*{-0.12\textwidth}
	\subfloat[Changing $\eta$]{\includegraphics[width=1.21\textwidth]{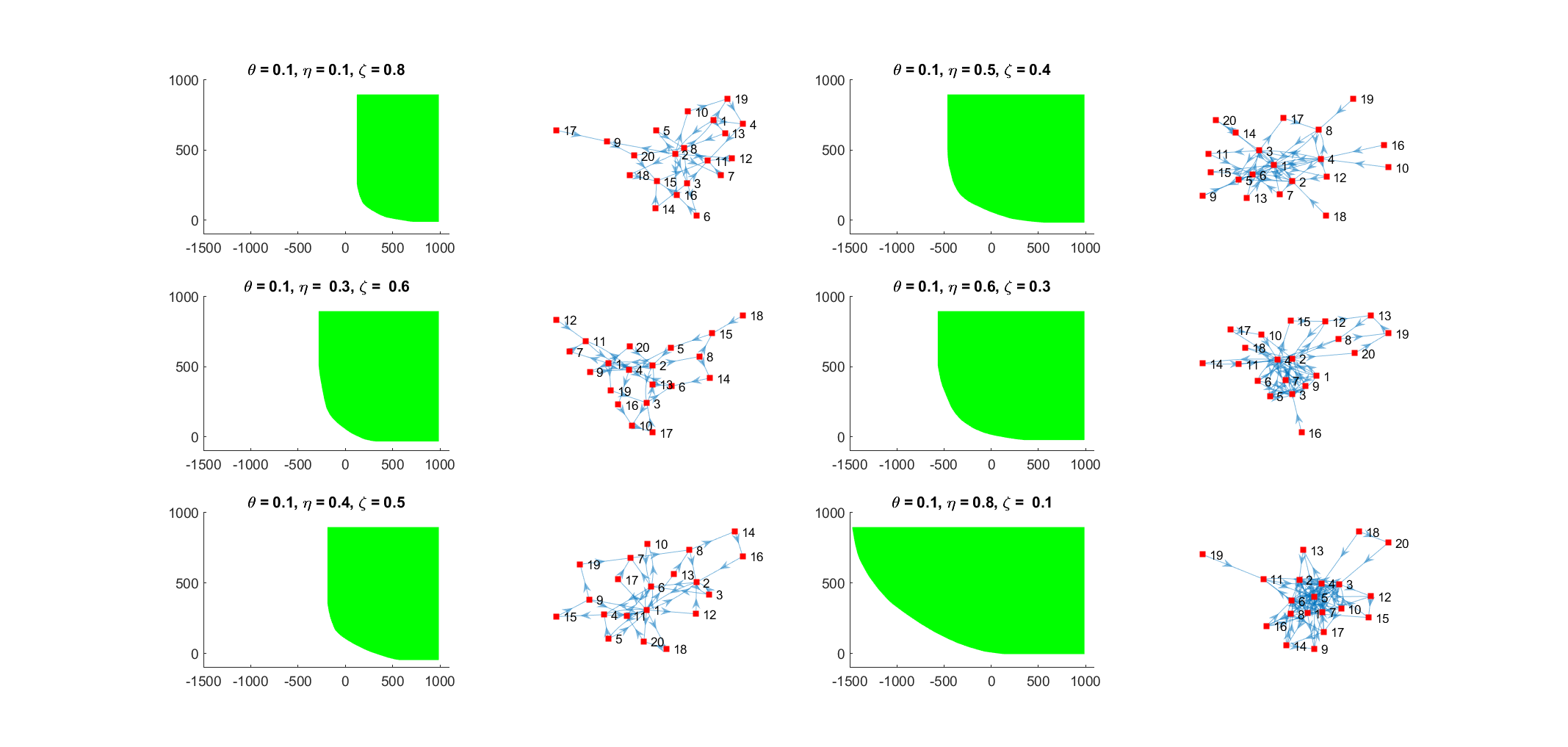}\label{fig:b11}}
	\hfill
	\hspace*{-0.12\textwidth}
	\subfloat[Reversing $\theta$ and $\zeta$]{\includegraphics[width=1.21\textwidth]{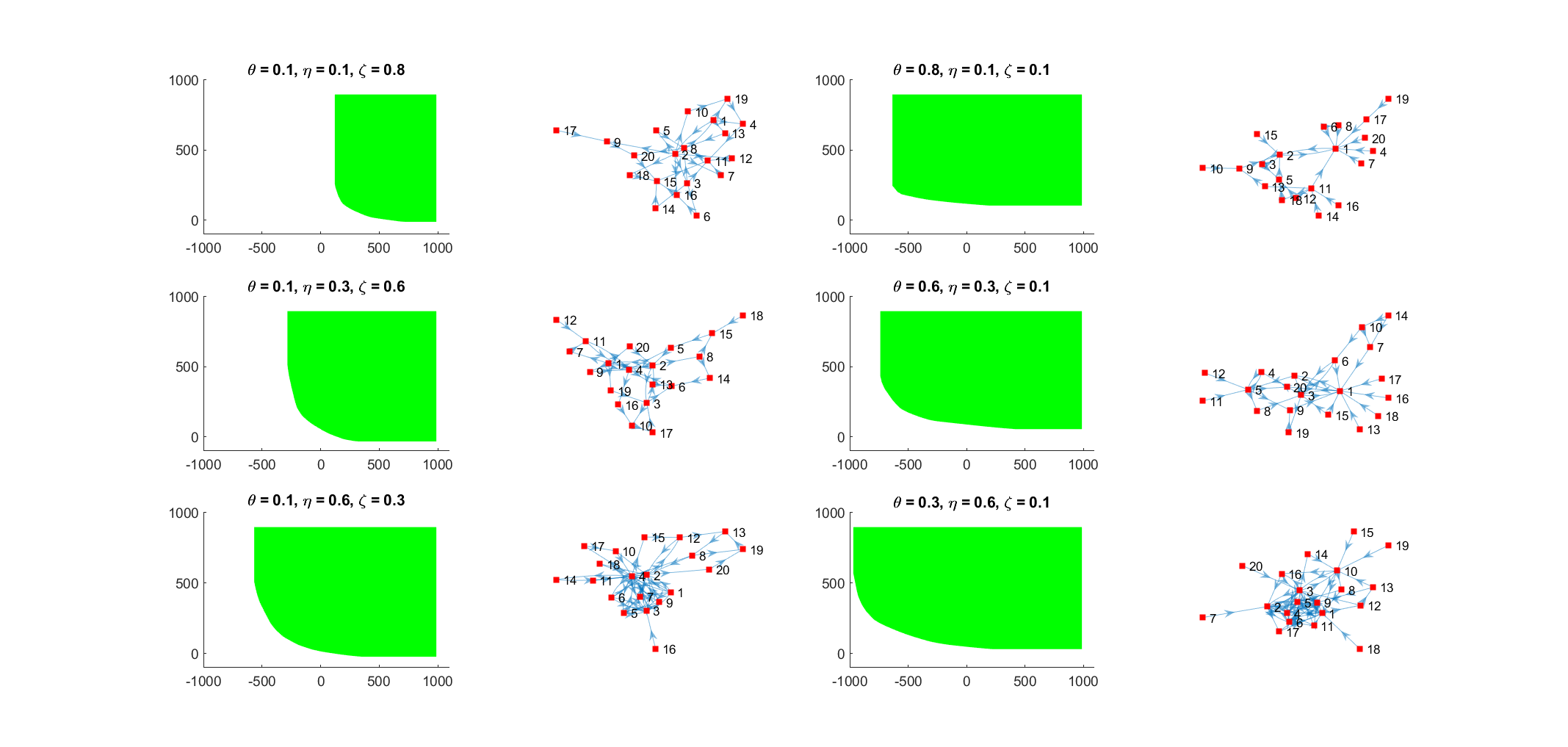}\label{fig:b22}}
	\caption{Approximations of sensitive systemic value-at-risk along with their corresponding networks.}
	\label{combinedRiskMeasures}
\end{figure}

In this subsection, we conduct a sensitivity analysis on the probabilities of the Bollobás network parameter, specifically focusing on $\theta$, $\eta$, and $\zeta$. The remaining parameters from \Cref{bench} are kept constant. Visual representations of systemic risk measures are found in \Cref{combinedRiskMeasures} for $\epsilon = 1$. The x-axis and y-axis in each subplot represent capital injections to core and peripheral banks, respectively, with an adjacent subplot displaying the corresponding network graph.

We analyze the following network statistics for each combination of Bollobás network parameter probabilities:

\begin{itemize}
	\item Average Degree (AD): Mean connections per node in the network, calculated as $\text{AD} = \frac{{\sum_{i=1}^{n} \sum_{j=1}^{n} A_{ij}}}{n}$, where $n$ is the total number of nodes, and $\bm{A}$ is the adjacency matrix.
	
	\item Network Density (ND): The ratio of total connections to total possible connections, given by $\text{ND} = \frac{{\sum_{i=1}^{n} \sum_{j=1}^{n} A_{ij}}}{n(n-1)}$.
	
	\item Total Clustering Coefficient (TCC): A metric measuring the tendency of nodes to form clusters. Computed as $\text{TCC} = \sum_{i=1}^{n} \text{CC}_i$, where $\text{CC}_i$ is the clustering coefficient of node $i$.
	
	\item Core-Periphery Error (CPE): Based on \cite{craig2014interbank}, CPE is a measure evaluating the disparity between observed and expected core-periphery configurations.
	
	\item Core-Periphery Index (CPI): A robust measure assessing the strength of the core-periphery structure. Defined as $\text{CPI} = \frac{{\sum_{i=1}^{n} \sum_{j=1}^{n} A_{ij}} - \sum_{i \in \mathcal{P}} \sum_{j \in \mathcal{P}}  A_{ij}}{{\sum_{i=1}^{n} \sum_{j=1}^{n} A_{ij}}}$, where $\mathcal{C}$ and $\mathcal{P}$ are the core and periphery sets, respectively.
\end{itemize}

The network statistics provide key insights into Bollobás network structural characteristics. Lower Core-Periphery Error (CPE) and higher Core-Periphery Index (CPI) values indicate a more pronounced core-periphery structure.

First, we aim to examine the impact of variations in the parameter $\eta$ on the sensitivity of the systemic risk measure. Specifically, we keep the parameter $\theta$ constant at a value of 0.1, and manipulate the parameter $\zeta$ to be equal to 1 minus the sum of 0.1 and $\eta$. The computational statistics for various values of the $\eta$ parameter, specifically $\eta \in \{0.1, 0.3, 0.4, 0.5, 0.6, 0.8\}$, are presented in \Cref{combinedNetworkStats}(left). The corresponding approximations are displayed in \Cref{fig:b11}.

Based on the data presented in \Cref{combinedNetworkStats}(left), it can be observed that there is a positive correlation between the variable $\eta$ and the average degree of each node, network density, and total clustering coefficient. As the value of $\eta$ increases, these network characteristics consistently increase. Furthermore, the observed rise in the $\eta$ value signifies a heightened presence of a core-periphery structure within the network. This finding is substantiated by the core-periphery index and core-periphery error measures discussed in \cite{craig2014interbank}.

The impact of different values of $\eta$ on the systemic set-valued risk measure is further illustrated in \Cref{fig:b11}. As the parameter $\eta$ increases, there is a noticeable downward trend in the measure, indicating that higher values of $\eta$ are associated with a reduction in systemic risk. The correlation between the statistical variations presented in \Cref{combinedNetworkStats}(left) highlights the significant impact of $\eta$ on both network dynamics and systemic risk.

In the subsequent analysis, we investigate the impact on the systemic risk measure when the parameters $\theta$ and $\zeta$ are reversed, while maintaining the $\eta$ parameter fixed. The following parameter pairs are taken into consideration:
\begin{itemize}
	\item $(\theta = 0.1, \eta = 0.1, \zeta = 0.8)$ and $(\theta = 0.8, \eta = 0.1, \zeta = 0.1)$.
	\item $(\theta = 0.1, \eta = 0.3, \zeta = 0.6)$ and $(\theta = 0.6, \eta = 0.3, \zeta = 0.1)$.
	\item $(\theta = 0.1, \eta = 0.6, \zeta = 0.3)$ and $(\theta = 0.3, \eta = 0.6, \zeta = 0.1)$.	
\end{itemize}

\Cref{combinedNetworkStats} presents the computational statistics for these different network parameter combinations, and \Cref{combinedRiskMeasures} provides the corresponding approximations.

\begin{table}[htbp]
	\centering
	\resizebox{0.4\textwidth}{!}{
			\begin{tabular}{|c|c|c|c|c|c|}
				\hline
				\multirow{2}{*}{\textbf{$\theta$, $\eta$, $\zeta$}} & \multicolumn{5}{c|}{\textbf{Network Statistics}} \\ \cline{2-6} 
				& \textbf{AD} & \textbf{ND} & \textbf{TCC} & \textbf{CPE} & \textbf{CPI} \\ \hline
				\textbf{0.1, 0.1, 0.8} & 1.7 & 0.0895 & 0.3333 & 0.67 & 0.6471 \\ \hline
				\textbf{0.1, 0.3, 0.6} & 1.95 & 0.1026 & 1.6333 & 0.4103 & 0.7179 \\ \hline
				\textbf{0.1, 0.4, 0.5} & 2 & 0.1053 & 4.4111 & 0.375 & 0.7251 \\ \hline
				\textbf{0.1, 0.5, 0.4} & 2.45 & 0.1289 & 7.1659 & 0.1837 & 0.8367 \\ \hline
				\textbf{0.1, 0.6, 0.3} & 2.8 & 0.1474 & 11.293 & 0.1786 & 0.8393 \\ \hline
				\textbf{0.1, 0.8, 0.1} & 3.55 & 0.1868 & 14.5913 & 0.1672 & 0.8528 \\ \hline
			\end{tabular}%
		}
		\quad 
		\resizebox{0.4\textwidth}{!}{
			\begin{tabular}{|c|c|c|c|c|c|}
				\hline
				\multirow{2}{*}{\textbf{$\theta$, $\eta$, $\zeta$}} & \multicolumn{5}{c|}{\textbf{Network Statistics}} \\ \cline{2-6} 
				& \textbf{AD} & \textbf{ND} & \textbf{TCC} & \textbf{CPE} & \textbf{CPI} \\ \hline
				\textbf{0.1, 0.1, 0.8} & 1.7 & 0.0895 & 0.3333 & 0.6765 & 0.6471 \\ \hline
				\textbf{0.8, 0.1, 0.1} & 1.3 & 0.0684 & 2 & 0.5385 & 0.8462 \\ \hline
				\textbf{0.1, 0.3, 0.6} & 1.95 & 0.1026 & 1.633 & 0.4103 & 0.7179 \\ \hline
				\textbf{0.6, 0.3, 0.1} & 1.65 & 0.0868 & 6.3 & 0.4042 & 0.8485 \\ \hline
				\textbf{0.1, 0.6, 0.3} & 2.8 & 0.1474 & 11.293 & 0.1786 & 0.8493 \\ \hline
				\textbf{0.3, 0.6, 0.1} & 3.2 & 0.1684 & 15.419 & 0.1744 & 0.8656 \\ \hline
			\end{tabular}%
		}
		\vspace{5pt}
		\caption{Network statistics for different Bollobás probability parameters: changing $\eta$ (left), and reversing $\theta$ and $\zeta$ (right).}
		\label{combinedNetworkStats}
\end{table}

Upon examination of \Cref{combinedNetworkStats}(right), it is observed that the reversing of parameters $\theta$ and $\zeta$ does not yield a consistent increase or decrease in both the average degree of each node and the network density. Nevertheless, the overall clustering coefficient, as well as the core-periphery index and core-periphery error measures, suggest a more evident core-periphery structure in the network when the value of $\theta$ exceeds $\zeta$.

The results presented in \Cref{fig:b22} support the conclusions drawn from \Cref{combinedNetworkStats}, indicating that the systemic set-valued risk measure exhibits a decrease in risk as $\theta$ is greater than $\zeta$. The observed behavior can be ascribed to the emergence of a core-periphery configuration within the network, which has implications for the interconnectivity and dissemination of risks.

Through the implementation of this sensitivity analysis, we are able to acquire significant insights regarding the impact of various network parameters on the measure of systemic risk. It is worth noting that the presence of a core-periphery structure within the network has a significant impact on decreasing the systemic set-valued risk measure. This suggests that there may be opportunities to develop risk mitigation strategies based on this observation.

\section{Conclusion}

In this paper, we investigate the convergence properties of sample-average approximations for set-valued systemic risk measures. We have demonstrated that the optimal solutions of the SAA scalarization problems converge almost surely to the true set-valued risk measure. Moreover, we have shown that the SAA (in)sensitive set-valued risk measures converge to the true ones almost surelt in the Wijsman and Hausdorff topologies. Our modified grid algorithm, employing weighted-sum and norm-minimizing scalarizations, proves effective in approximating these systemic risk measures. Overall, this study enhances the understanding of systemic risk assessment and offers robust convergence guarantees for SAA approximations. These findings are valuable for developing more effective risk management strategies in the financial industry.

\bibliographystyle{plainnat}
\bibliography{AA_OPRE_arXiv}

\ECSwitch

\ECHead{Additional Results and Proofs}

\section{Projection Characterization of Set of All Eisenberg-Noe Clearing Vectors} \label{app:all}

Let $(\bm{\pi}, \bar{\bm{p}}, \bm{x})$ be an Eisenberg-Noe network with $d$ nodes. Recall that \Cref{lem:mp} shows that the maximal clearing vector of the network can be found by solving a continuous optimization problem, e.g., a linear programming problem. In this section, show that the calculation of the set of \emph{all} clearing vectors can be formulated as a mixed-integer polyhedral projection problem. In the recent literature, it has been shown that linear/convex vector optimization problems are closely related to polyhedral/convex projection problems; see \cite{polyhedralprojection}, \cite{convexprojection}. Hence, our result can be considered as a motivating example for studying the potential connection between mixed-integer linear vector optimization problems and mixed-integer polyhedral projection problems, which we leave for future research.

\begin{theorem}\label{thm:all}
	Let $C(\bm{x})$ denote the set of all clearing vectors of $(\bm{\pi}, \bar{\bm{p}}, \bm{x})$. Then,
	\begin{equation}\label{set}
		C(\bm{x}) = \cb{\bm{p} \in [\bm{0}_d,\bar{\bm{p}}] \mid \bm{p} \leq \bm{x} + \bm{\pi}^{\mathsf{T}}\bm{p},\  \bm{p} \geq \bar{\bm{p}}\bm{y},\ \bm{p} \geq \bm{x} + \bm{\pi}^{\mathsf{T}}\bm{p} -\bm{y}Q,\ \bm{y} \in \{0,1\}^d}, 
	\end{equation}
	where $Q \geq \max_{i\in[d]}((\bm{\pi}^{\mathsf{T}}\bar{\bm{p}})_i+x_i)$ is a constant.
\end{theorem}


\proof{Proof.}
Let $\tilde{C}(\bm{x})$ denote the set on the right of \eqref{set}. Let $\bm{p}\in [\bm{0}_d,\bar{\bm{p}}]\in C(\bm{x})$. We claim that $\bm{p} \in \tilde{C}(\bm{x})$. Since $\bm{p}$ is a clearing vector, we have $\bm{p} \leq \bm{\pi}^{\top}\bm{p} + \bm{x}$ by limited liability. Let $\bm{y}\in \{0,1\}^d$ be defined by
	\begin{equation} \label{indicator}
		y_i = 
		\begin{cases}
			0, & \text{if $p_{i} < \bar{p}_{i}$}, \\
			1, & \text{if $p_{i} = \bar{p}_{i}$}, \\
		\end{cases}
	\end{equation}
	for each $i\in[d]$. Let us fix $i\in[d]$. We consider the following two cases:
	\begin{itemize}
		\item Assume that $p_{i} = \bar{p_i}$. Then, by \eqref{indicator}, $y_{i} = 1$. Thus, the constraint $p_i\geq\bar{p_i}y_i$ is satisfied and the constraint $p_i \geq \sum_{j=1}^{d} \pi_{ji}p_j + x_i - y_iQ$ is trivially satisfied by the choice of $Q$. Hence, $(p_i, y_i)$ satisfies all the constraints in \eqref{set}.
		\item Assume that $p_i = \sum_{j=1}^{d} \pi_{ji}p_j + x_i < \bar{p_i}$. Then, by \eqref{indicator}, $y_{i} = 0$. Thus, the constraint $p_i\geq\bar{p_i}y_i$ is trivially satisfied and the constraint $p_i \geq \sum_{j=1}^{d} \pi_{ji}p_j + x_i - y_iQ$ yields that $p_i = \sum_{j=1}^{d} \pi_{ji}p_j + x_i$. Hence, $(p_i, y_i)$ satisfies all the constraints in \eqref{set}.
	\end{itemize}
	Therefore, $\bm{p}\in \tilde{C}(\bm{x})$.
	
	Conversely, let $\bm{p} \in \tilde{C}(\bm{x})$. We claim that $\bm{p}$ is a clearing vector. Since $\bm{p} \in C(\bm{x})$, there exists $\bm{y}\in\{0,1\}^d$ such that $(\bm{p}, \bm{y})$ satisfies all the constraints in \eqref{set}. In order to show that $\bm{p}$ is a clearing vector, we verify limited liability and absolute priority.
	
	The fact that $\bm{p} \in C(\bm{x})$ enforces $\bm{p}$ to satisfy the limited liability condition by the constraint $\bm{p} \leq \bm{\pi}^{\top}\bm{p} + \bm{x}$. To show that $\bm{p}$ must also satisfy the absolute priority condition, we consider the following two cases for each $i\in[d]$.
	\begin{itemize}
		\item Assume that $y_{i} = 1$. Then, $p_i = \bar{p_i} \leq \sum_{j=1}^{d}\pi_{ij}p_{j} + x_{i}$ by the constraints $p_i \leq \bar{p_i}$, $p_i \geq \bar{p_i}y_i$ and $p_i \leq \sum_{j=1}^{d} \pi_{ji}p_j + x_i$. Therefore, $p_i = \bar{p_i} \wedge (\sum_{j=1}^{d}\pi_{ij}p_{j} + x_{i})$, and thus the absolute priority condition holds.
		\item Assume that $y_{i} = 0$. Then, $p_i = \sum_{j=1}^{d}\pi_{ji} p_{j} + x_{i} \leq \bar{p_i}$ by the constraints $p_i \geq \sum_{j=1}^{d} \pi_{ji}p_j + x_i - y_iQ$, $p_i \leq \sum_{j=1}^{d} \pi_{ji}p_j + x_i$ and $p_i \leq \bar{p_i}$. Therefore, $p_i = \bar{p_i} \wedge (\sum_{j=1}^{d}\pi_{ij}p_{j} + x_{i})$, and thus the absolute priority condition holds. 
	\end{itemize}
	Hence, $\bm{p}\in C(\bm{x})$.
\Halmos
\endproof

\section{Proofs of the Results in \Cref{sec:ENM}}\label{app:ENM}

\proof{Proof of \Cref{lem:ins}.}
Let $\bm{X}^1,\bm{X}^2\in\X_\Lambda$ be with $\bm{X}^1\geq \bm{X}^2$. Then, $\Lambda(\bm{X}^1)\geq \Lambda(\bm{X}^2)$ by the monotonicity of $\Lambda$. Moreover, $\Lambda(\bm{X}^1),\Lambda(\bm{X}^2)\in\Y$ by \Cref{asmp:domain}. Hence, $r(\bm{X}^1)=\rho(\Lambda(\bm{X}^1))\leq \rho(\Lambda(\bm{X}^2))=r(\bm{X}^2)$ by the monotonicity of $\rho$.
\Halmos
\endproof

\proof{Proof of \Cref{lem:systrisk}.}
Let $\bm{X}^1,\bm{X}^2\in\X$ with $\bm{X}^1\geq \bm{X}^2$. Let $\bm{z}\in R(\bm{X}^2)$. Since $\bm{X}^2+\bm{B}^{\mathsf{T}}\bm{z}\in\X_\Lambda$ and $\Lambda$ is monotone, we have 
\[
\mathbb{P}\{\Lambda(\bm{X}^1+\bm{B}^{\mathsf{T}}\bm{z})>-\infty\}\geq \mathbb{P}\{\Lambda(\bm{X}^2+\bm{B}^{\mathsf{T}}\bm{z})>-\infty\}=1
\]
so that $\bm{X}^1+\bm{B}^{\mathsf{T}}\bm{z}\in\X_\Lambda$. Moreover, since $\Lambda(\bm{X}^2+\bm{B}^{\mathsf{T}}\bm{z})\in\A$ and by the monotonicity of $\rho$ and $\Lambda$, we have $\rho(\Lambda(\bm{X}^1+\bm{B}^{\mathsf{T}}\bm{z}))\leq \rho(\Lambda(\bm{X}^2+\bm{B}^{\mathsf{T}}\bm{z}))\leq 0$ so that $\Lambda(\bm{X}^1+\bm{B}^{\mathsf{T}}\bm{z})\in\A$. Therefore, $\bm{z}\in R(\bm{X}^1)$, which completes the proof of monotonicity.

Next, let $\bm{X}\in\X$ and $\bm{z}\in\R^g$. Then, by a simple translation of sets, we obtain
\begin{align}
	R(\bm{X}+\bm{B}^{\mathsf{T}}\bm{z})&=\{\bm{v}\in\R^g\mid \Lambda(\bm{X}+\bm{B}^{\mathsf{T}}(\bm{z}+\bm{v}))\in\A,\ \bm{X}+\bm{B}^{\mathsf{T}}(\bm{z}+\bm{v})\in \X_\Lambda\}
	\notag \\
	&=\{\bar{\bm{v}}\in\R^g\mid \Lambda(\bm{X}+\bm{B}^{\mathsf{T}}\bar{\bm{v}})\in\A,\ \bm{X}+\bm{B}^{\mathsf{T}}\bar{\bm{v}}\in \X_\Lambda\}-\bm{z}=R(\bm{X})-\bm{z},
	\notag 
\end{align}
which completes the proof of translativity.

Finally, let $\bm{X}\in\X$. If $R(\bm{X})=\emptyset$, then the upper set property holds trivially. Suppose that $R(\bm{X})\neq\emptyset$. Clearly, $R(\bm{X})\subseteq R(\bm{X})+\R^g_+$. Let $\bm{z}\in\R^g_+$. Since $\bm{X}\geq \bm{X}-\bm{B}^{\mathsf{T}}\bm{z}$, by the monotonicity and translativity of $R$, we have $R(\bm{X})+\bm{z}=R(\bm{X}-\bm{B}^{\mathsf{T}}\bm{z})\subseteq R(\bm{X})$. It follows that $R(\bm{X})+\R^g_+\subseteq R(\bm{X})$.
\Halmos
\endproof

\section{SAA for the Insensitive Systemic Value-at-Risk}\label{app:ins}

In this section, we prove a convergence result for the insensitive systemic value-at-risk defined in \Cref{defn:syst}(i).

Let us fix an aggregation function $\Lambda$ that satisfies \Cref{asmp:dom} and \Cref{asmp:domain}. We also fix $\alpha\in\R$ and $\lambda\in (0,1)$. Let $\bm{X}\in\X_\Lambda$ with a sequence $(\bm{X}^N)_{N\in\N}$ of its independent copies.

Let $N\in\N$. For each $\bm{x}^1,\ldots,\bm{x}^N\in \R^d$, we define
\[
r_{\alpha,\lambda}^N(\bm{x}^1,\ldots,\bm{x}^N)\coloneqq \inf\cb{y\in\R\mid \frac{1}{N}\sum_{n=1}^N 1_{[-\infty,\alpha)}(\Lambda(\bm{x}^n)+y)\leq \lambda}.
\]
Clearly, when $\bm{x}^1,\ldots,\bm{x}^N\in \dom \Lambda$, this is a special case of the insensitive risk measure for a random vector that is uniformly distributed over the finite set $\{\bm{x}^1,\ldots,\bm{x}^N\}$ (assuming distinct values; otherwise, another discrete distribution where probabilities are integer multiples of $\frac1N$). Moreover, since $r^N_{\alpha,\lambda}=r^N_{0,\lambda}+\alpha$, the calculation of $r^N_{\alpha,\lambda}$ reduces down to a simple quantile calculation based on a histogram. The next lemma gives simple bounds on the values of $r^N_{\alpha,\lambda}$.

\begin{lemma}\label{lem:bound}
	Let $\bm{x}^1,\ldots,\bm{x}^N\in \dom \Lambda$, where $N\in\N$. Then,
	\[
	\alpha-\max_{n\in[N]}\Lambda(\bm{x}^n)\leq r_{\alpha,\lambda}^N(\bm{x}^1,\ldots,\bm{x}^N)\leq \alpha-\min_{n\in[N]}\Lambda(\bm{x}^n).
	\]
\end{lemma}

\proof{Proof of \Cref{lem:bound}.}
Since $y\mapsto \frac{1}{N}\sum_{n=1}^N 1_{[-\infty,\alpha)}(\Lambda(\bm{x}^n)+y)$ is a decreasing function on $\R$, the set in the definition of $r_{\alpha,\lambda}^N(\bm{x}^1,\ldots,\bm{x}^N)$ is an interval that is unbounded from above. Taking $y_1\coloneqq \alpha-\max_{n\in[N]}\Lambda(\bm{x}^n)$ yields $ \frac{1}{N}\sum_{n=1}^N 1_{[-\infty,\alpha)}(\Lambda(\bm{x}^n)+y_1)=1>\lambda$ so that the first inequality in the lemma follows. Taking $y_2\coloneqq \alpha-\min_{n\in[N]}\Lambda(\bm{x}^n)$ yields $ \frac{1}{N}\sum_{n=1}^N 1_{[-\infty,\alpha)}(\Lambda(\bm{x}^n)+y_2)=0\leq\lambda$ so that the second inequality in the lemma follows.
\Halmos
\endproof

The measurability of $r^N_{\alpha,\lambda}$ is checked in the next lemma.

\begin{lemma}\label{lem:measurable}
	Let $N\in\N$. The function $r^N_{\alpha,\lambda}\colon \R^{dN}\to \R\cup\{-\infty\}$ is measurable.
\end{lemma}

\proof{Proof of \Cref{lem:measurable}.}
Let $g(y,\bm{x})\coloneqq \Lambda(\bm{x})+y$ for each $\bm{x}\in\R^d$ and $y\in\R$. \Cref{asmp:dom} guarantees that $\Lambda$ is measurable. Hence, $g$ is a Carath\'{e}odory function. Then, by \Cref{prop:technical}(i), the function
\[
(y;\bm{x}^1,\ldots,\bm{x}^N)\mapsto p^N(y;\bm{x}^1,\ldots,\bm{x}^N)\coloneqq \frac{1}{N}\sum_{n=1}^N1_{[-\infty,\alpha)}(\Lambda(\bm{x}^n)+y)
\]
is a normal integrand on $\R\times\R^{dN}$. In particular, the multifunction $(\bm{x}^1,\ldots,\bm{x}^N)\mapsto \{y\in\R\mid p^N(y;\bm{x}^1,\ldots,\bm{x}^N)\leq \lambda\}$ is closed-valued and measurable by \citet[Proposition~14.33]{rockafellar2009variational}. Then, as the pointwise infimum of this multifunction, $r^N_{\alpha,\lambda}$ is measurable by \citet[Example~14.51]{rockafellar2009variational}.
\Halmos
\endproof

In view of \Cref{lem:measurable}, composing the function $r^N_{\alpha,\lambda}$ with the first $N$ independent copies of $\bm{X}$ yields the random variable $r^N_{\alpha,\lambda}(\bm{X}^1,\ldots,\bm{X}^N)$. The first main result of the paper addresses the convergence of this random variable to the true value $r_{\alpha,\lambda}(\bm{X})$ as $N\rightarrow\infty$.

\begin{theorem}\label{thm:ins}
	Assume that $\lambda$ is not a flat value of the cumulative distribution function of $\Lambda(\bm{X})$, i.e., the set $\{y\in\R\mid \mathbb{P}\{\Lambda(\bm{X})\leq y\}=\lambda\}$ is either the empty set or a singleton. Then, the sequence $(r^N_{\alpha,\lambda}(\bm{X}^1,\ldots,\bm{X}^N))_{N\in\N}$ converges to $r_{\alpha,\lambda}(\bm{X})$ almost surely.
\end{theorem}

\proof{Proof of \Cref{thm:ins}.}
As a first step, suppose that there exists $y\in\R$ such that $\mathbb{P}\{\Lambda(\bm{X})+y<\alpha\}<\lambda$. Let
\[
\O^{\text{LLN}}_y\coloneqq\cb{\lim_{N\rightarrow\infty}\frac1N\sum_{n=1}^N 1_{[-\infty,\alpha)}(\Lambda(\bm{X}^n)+y)=\mathbb{P}\{\Lambda(\bm{X})+y<\alpha\}}.
\]
By the strong law of large numbers, we have $\mathbb{P}(\O^{\text{LLN}}_y)=1$. Let $\o\in\O^{\text{LLN}}_y$. Since $\mathbb{P}\{\Lambda(\bm{X})+y<\alpha\}<\lambda$, there exists $N_{\o,y}\in\N$ such that
\[
\frac1N\sum_{n=1}^N 1_{[-\infty,\alpha)}(\Lambda(\bm{X}^n(\o))+y)<\lambda
\]
holds for every $N\geq N_{\o,y}$. In particular, $y$ is feasible for the calculation of $r^N_{\alpha,\lambda}(\bm{X}^1(\o),\ldots,\bm{X}^N(\o))$ so that $r^N_{\alpha,\lambda}(\bm{X}^1(\o),\ldots,\bm{X}^N(\o))\leq y$ for every $N\geq N_{\o,y}$. Hence,
\begin{equation}\label{eq:limsup}
	\limsup_{N\rightarrow\infty} r^N_{\alpha,\lambda}(\bm{X}^1(\o),\ldots,\bm{X}^N(\o))\leq y.
\end{equation}

Next, note that $\mathbb{P}\{\Lambda(\bm{X})>-\infty\}=1$, which implies that $\bar{y}\coloneqq r_{\alpha,\lambda}(\bm{X})=\var_{\alpha,\lambda}(\Lambda(\bm{X}))\in \R$ and it satisfies $\mathbb{P}\{\Lambda(\bm{X})+\bar{y}<\alpha\}\leq \lambda$. Moreover, since the cardinality of $\{y\in\R\mid \mathbb{P}\{\Lambda(\bm{X})\leq y\}=\lambda\}$ is at most one, the cardinality of $\{y\in\R\mid \mathbb{P}\{\Lambda(\bm{X})< y\}=\lambda\}$ is also at most one. Hence, we have $\mathbb{P}\{\Lambda(\bm{X})+y< \alpha\}<\lambda$ for every $y<\bar{y}$. Then, there exists a sequence $(y^\ell)_{\ell\in\N}$ that converges to $\bar{y}$ and satisfies $\mathbb{P}\{\Lambda(\bm{X})+y^\ell< \alpha\}<\lambda$ for every $\ell\in\N$. Let $\O_{\bar{y}}\coloneqq \bigcap_{\ell=1}^\infty \O^{\text{LLN}}_{y^\ell}$. Note that $\mathbb{P}(\O_{\bar{y}})=1$. Given $\o\in\O_{\bar{y}}$, by taking $y=y^\ell$ in \eqref{eq:limsup} for each $\ell\in\N$ and letting $\ell\rightarrow\infty$, we get
\[
\limsup_{N\rightarrow\infty} r^N_{\alpha,\lambda}(\bm{X}^1(\o),\ldots,\bm{X}^N(\o))\leq \bar{y}=r_{\alpha,\lambda}(\bm{X}).
\]
Hence, $\limsup_{N\rightarrow\infty} r^N_{\alpha,\lambda}(\bm{X}^1,\ldots,\bm{X}^N)\leq r_{\alpha,\lambda}(\bm{X})$ almost surely.

Finally, let us define $\O^\Lambda\coloneqq \bigcap_{N=1}^\infty \{\bm{X}^N\in\dom\Lambda\}$. Since $\bm{X}^N\in\X_\Lambda$ for each $N\in\N$, we have $\mathbb{P}(\O^\Lambda)=1$. Moreover, let $g(y,\bm{x})\coloneqq\Lambda(\bm{x})+y$ for each $\bm{x}\in\R^d$ and $y\in\R$. Since $g$ is a Carath\'eodory function, by \Cref{prop:technical}(iii), there exists $\O^{\text{EPI}}\in\mathcal{F}$ with $\mathbb{P}(\O^{\text{EPI}})=1$ such that the sequence  $(p^N(\cdot;\bm{X}^1(\o),\ldots,\bm{X}^N(\o)))_{N\in\N}$ epi-converges to $p(\cdot)$ for each $\o\in\O^{\text{EPI}}$, where $(p^N)_{N\in\N}$ and $p$ are defined in this proposition in terms of $g$. In particular, $\mathbb{P}(\O^{\Lambda}\cap \O^{\text{EPI}})=1$. Let $\o\in\O^{\Lambda}\cap \O^{\text{EPI}}$ and $N\in\N$. Let us define $y^N(\o)\coloneqq r_{\alpha,\lambda}^N(\bm{X}^1(\o),\ldots,\bm{X}^N(\o))$. Since $\Lambda(\bm{X}^n(\o))>-\infty$ for each $n\in[N]$, we have $y^N(\o)\in\R$ and
\[
\frac{1}{N}\sum_{n=1}^N 1_{[-\infty,\alpha)}(\Lambda(\bm{X}^n(\o))+y^N(\o))\leq \lambda.
\]
Moreover, \Cref{lem:bound} and the boundedness of $\Lambda$ on $\dom\Lambda$ imply that $(y^N(\o))_{N\in\N}$ is a bounded sequence. Hence, it has a subsequence $(y^{N_k(\o)}(\o))_{k\in\N}$ that converges to some $y(\o)\in\R$. Then, by the epi-convergence property in \Cref{prop:technical}(iii), we get
\[
\lambda \geq \liminf_{k\rightarrow\infty}\frac{1}{N_k(\o)}\sum_{n=1}^{N_k(\o)}1_{[-\infty,\alpha)}(\Lambda(\bm{X}^n(\o))+y^{N_k(\o)}(\o))\geq \mathbb{P}\{\Lambda(\bm{X})+y(\o)<\alpha\}.
\]
Hence, $y(\o)$ is feasible for the calculation of $r_{\alpha,\lambda}(\bm{X})$ so that
\[
r_{\alpha,\lambda}(\bm{X})\leq y(\o)=\lim_{k\rightarrow\infty}y^{N_k(\o)}(\o) = \lim_{k\rightarrow\infty}r_{\alpha,\lambda}^{N_k(\o)}(\bm{X}^1(\o),\ldots,\bm{X}^{N_k(\o)}(\o)).
\]
It follows that
\[
r_{\alpha,\lambda}(\bm{X})\leq \liminf_{N\rightarrow\infty}r_{\alpha,\lambda}^N(\bm{X}^1(\o),\ldots,\bm{X}^N(\o)).
\]
Therefore, $r_{\alpha,\lambda}(\bm{X})\leq \liminf_{N\rightarrow\infty} r^N_{\alpha,\lambda}(\bm{X}^1,\ldots,\bm{X}^N)$ almost surely, which completes the proof.
\Halmos
\endproof

\section{Proofs of the Results in \Cref{sec:theory}}\label{app:theory}

\proof{Proof of \Cref{prop:technical}.}
(i) Since $D$ is open, the function $1_D$ is lower semicontinuous on $[-\infty,+\infty]$ (on $\R^{d^\prime}$). Since $G$ is a Carath\'{e}odory function, $(\bm{y},\bm{x})\mapsto 1_D(g(\bm{y},\bm{x}))$ is a normal integrand on $\R^m\times\R^d$ by \citet[Proposition~14.45]{rockafellar2009variational}. It follows that $p^N$ is a normal integrand on $\R^m\times \R^{dN}$.\\
(ii) Since $(\bm{y},\bm{x})\mapsto 1_D(g(\bm{y},\bm{x}))$ is a normal integrand, by Fatou's lemma and monotonicity of expectation, we have
\[
\liminf_{\bm{y}^\prime\rightarrow\bm{y}}p(\bm{y}^{\prime})=\liminf_{\bm{y}^\prime \rightarrow\bm{y}}\mathbb{E}[1_D(g(\bm{y}^{\prime},\bm{Z}))]
\geq \mathbb{E}\sqb{ \liminf_{\bm{y}^\prime\rightarrow\bm{y}}1_D(g(\bm{y}^\prime,\bm{Z}))}
\geq \mathbb{E}[1_D(g(\bm{y},\bm{Z}))]=p(\bm{y})
\]
for each $\bm{y}\in\R^m$, which shows the lower semicontinuity of $p$.\\
(iii) Since $(\bm{y},\bm{x})\mapsto 1_D(g(\bm{y},\bm{x}))$ is a normal integrand and it is a bounded function, the epi-convergence result is directly given by \citet[Theorem~2.3]{artstein}.
\Halmos
\endproof

\proof{Proof of \Cref{lem:closed}.}
Note that $\dom\Lambda$ is closed and $\Lambda$ is continuous on $\dom\Lambda$ by \Cref{asmp:dom}. Hence, by Tietze extension theorem, there exists a continuous function $\tilde{\Lambda}\colon \R^d\to\R$ such that $\Lambda(\bm{x})=\tilde{\Lambda}(\bm{x})$ for every $\bm{x}\in\dom\Lambda$. Then, we may write
\[
R_{\alpha,\lambda}(\bm{X})=\cb{\bm{z}\in\R^g\mid \mathbb{P}\{\tilde{\Lambda}(\bm{X}+\bm{B}^{\mathsf{T}}\bm{z})\in (-\infty,\alpha)\}\leq\lambda, \mathbb{P}\{\bm{X}+\bm{B}^{\mathsf{T}}\bm{z}\in (\dom\Lambda)^c\}\leq 0}.
\]
Let $g_1(\bm{z},\bm{x})\coloneqq \tilde{\Lambda}(\bm{x}+\bm{B}^{\mathsf{T}}\bm{z})$ and $g_2(\bm{z},\bm{x})\coloneqq \bm{x}+\bm{B}^{\mathsf{T}}\bm{z}$ for each $\bm{z}\in\R^g$ and $\bm{x}\in\R^d$. Since $\tilde{\Lambda}$ is continuous, it is also measurable. Hence, $g_1$ is a Carath\'{e}odory function. Clearly, $g_2$ is also a Carath\'{e}odory function. Then, by \Cref{prop:technical}(i), the functions
\begin{align}
	&\bm{z}\mapsto p_1(\bm{z})\coloneqq \mathbb{P}\{\tilde{\Lambda}(\bm{X}+\bm{B}^{\mathsf{T}}\bm{z})\in (-\infty,\alpha)\},\notag \\ &\bm{z}\mapsto p_2(\bm{z})\coloneqq \mathbb{P}\{\bm{X}+\bm{B}^{\mathsf{T}}\bm{z}\in (\dom\Lambda)^c\}\notag 
\end{align}
are lower semicontinuous on $\R^g$. In particular, the set
\[
R_{\alpha,\lambda}(\bm{X})=\cb{\bm{z}\in\R^g\mid p_1(\bm{z})\leq \lambda,\ p_2(\bm{z})\leq 0}
\]
is closed. Finally, the property $R_{\alpha,\lambda}(\bm{X})=\cl(R_{\alpha,\lambda}(\bm{X})+\R^g_+)$ follows from closedness combined with the upper set property in \Cref{lem:systrisk}.
\Halmos
\endproof 

\proof{Proof of \Cref{lem:meas-sen}.}
Let $\tilde{\Lambda}, g_1, g_2$ be as in the proof of \Cref{lem:closed}. For each $\bm{x}^1,\ldots,\bm{x}^N\in\R^d$, we may write
\[
R^N_{\alpha,\lambda}(\bm{x}^1,\ldots,\bm{x}^N)\negthinspace=\negthinspace \cb{\bm{z}\in\R^g\mid \frac{1}{N}\sum_{n=1}^N 1_{(-\infty,\alpha)}(\tilde{\Lambda}(\bm{x}^n+\bm{B}^{\mathsf{T}}\bm{z}))\leq\lambda, \frac{1}{N}\sum_{n=1}^N 1_{(\dom\Lambda)^c}(\bm{x}^n+\bm{B}^{\mathsf{T}}\bm{z})\leq 0}.
\]
Since $g_1,g_2$ are Carath\'{e}ordory functions, by \Cref{prop:technical}(i), the functions
\begin{align}
	&(\bm{z};\bm{x}^1,\ldots,\bm{x}^N)\mapsto p^N_1(\bm{z};\bm{x}^1,\ldots,\bm{x}^N)\coloneqq \frac{1}{N}\sum_{n=1}^N 1_{(-\infty,\alpha)}(\tilde{\Lambda}(\bm{x}^n+\bm{B}^{\mathsf{T}}\bm{z})),\notag \\ &(\bm{z};\bm{x}^1,\ldots,\bm{x}^N)\mapsto p^N_2(\bm{z};\bm{x}^1,\ldots,\bm{x}^N)\coloneqq \frac{1}{N}\sum_{n=1}^N 1_{(\dom\Lambda)^c}(\bm{x}^n+\bm{B}^{\mathsf{T}}\bm{z})\notag 
\end{align}
are normal integrands on $\R^g\times\R^{dN}$. In particular, the multifunction
\begin{equation}\label{eq:p1p2}
	(\bm{x}^1,\ldots,\bm{x}^N)\negthinspace\mapsto\negthinspace R^N_{\alpha,\lambda}(\bm{x}^1,\ldots,\bm{x}^N)\negthinspace=\negthinspace\{\bm{z}\in\R^g\mid p^N_1(\bm{z};\bm{x}^1,\ldots,\bm{x}^N)\leq \lambda,  p^N_2(\bm{z};\bm{x}^1,\ldots,\bm{x}^N)\leq 0 \}
\end{equation}
is closed-valued and measurable by \citet[Proposition~14.33]{rockafellar2009variational}.
\Halmos
\endproof

\proof{Proof of \Cref{lem:meas-sc}.}
Since $\varphi$ is a continuous function, $R^N_{\alpha,\lambda}$ is closed-valued and measurable, and $\mathcal{Z}$ is a closed set, the measurability of the scalarization $(\bm{x}^1,\ldots,\bm{x}^N) \mapsto \mathsf{u}(\varphi,R^N_{\alpha,\lambda}(\bm{x}^1,\ldots,\bm{x}^N)\cap\mathcal{Z})$ follows as a direct consequence of \citet[Example~14.32]{rockafellar2009variational}.
\Halmos
\endproof

\proof{Proof of \Cref{thm:generalconv}.}
First, let us consider an arbitrary vector $\bm{z}\in \mathcal{Z}$ such that $\mathbb{P}\{\Lambda(\bm{X}+\bm{B}^{\mathsf{T}}\bm{z})<\alpha\}<\lambda$ and $\mathbb{P}\{\bm{X}+\bm{B}^{\mathsf{T}}\bm{z}\in\dom \Lambda\}=1$. Let $\O_{\bm{z}}^{\Lambda}\coloneqq\bigcap_{N=1}^\infty\{\bm{X}^N+\bm{B}^{\mathsf{T}}\bm{z}\in\dom\Lambda\}$. Clearly, $\mathbb{P}(\O^\Lambda_{\bm{z}})=1$. Moreover, let
\[
\O^{\text{LLN}}_{\bm{z}}\coloneqq \cb{\lim_{N\rightarrow\infty}\frac{1}{N}\sum_{n=1}^N 1_{[-\infty,\alpha)}(\Lambda(\bm{X}^n+\bm{B}^{\mathsf{T}}\bm{z}))=\mathbb{P}\{\Lambda(\bm{X}+\bm{B}^{\mathsf{T}}\bm{z})<\alpha\}}.
\]
By the strong law of large numbers, we have $\mathbb{P}(\O^{\text{LLN}}_{\bm{z}})=1$. In particular, as $\mathbb{P}\{\Lambda(\bm{X}+\bm{B}^{\mathsf{T}}\bm{z})<\alpha\}<\lambda$, for every $\o\in\O^{\text{LLN}}_{\bm{z}}$, there exists $N_{\o,\bm{z}}\in\N$ such that
\[
\frac{1}{N}\sum_{n=1}^N 1_{[-\infty,\alpha)}(\Lambda(\bm{X}^n(\o)+\bm{z}))<\lambda
\]
holds for every $N\geq N_{\o,\bm{z}}$. Therefore, for every $\o\in \O^\Lambda_{\bm{z}}\cap\O^{\text{LLN}}_{\bm{z}}$, $\bm{z}$ is feasible for the calculation of $\mathsf{u}(\varphi,R^N_{\alpha,\lambda}(\bm{X}^1(\o),\ldots,\bm{X}^N(\o))\cap\mathcal{Z})$ for every $N\geq N_{\o,\bm{z}}$. Hence,
\begin{equation}\label{eq:limsup-sen}
	\limsup_{N\rightarrow\infty} \mathsf{u}(\varphi,R^N_{\alpha,\lambda}(\bm{X}^1(\o),\ldots,\bm{X}^N(\o))\cap\mathcal{Z}) \leq \varphi(\bm{z}).
\end{equation}

Next, let $\bar{\bm{z}}$ and $(\bm{z}^\ell)_{\ell\in\N}$ be given by \Cref{asmp:slater}. For each $\ell\in\N$, we have $\mathbb{P}\{\Lambda(\bm{X}+\bm{B}^{\mathsf{T}}\bm{z}^\ell)<\alpha\}<\lambda$ and $\mathbb{P}\{\bm{X}+\bm{B}^{\mathsf{T}}
\bm{z}^\ell\in \dom\Lambda\}=1$. Let us define
\[
\O_{\bar{\bm{z}}}\coloneqq \bigcap_{\ell=1}^\infty (\O^{\Lambda}_{\bm{z}^\ell}\cap \O^{\text{LLN}}_{\bm{z}^\ell}).
\]
Note that $\mathbb{P}(\O_{\bar{\bm{z}}})=1$. Given $\o\in\O_{\bar{\bm{z}}}$, taking $\bm{z}=\bm{z}^\ell$ in \eqref{eq:limsup-sen} for each $\ell\in\N$ and letting $\ell\rightarrow\infty$ gives
\[
\limsup_{N\rightarrow\infty} \mathsf{u}(\varphi,R^N_{\alpha,\lambda}(\bm{X}^1(\o),\ldots,\bm{X}^N(\o))\cap\mathcal{Z}) \leq \liminf_{\ell\rightarrow\infty}\varphi(\bm{z}^\ell)=\varphi(\bar{\bm{z}})=\mathsf{u}(\varphi,R_{\alpha,\lambda}(\bm{X})\cap\mathcal{Z})
\]
since $(\bm{z}^\ell)_{\ell\in\N}$ converges to $\bar{\bm{z}}$ and $\varphi$ is continuous. Hence, $	\limsup_{N\rightarrow\infty} \mathsf{u}(\varphi,R^N_{\alpha,\lambda}(\bm{X}^1,\ldots,\bm{X}^N)\cap\mathcal{Z}) \leq \mathsf{u}(\varphi,R_{\alpha,\lambda}(\bm{X})\cap\mathcal{Z})$ almost surely.

Conversely, let $\tilde{\Lambda},g_1,g_2,(p^N_1)_{N\in\N},(p^N_2)_{N\in\N}$ be defined as in the proof of \Cref{lem:meas-sen} and recall that we have \eqref{eq:p1p2} for each $\bm{x}^1,\ldots,\bm{x}^N\in\R^d$ and $N\in\N$. Let $p_1(\bm{z})\coloneqq \mathbb{P}\{g_1(\bm{z},\bm{X})<\alpha\}$ and $p_2(\bm{z})\coloneqq \mathbb{P}\{g_2(\bm{z},\bm{X})\in (\dom \Lambda)^c\}$ for each $\bm{z}\in\R^g$. Similar to \eqref{eq:p1p2}, note that we also have
\[
R_{\alpha,\lambda}(\bm{X})=\{\bm{z}\in\R^g\mid p_1(\bm{z})\leq\lambda,\ p_2(\bm{z})\leq 0\}.
\] 
Moreover, by \Cref{prop:technical}(iii), there exists $\O^{\text{EPI}}\in\mathcal{F}$ with $\mathbb{P}(\O^{\text{EPI}})=1$ such that the sequences $(p_1^N(\cdot;\bm{X}^1(\o),\ldots,\bm{X}^N(\o)))_{N\in\N}$ and $(p_2^N(\cdot;\bm{X}^1(\o),\ldots,\bm{X}^N(\o)))_{N\in\N}$ epi-converge to $p_1(\cdot)$, $p_2(\cdot)$, respectively, for each $\o\in\O^{\text{EPI}}$. In particular, $\mathbb{P}(\O^\Lambda_{\bar{\bm{z}}}\cap\O^{\text{LLN}}_{\bar{\bm{z}}}\cap \O^{\text{EPI}})=1$.

Let $\o\in\O^\Lambda_{\bar{\bm{z}}}\cap\O^{\text{LLN}}_{\bar{\bm{z}}}\cap \O^{\text{EPI}}$ and $N\in\N$. By the first part of the proof, $\bar{\bm{z}}$ is feasible for the calculation of $\mathsf{u}(\varphi,R^N_{\alpha,\lambda}(\bm{X}^1(\o),\ldots,\bm{X}^N(\o))\cap\mathcal{Z})$ for every $N\geq N_{\o,\bar{\bm{z}}}$. Since $R^N_{\alpha,\lambda}(\bm{X}^1(\o),\ldots,\bm{X}^N(\o))$ is a closed set by \Cref{lem:meas-sen} and $\mathcal{Z}$ is a compact set, it follows that there exists an optimal solution $\bm{z}^N(\o)$, i.e., $\bm{z}^N(\o)\in R^N_{\alpha,\lambda}(\bm{X}^1(\o),\ldots,\bm{X}^N(\o))\cap\mathcal{Z}$ and
\[
\varphi(\bm{z}^N(\o))=\mathsf{u}(\varphi,R^N_{\alpha,\lambda}(\bm{X}^1(\o),\ldots,\bm{X}^N(\o))\cap\mathcal{Z})
\]
for every $N\geq N_{\o,\bar{\bm{z}}}$. Since $\mathcal{Z}$ is compact, there exists a subsequence $(\bm{z}^{N_k(\o)}(\o))_{k\in\N}$ that converges to some $\bm{z}(\o)\in\mathcal{Z}$. Moreover, the epi-convergence property given by \Cref{prop:technical}(iii) yields that
\[
\lambda\geq \liminf_{k\rightarrow\infty}\frac{1}{N_k(\o)}\sum_{n=1}^{N_k(\o)}1_{(-\infty,\alpha)}(\tilde{\Lambda}(\bm{X}^n(\o)+\bm{B}^{\mathsf{T}}\bm{z}^n(\o)))\geq \mathbb{P}\{\tilde{\Lambda}(\bm{X}+\bm{B}^{\mathsf{T}}\bm{z}(\o))<\alpha\}
\]
as well as
\[
0\geq  \liminf_{k\rightarrow\infty}\frac{1}{N_k(\o)}\sum_{n=1}^{N_k(\o)}1_{(\dom\Lambda)^c}(\bm{X}^n(\o)+\bm{B}^{\mathsf{T}}\bm{z}^n(\o))\geq \mathbb{P}\{\bm{X}+\bm{B}^{\mathsf{T}}\bm{z}(\o)\notin\dom\Lambda\}
\]
In other words, we obtain
\begin{align}
	& p_1(\bm{z}(\o)) =\mathbb{P}\{\tilde{\Lambda}(\bm{X}+\bm{B}^{\mathsf{T}}\bm{z}(\o))<\alpha\}\leq\lambda,\notag\\
	&p_2(\bm{z}(\o)) =\mathbb{P}\{\bm{X}+\bm{B}^{\mathsf{T}}\bm{z}(\o)\in\dom\Lambda\}\leq 0.\notag
\end{align}
Hence, $\bm{z}(\o)$ is feasible for the calculation of $\mathsf{u}(\varphi,R_{\alpha,\lambda}(\bm{X})\cap\mathcal{Z})$ so that
\[
\mathsf{u}(\varphi,R_{\alpha,\lambda}(\bm{X})\cap\mathcal{Z})\leq \varphi(\bm{z}(\o))=\lim_{k\rightarrow\infty}\varphi(\bm{z}^{N_k(\o)}(\o))=\lim_{k\rightarrow\infty}\mathsf{u}(\varphi,R^N_{\alpha,\lambda}(\bm{X}^1(\o),\ldots,\bm{X}^{N_k(\o)}(\o))\cap\mathcal{Z}).
\]
It follows that
\[
\mathsf{u}(\varphi,R_{\alpha,\lambda}(\bm{X})\cap\mathcal{Z})\leq \liminf_{N\rightarrow\infty} \mathsf{u}(\varphi,R^N_{\alpha,\lambda}(\bm{X}^1(\o),\ldots,\bm{X}^N(\o))\cap\mathcal{Z}).
\]
Therefore, $\mathsf{u}(\varphi,R_{\alpha,\lambda}(\bm{X})\cap\mathcal{Z})\leq \liminf_{N\rightarrow\infty} \mathsf{u}(\varphi,R^N_{\alpha,\lambda}(\bm{X}^1,\ldots,\bm{X}^N)\cap\mathcal{Z}) $ almost surely.
\Halmos
\endproof

\proof{Proof of \Cref{thm:scalar}.}
Let $\bar{\bm{z}}, (\bm{z}^\ell)_{\ell\in\N}$ be as given by \Cref{asmp:scalar}. Note that we have 
\begin{align}
	\mathsf{s}(\bm{w},R_{\alpha,\lambda}(\bm{X}))&=\inf_{\bm{z}\in R_{\alpha,\lambda}(\bm{X})\cap\mathcal{Z}+\R^g_+}\bm{w}^{\mathsf{T}}\bm{z}\notag \\
	&=\inf_{\bm{z}\in R_{\alpha,\lambda}(\bm{X})\cap\mathcal{Z}}\bm{w}^{\mathsf{T}}\bm{z}+\inf_{\bm{z}\in \R^g_+}\bm{w}^{\mathsf{T}}\bm{z}\notag \\
	&=\inf_{\bm{z}\in R_{\alpha,\lambda}(\bm{X})\cap\mathcal{Z}}\bm{w}^{\mathsf{T}}\bm{z}=\mathsf{s}(\bm{w},R_{\alpha,\lambda}(\bm{X})\cap \mathcal{Z}).\notag 
\end{align}
Moreover, we have $\bar{\bm{z}}\in\mathcal{Z}$ since $\bm{z}^\ell\in \mathcal{Z}$ for each $\ell\in\N$. Let $\varphi(\bm{z})\coloneqq \bm{w}^{\mathsf{T}}\bm{z}$ for each $\bm{z}\in\R^g$. Then, the sequence $(\bm{z}^\ell)_{\ell\in\N}$ satisfies \Cref{asmp:slater} for this choice of $\varphi$. Therefore, by \Cref{thm:generalconv}, $(\mathsf{s}(\bm{w},R^N_{\alpha,\lambda}(\bm{X}^1,\ldots,\bm{X}^N))\cap \mathcal{Z})_{N\in\N}$ converges to $\mathsf{s}(\bm{w},R_{\alpha,\lambda}(\bm{X})\cap\mathcal{Z})=\mathsf{s}(\bm{w},R_{\alpha,\lambda}(\bm{X}))$ almost surely. 
\Halmos
\endproof

\proof{Proof of \Cref{lem:bdd}.}
Let $\bm{z}^{\text{LB}}$ be as in \Cref{asmp:dom2}. Since $\bm{X}\in\X_\Lambda$, we have
\[
1=\mathbb{P}\{\bm{X}\in\dom\Lambda\}\leq \mathbb{P}\{\bm{X}\geq \bm{x}^{\text{LB}}\}.
\]
In particular, letting $\essinf X_i\coloneqq \sup\{c>0\mid \mathbb{P}\{X_i\geq c\}=1\}$ for each $i\in[d]$ and writing $\essinf\bm{X}\coloneqq (\essinf X_1,\ldots,\essinf X_d)^{\mathsf{T}}$, we have $\essinf\bm{X}\geq \bm{x}^{\text{LB}}$ so that $\essinf\bm{X}\in\R^d$. Let $\bm{z}\in R_{\alpha,\lambda}(\bm{X})$. Note that we have
\[
1=\mathbb{P}\{\bm{X}+\bm{B}^{\mathsf{T}}\bm{z}\in \dom \Lambda\}\leq \mathbb{P}\{\bm{X}\geq \bm{x}^{\text{LB}}-\bm{B}^{\mathsf{T}}\bm{z}\}.
\]
In particular, $\essinf\bm{X}\geq \bm{x}^{\text{LB}}-\bm{B}^{\mathsf{T}}\bm{z}$, i.e., $\bm{B}^{\mathsf{T}}\bm{z}\geq \bm{x}^{\text{LB}}-\essinf\bm{X}$. This implies that $\bm{z}\geq \bm{z}^{\text{LB}}$, where $\bm{z}^{\text{LB}}\in\R^g$ is defined by
\[
z^{\text{LB}}_j\coloneqq \max_{i\in[d]\colon B_{ji}=1} (x^{\text{LB}}_i-\essinf X_i).
\]
for each $j\in [g]$. Hence, $
R_{\alpha,\lambda}(\bm{X})\subseteq \bm{z}^{\text{LB}}+\R^g_+$ follows. The requirement $\bm{z}^{\text{LB}}\notin R_{\alpha,\lambda}(\bm{X})$ can be satisfied by replacing $\bm{z}^{\text{LB}}$ by $\bm{z}^{\text{LB}}-\bm{1}_g$ if necessary.
\Halmos
\endproof

\proof{Proof of \Cref{thm:wijsman}.}
Let $\bm{v}\in\bm{z}^{\text{LB}}+\R^g_+$. Let $\varepsilon\coloneqq \mathsf{d}(\bm{z}^{\textup{LB}},R_{\alpha,\lambda}(\bm{X}))>0$ be as in \Cref{asmp:wijsman}. We claim that
\begin{equation}\label{eq:maxdist}
	\mathsf{d}(\bm{v},R_{\alpha,\lambda}(\bm{X}))\leq \mathsf{d}(\bm{z}^{\text{LB}},R_{\alpha,\lambda}(\bm{X})).
\end{equation}
By \Cref{lem:dopt}, there exists $\bm{y}^\text{LB}\in R_{\alpha,\lambda}(\bm{X})$ such that 
\[
|\bm{z}^{\text{LB}}-\bm{y}^{\text{LB}}|_2=\mathsf{d}(\bm{z}^{\text{LB}},R_{\alpha,\lambda}(\bm{X}))
\]
and we have $\bm{z}^{\text{LB}}\leq \bm{y}^{\text{LB}}$. If $\bm{v}\in[\bm{z}^{\text{LB}},\bm{y}^{\text{LB}}]$, then we have
\[
\mathsf{d}(\bm{z}^{\text{LB}},R_{\alpha,\lambda}(\bm{X}))=|\bm{z}^{\text{LB}}-\bm{y}^{\text{LB}}|_2\geq |\bm{v}-\bm{y}^{\text{LB}}|_2\geq \mathsf{d}(\bm{v},R_{\alpha,\lambda}(\bm{X}))
\]
since $\bm{y}^{\text{LB}} \in R_{\alpha,\lambda}(\bm{X})$; hence, \eqref{eq:maxdist} holds in this case. In general, since $\bm{y}^{\text{LB}}+\R^g_+\subseteq R_{\alpha,\lambda}(\bm{X})$, we have
\[
|\bm{v}-\bm{v}\vee \bm{y}^{\text{LB}}|_2=\mathsf{d}(\bm{v},\bm{y}^{\text{LB}}+\R^g_+)\geq \mathsf{d}(\bm{v},R_{\alpha,\lambda}(\bm{X})),
\]
where the equality follows as a result of an elementary calculation. Let $\bm{y}^{\bm{v}}\coloneqq \bm{v}-\bm{v}\vee \bm{y}^{\text{LB}}+\bm{y}^{\text{LB}}$. Note that $\bm{y}^{\bm{v}}\in[\bm{z}^{\text{LB}},\bm{y}^{\text{LB}}]$ so that
\[
\mathsf{d}(\bm{z}^{\text{LB}},R_{\alpha,\lambda}(\bm{X}))=|\bm{z}^{\text{LB}}-\bm{y}^{\text{LB}}|\geq |\bm{y}^{\bm{v}}-\bm{y}^{\text{LB}}|_2=|\bm{v}-\bm{v}\vee \bm{y}^{\text{LB}}|_2
\]
by the previous case. Hence, \eqref{eq:maxdist} holds in the general case as well.

Let $\bm{z}^{\bm{v}}, (\bm{z}^{\bm{v},\ell})_{\ell\in\N}$ be given by \Cref{asmp:wijsman}. In particular, $|\bm{v}-\bm{z}^{\bm{v}}|_2=\mathsf{d}(\bm{v},R_{\alpha,\lambda}(\bm{X}))$. Thanks to \eqref{eq:maxdist}, we have $|\bm{v}-\bm{z}^{\bm{v}}|_2\leq \varepsilon<2\varepsilon$ so that $\bm{z}^{\bm{v}}\in\Int\mathbb{B}({\bm{v},2\varepsilon})$. Then, the definition of the distance function $\mathsf{d}$ guarantees that
\[
\mathsf{d}(\bm{v},R_{\alpha,\lambda}(\bm{X}))= \mathsf{d}(\bm{v},R_{\alpha,\lambda}(\bm{X})\cap \mathbb{B}(\bm{v},2\varepsilon)).
\]
Since $\bm{z}^{\bm{v}}\in\Int\mathbb{B}(\bm{v},2\varepsilon)$ and $(\bm{z}^{\bm{v},\ell})_{\ell\in\N}$ converges to $\bm{z}^{\bm{v}}$, there exists $\ell_0\in\N$ such that $\bm{z}^{\bm{v},\ell}\in\mathbb{B}(\bm{v},2\varepsilon)$ for every $\ell\geq \ell_0$. Let $\varphi(\bm{z})\coloneqq |\bm{v}-\bm{z}|_2$ for each $\bm{z}\in\R^g$. Then, the sequence $(\bm{z}^{\bm{v},\ell})_{\ell\geq \ell_0}$ satisfies \Cref{asmp:slater} for $\varphi$. Hence, by \Cref{thm:generalconv}, $(\mathsf{d}(\bm{v},R^N_{\alpha,\lambda}(\bm{X}^1,\ldots,\bm{X}^N)\cap \mathbb{B}(\bm{v},2\varepsilon)))_{N\in\N}$ converges to $\mathsf{d}(\bm{v},R_{\alpha,\lambda}(\bm{X})\cap \mathbb{B}(\bm{v},2\varepsilon))$ almost surely.

Let
\[
\O^{\bm{v}}\coloneqq\cb{\lim_{N\rightarrow\infty}\mathsf{d}(\bm{v},R^N_{\alpha,\lambda}(\bm{X}^1,\ldots,\bm{X}^N)\cap \mathbb{B}(\bm{v},2\varepsilon))=\mathsf{d}(\bm{v},R_{\alpha,\lambda}(\bm{X})\cap \mathbb{B}(\bm{v},2\varepsilon))}.
\]
Note that $\mathbb{P}(\O^{\bm{v}})=1$. Let $\o\in\O^{\bm{v}}$. We have $\mathsf{d}(\bm{v},R_{\alpha,\lambda}(\bm{X})\cap \mathbb{B}(\bm{v},2\varepsilon))= \mathsf{d}(\bm{v},R_{\alpha,\lambda}(\bm{X}))<+\infty$. Hence, there exists $N_0(\o)\in\N$ such that $\mathsf{d}(\bm{v},R^N_{\alpha,\lambda}(\bm{X}^1(\o),\ldots,\bm{X}^N(\o))\cap \mathbb{B}(\bm{v},2\varepsilon))<+\infty$ for every $N\geq N_0(\o)$. Let $N\geq N_0(\o)$. Then, there exists $\bm{z}^N\in \mathbb{B}(\bm{v},2\varepsilon)$ such that $\bm{z}^N\in R_{\alpha,\lambda}^N(\bm{X}^1(\o),\ldots,\bm{X}^N(\o))$. In particular, since $R_{\alpha,\lambda}^N(\bm{X}^1(\o),\ldots,\bm{X}^N(\o))\neq\emptyset$, by \Cref{lem:dopt}, there exists $\bm{z}^{N,\ast}\in R_{\alpha,\lambda}^N(\bm{X}^1(\o),\ldots,\bm{X}^N(\o))$ such that $|\bm{v}-\bm{z}^{N,\ast}|_2=\mathsf{d}(\bm{v},R_{\alpha,\lambda}^N(\bm{X}^1(\o),\ldots,\bm{X}^N(\o)))$. Then,
\[
|\bm{v}-\bm{z}^{N,\ast}|_2\leq |\bm{v}-\bm{z}^{N}|_2\leq 2\varepsilon
\]
so that $\bm{z}^{N,\ast}\in \mathbb{B}(\bm{v},2\varepsilon)$. By the definition of the distance function $\mathsf{d}$, it follows that
\[
\mathsf{d}(\bm{v},R_{\alpha,\lambda}^N(\bm{X}^1(\o),\ldots,\bm{X}^N(\o)))=\mathsf{d}(\bm{v},R_{\alpha,\lambda}^N(\bm{X}^1(\o),\ldots,\bm{X}^N(\o))\cap\mathbb{B}(\bm{v},2\varepsilon)).
\]
Since this holds for every $N\geq N_0(\o)$, we conclude that
\[
\lim_{N\rightarrow\infty}\mathsf{d}(\bm{v},R^N_{\alpha,\lambda}(\bm{X}^1(\o),\ldots,\bm{X}^N(\o)))=\mathsf{d}(\bm{v},R_{\alpha,\lambda}(\bm{X})).
\]
Hence, $(\mathsf{d}(\bm{v},R^N_{\alpha,\lambda}(\bm{X}^1,\ldots,\bm{X}^N))_{N\in\N}$ converges to $\mathsf{d}(\bm{v},R_{\alpha,\lambda}(\bm{X}))$ almost surely.

Finally, by choosing a countable dense subset of $\bm{z}^{\text{LB}}+\R^g_+$ and using the continuity of $\mathsf{d}(\cdot,R_{\alpha,\lambda}(\bm{X}))$ and $\mathsf{d}(\cdot,R^N_{\alpha,\bm{X}}(\bm{X}^1,\ldots,\bm{X}^N))$ for each $N\in\N$, we conclude that $(R_{\alpha,\lambda}^N(\bm{X}^1,\ldots,\bm{X}^N))_{N\in\N}$ Wijsman-converges to $R_{\alpha,\lambda}(\bm{X})$ almost surely.
\Halmos
\endproof

\proof{Proof of \Cref{lem:dopt}.}
Let us fix $\hat{\bm{z}}\in R_{\alpha,\lambda}(\bm{X})$, which exists by assumption. Let $\varepsilon\coloneqq |\bm{v}-\hat{\bm{z}}|_2\geq 0$. Due to the definition of the distance function $\mathsf{d}$, we have 
\[
\mathsf{d}(\bm{v},R_{\alpha,\lambda}(\bm{X}))=\mathsf{d}(\bm{v},R_{\alpha,\lambda}(\bm{X})\cap \mathbb{B}(\bm{v},\varepsilon)).
\]
Since $R_{\alpha,\lambda}(\bm{X})\cap \mathbb{B}(\bm{v},\varepsilon)$ is a nonempty compact set, it follows that there exists $\bm{z}^\ast\in R_{\alpha,\lambda}(\bm{X})\cap \mathbb{B}(\bm{v},\varepsilon)$ such that $\mathsf{d}(\bm{v},R_{\alpha,\lambda}(\bm{X}))=|\bm{v}-\bm{z}^\ast|_2$. We claim that 
$\bm{v}\leq\bm{z}^\ast$. Indeed, since $R_{\alpha,\lambda}(\bm{X})=R_{\alpha,\lambda}(\bm{X})+\R^g_+$, we have
\[
\inf_{\bm{y}\in\R^g_+}|\bm{v}-\bm{z}^\ast-\bm{y}|^2_2\geq (\mathsf{d}(\bm{v},R_{\alpha,\lambda}(\bm{X})))^2= |\bm{v}-\bm{z}^\ast|_2^2.
\]
Then, an elementary calculation to evaluate the above infimum yields
\[
|(\bm{v}-\bm{z}^\ast)^-|_2^2 \geq |\bm{v}-\bm{z}^\ast|_2^2,
\]
which implies that $\bm{v}\leq \bm{z}^\ast$.
\Halmos
\endproof

\proof{Proof of \Cref{lem:d}.}
If $\bm{v}\in R_{\alpha,\lambda}(\bm{X})$, then we have $	\mathsf{d}(\bm{v},R_{\alpha,\lambda}(\bm{X}))=\mathsf{d}(\bm{v},R_{\alpha,\lambda}(\bm{X})\cap \mathcal{Z})=0$ trivially. Let us assume that $\bm{v}\notin R_{\alpha,\lambda}(\bm{X})$ for the rest of the proof.

The $\leq$ part of the claimed equality is obvious. For the $\geq$ part, by \Cref{lem:dopt}, note that there exists $\bm{z}^\ast\in R_{\alpha,\lambda}(\bm{X})$ such that $|\bm{v}-\bm{z}^\ast|_2=\mathsf{d}(\bm{v},R_{\alpha,\lambda}(\bm{X}))$ and we also have $\bm{v}\leq \bm{z}^\ast$.	On the other hand, since $\bm{v}\notin R_{\alpha,\lambda}(\bm{X})$, we have $\bm{z}^\ast\in\bd R_{\alpha,\lambda}(\bm{X})$ as otherwise one could construct a point in a neighborhood of $\bm{z}^\ast$ that is strictly closer to $\bm{v}$ than $\bm{z}^\ast$. Then, by \citet[Definition 1.45, Corollary 1.48(iv)]{lohne}, $\bm{z}^\ast$ is a weakly minimal point of $R_{\alpha,\lambda}(\bm{X})$. 
%
%
Hence,
\begin{equation}\label{eq:wmin}
	(\bm{z}^\ast-\R^g_{++})\cap R_{\alpha,\lambda}(\bm{X})=\emptyset.
\end{equation}
Since we also have $(\bm{z}^\ast-\R^g_+)\cap R_{\alpha,\lambda}(\bm{X})\cap \mathcal{Z}\neq \emptyset$ by \eqref{eq:cpc}, there exists $\bm{y}\in\R^g_+$ such that $\bm{z}^\ast-\bm{y}\in R_{\alpha,\lambda}(\bm{X})\cap \mathcal{Z}$. Then, \eqref{eq:wmin} implies that $\bm{y}\in \bd \R^g_+$. If $\bm{y}=\bm{0}_g$, then $\bm{z}^\ast \in R_{\alpha,\lambda}(\bm{X})\cap \mathcal{Z}$ so that $|\bm{v}-\bm{z}^\ast|_2\geq \mathsf{d}(\bm{v},R_{\alpha,\lambda}(\bm{X})\cap \mathcal{Z})$, which completes the proof in this case.

Suppose that $\bm{y}\neq \bm{0}_g$. Let $\mathcal{J}\coloneqq \{j\in[g]\mid y_j=0\}$. Then, we have $\mathcal{J}\neq\emptyset$ and $\mathcal{J}^c\coloneqq \{j\in[g]\mid y_j>0\}\neq\emptyset$. Let us define a point $\hat{\bm{z}}\in\R^g$ by
\[
\hat{z}_j\coloneqq \begin{cases}
	z^\ast_j& \text{if }j\in\mathcal{J}\\ (z^\ast_j-y_j)\vee v_j&\text{if }j\in \mathcal{J}^c.
\end{cases}
\]
Clearly, $\hat{\bm{z}}\geq \bm{z}^\ast -\bm{y}$. Since $\bm{z}^\ast -\bm{y}\in R_{\alpha,\lambda}(\bm{X})$, we also have $\hat{\bm{z}}\in R_{\alpha,\lambda}(\bm{X})$. Since $\bm{z}^\ast-\bm{y}\in\mathcal{Z}$ and $\bm{v}\in\mathcal{Z}$, we also have $\hat{\bm{z}}\in\mathcal{Z}$. Since $\bm{v}\leq\bm{z}^\ast$, for each $j\in\mathcal{J}$, we have $\hat{z}_j-v_j=z^\ast_j-v_j\geq 0$; for each $j\in\mathcal{J}^c$, we have
\[
0\leq \hat{z}_j-v_j= \of{(z^\ast_j-y_j)\vee v_j}-v_j=(z^\ast_j-y_j-v_j)\vee 0 \leq (z^\ast_j-v_j)\vee 0=z^\ast_j-v_j.
\]
It follows that
\[
|\hat{\bm{z}}-\bm{v}_j|^2_2 =\sum_{j=1}^g(\hat{z}_j-v_j)^2
\leq \sum_{j=1}^g (z^\ast_j-v_j)^2 = |\bm{z}^\ast-\bm{v}|_2^2=(\mathsf{d}(\bm{v},R_{\alpha,\lambda}(\bm{X})))^2.
\]
Since $\hat{\bm{z}}\in R_{\alpha,\lambda}(\bm{X})\cap\mathcal{Z}$, we conclude that $\mathsf{d}(\bm{v},R_{\alpha,\lambda}(\bm{X})\cap \mathcal{Z})\leq \mathsf{d}(\bm{v},R_{\alpha,\lambda}(\bm{X}))$, which completes the proof in this case too.
\Halmos
\endproof

\proof{Proof of \Cref{thm:hausdorff}.}
First, we show that $(R^N_{\alpha,\lambda}(\bm{X}^1,\ldots,\bm{X}^N)\cap\mathcal{Z})_{N\in\N}$ Wijsman-converges to $R_{\alpha,\lambda}(\bm{X})\cap\mathcal{Z}$ almost surely. Let $\bm{v}\in\mathcal{Z}$. By \Cref{lem:d}, we have
\[
\mathsf{d}(\bm{v},R_{\alpha,\lambda}(\bm{X}))=\mathsf{d}(\bm{v},R_{\alpha,\lambda}(\bm{X})\cap \mathcal{Z}).
\]
Let $\varphi(\bm{z})\coloneqq |\bm{v}-\bm{z}|_2$ for each $\bm{z}\in\R^g$. Then, the sequence $(\bm{z}^\ell)_{\ell\in\N}$ and the point $\bar{\bm{z}}$ given by \Cref{asmp:hausdorff} satisfy \Cref{asmp:slater} for $\varphi$. Hence, by \Cref{thm:generalconv}, $(\mathsf{d}(\bm{v},R^N_{\alpha,\lambda}(\bm{X}^1,\ldots,\bm{X}^N))\cap \mathcal{Z})_{N\in\N}$ converges to $\mathsf{d}(\bm{v},R_{\alpha,\lambda}(\bm{X})\cap \mathcal{Z})$ almost surely. By choosing a countable dense subset of $\mathcal{Z}$ and using the continuity of $\mathsf{d}(\cdot,R_{\alpha,\lambda}(\bm{X})\cap\mathcal{Z})$ and $\mathsf{d}(\cdot,R^N_{\alpha,\bm{X}}(\bm{X}^1,\ldots,\bm{X}^N)\cap\mathcal{Z})$ for each $N\in\N$, we conclude that $(R^N_{\alpha,\lambda}(\bm{X}^1,\ldots,\bm{X}^N)\cap\mathcal{Z})_{N\in\N}$ Wijsman-converges to $R_{\alpha,\lambda}(\bm{X})\cap\mathcal{Z}$ almost surely. Since $\mathcal{Z}$ is a compact metric space, by \citet[Theorem~2.5]{beer1987metric}, Wijsman convergence coincides with Hausdorff convergence for closed subsets of $\mathcal{Z}$. Hence, $(R^N_{\alpha,\lambda}(\bm{X}^1,\ldots,\bm{X}^N)\cap\mathcal{Z})_{N\in\N}$ Hausdorff-converges to $R_{\alpha,\lambda}(\bm{X})\cap\mathcal{Z}$ almost surely as well, i.e., there exists $\O^{\mathsf{h}}\in\mathcal{F}$ with $\mathbb{P}(\O^{\mathbb{H}})=1$ such that
\begin{equation}\label{eq:h}
	\lim_{N\rightarrow\infty}\mathsf{h}(R^N_{\alpha,\lambda}(\bm{X}^1(\o),\ldots,\bm{X}^N(\o))\cap\mathcal{Z},R_{\alpha,\lambda}(\bm{X})\cap\mathcal{Z})=0
\end{equation}
for every $\o\in\O^{\mathsf{h}}$.

Let us define
\[
\O^{\mathcal{Z}}\coloneqq\bigcap_{N=1}^\infty \{	R^N_{\alpha,\lambda}(\bm{X}^1,\ldots,\bm{X}^N)=R^N_{\alpha,\lambda}(\bm{X}^1,\ldots,\bm{X}^N)\cap\mathcal{Z}+\R^g_+\}.
\]
Then, $\mathbb{P}(\O^{\mathcal{Z}})=1$. In particular, $\mathbb{P}(\O^{\mathsf{h}}\cap \O^{\mathcal{Z}})=1$. Let $\o\in \O^{\mathsf{h}}\cap \O^{\mathcal{Z}}$. By the definition of $\O^{\mathcal{Z}}$ and the properties of Hausdorff distance, we have
\begin{align}
	&\mathsf{h}(R^N_{\alpha,\lambda}(\bm{X}^1(\o),\ldots,\bm{X}^N(\o)), R_{\alpha,\lambda}(\bm{X}))\notag \\
	&=\mathsf{h}(R^N_{\alpha,\lambda}(\bm{X}^1(\o),\ldots,\bm{X}^N(\o))\cap\mathcal{Z}+\R^g_+, R_{\alpha,\lambda}(\bm{X})
	\cap\mathcal{Z}+\R^g_+) \notag\\
	&\leq \mathsf{h}(R^N_{\alpha,\lambda}(\bm{X}^1(\o),\ldots,\bm{X}^N(\o))\cap\mathcal{Z}, R_{\alpha,\lambda}(\bm{X})
	\cap\mathcal{Z}) +\mathsf{h}(\R^g_+,
	\R^g_+)\notag \\
	&= \mathsf{h}(R^N_{\alpha,\lambda}(\bm{X}^1(\o),\ldots,\bm{X}^N(\o))\cap\mathcal{Z}, R_{\alpha,\lambda}(\bm{X})
	\cap\mathcal{Z}) \notag 
\end{align}
for every $N\in\N$. Then, letting $N\rightarrow\infty$ and using \eqref{eq:h} yields
\[
\lim_{N\rightarrow\infty}\mathsf{h}(R^N_{\alpha,\lambda}(\bm{X}^1(\o),\ldots,\bm{X}^N(\o)), R_{\alpha,\lambda}(\bm{X}))=0.
\]
Therefore, the sequence $(R^N_{\alpha,\lambda}(\bm{X}^1,\ldots,\bm{X}^N))_{N\in\N}$ Hausdorff-converges to $R_{\alpha,\lambda}(\bm{X})$ almost surely. Since Hausdorff convergence implies Wijsman convergence in general, the proof is complete.
\Halmos
\endproof

\section{Proofs of the Results in \Cref{sec:systemicRM}}\label{app:systemicRM}

\proof{Proof of \Cref{lem:enmon}.}
Let $\bm{x}\in\R^d_+$. Note that $\bm{0}_d$ is always a feasible solution of the linear programming problem that defines $\Lambda^{\text{EN}}(\bm{x})$. Hence, $\Lambda^{\text{EN}}(\bm{x})>-\infty$. Since the feasible region of this problem is a subset of $[\bm{0}_d,\bar{\bm{p}}]$, we have $0\leq \Lambda^{\text{EN}}(\bm{x})\leq \bm{1}_d^{\mathsf{T}}\bar{\bm{p}}$. In particular, $\dom \Lambda^{\text{EN}}=\R^d_+$ so that \Cref{asmp:dom2} is satisfied. Moreover, $\Lambda^{\text{EN}}$ is bounded on $\dom\Lambda^{\text{EN}}$.

Note that the corresponding dual linear programming problem is given by
\begin{equation} \label{dual}
	\text{minimize} \quad \bm{\lambda}^{\mathsf{T}}\bar{\bm{p}} + \bm{\mu}^{\mathsf{T}}\bm{x}\quad
	\text{subject to} \quad \bm{\lambda} +(\bm{I}_{d} - \bm{\pi})^{\mathsf{T}} \bm{\mu} \geq \bm{1}_d,\quad
	\bm{\lambda} \geq \bm{0}_d,\quad
	\bm{\mu} \geq \bm{0}_d.
\end{equation}
By strong duality, the optimal value of this problem is equal to $\Lambda^{\text{EN}}(\bm{x})$. Hence, on $\dom\Lambda^{\text{EN}}$, the function $\Lambda^{\text{EN}}$ can be seen as the pointwise infimum of a family of affine functions; hence it is concave.  As a finite-valued concave function, $\Lambda^{\text{EN}}$ is also continuous on $\dom \Lambda^{\text{EN}}$. From the structure of the dual problem, it is obvious that $\Lambda^{\text{EN}}$ is increasing. Hence, \Cref{asmp:dom} is also satisfied.
\Halmos
\endproof

\proof{Proof of \Cref{prop:nonempty}.}
Assume that $\alpha \leq \bm{1}_d^{\mathsf{T}}\bar{\bm{p}}$. Let $\bm{z} = |\bar{\bm{p}}|_2 \bm{1}_g$. We claim that $\bm{z} \in R^{\text{EN}}_{\alpha,\lambda}(\bm{X})$. Clearly, $\mathbb{P}\{\bm{X}+\bm{B}^{\mathsf{T}}\bm{z}\geq 0\}=1$ since $\bm{X}\in L^0(\R^d_+)$ and $\bm{z}\in\R^g_+$. For every $\bm{x}\in\R^d_+$, note that
\begin{equation} \label{nonempty}
	\Lambda^{\EN}(\bm{x}+ \bm{B}^{\mathsf{T}}(|\bar{\bm{p}}|_2 \bm{1}_g))= \sup \{ \bm{1^{\mathsf{T}}}_d\bm{p} \mid \bm{p} \leq \bm{\pi}^{\mathsf{T}}\bm{p} + \bm{x} + |\bar{\bm{p}}|_2 \bm{1}_d,\ \bm{p} \in [\bm{0}_d,\bar{\bm{p}}] \}.
\end{equation}
It is clear that $\bm{p} = \bar{\bm{p}}$ is a feasible solution to \eqref{nonempty} as  $\bar{\bm{p}}\leq |\bar{\bm{p}}|_2 \bm{1}_d$. Then, since the objective function is strictly increasing, $\bar{\bm{p}}$ is indeed the unique optimal solution to \eqref{nonempty} so that $\Lambda^{\EN}(\bm{x}+ \bm{B}^{\mathsf{T}}(|\bar{\bm{p}}|_2 \bm{1}_g)) = \bm{1}_d^{\mathsf{T}}\bar{\bm{p}}$. Thus,
\[
\mathbb{P}\{\Lambda^{\text{EN}}(\bm{X}+\bm{B}^{\mathsf{T}}\bm{z})<\alpha\}=\mathbb{P}\{\bm{1}_d^{\mathsf{T}}\bar{\bm{p}}<\alpha\}=0\leq \lambda.
\]
Hence, $\bm{z} \in R^{\text{EN}}_{\alpha,\lambda}(\bm{X})$ so that $R^{\text{EN}}_{\alpha,\lambda}(\bm{X})\neq\emptyset$.

Assume that $\alpha>\bm{1}_d^{\mathsf{T}}\bar{\bm{p}}$. Then, since $\Lambda^{\text{EN}}$ is bounded from above by $\bm{1}_d^{\mathsf{T}}\bar{\bm{p}}$ and $\lambda<1$, we have $\mathbb{P}\{\Lambda^{\text{EN}}(\bm{X} + \bm{B}^{\mathsf{T}}\bm{z}) < \alpha\} = 1 > \lambda$. Hence, $\bm{z} \in R^{\text{EN}}_{\alpha,\lambda}(\bm{X}) = \emptyset$.
\Halmos
\endproof

\proof{Proof of \Cref{prop:compact}.}
By the upper set property in \Cref{lem:systrisk}, it is clear that $R^{\textup{EN}}_{\alpha,\lambda}(\bm{X}) \supseteq R^{\textup{EN}}_{\alpha,\lambda}(\bm{X})\cap \mathcal{Z} + \R^g_+$. For the reverse inclusion, let $\bm{z}\in R^{\textup{EN}}_{\alpha,\lambda}(\bm{X})$. Let us define $\bm{z}^\prime\coloneqq \bm{z}\wedge \bm{z}^{\text{UB}}$. Clearly, $\bm{z}\in\bm{z}^\prime+\R^g_+$ and $\bm{z}^\prime\in\mathcal{Z}$. To finish the proof, it is enough to show that $\bm{z}^\prime \in R^{\text{EN}}_{\alpha,\lambda}(\bm{X})$.

Let $\bm{x}\in\R^d_+$. We claim that $\Lambda^{\text{EN}}(\bm{x}+\bm{B}^{\mathsf{T}}\bm{z})=\Lambda^{\text{EN}}(\bm{x}+\bm{B}^{\mathsf{T}}\bm{z}^\prime)$. Indeed, since $\Lambda^{\text{EN}}$ is monotone and $\bm{z}^\prime\leq \bm{z}$, we have $\Lambda^{\text{EN}}(\bm{x}+\bm{B}^{\mathsf{T}}\bm{z})\geq \Lambda^{\text{EN}}(\bm{x}+\bm{B}^{\mathsf{T}}\bm{z}^\prime)$. To prove the reverse inequality, let $\bm{p}\in[\bm{0}_d,\bar{\bm{p}}]$ be a feasible solution for the problem of calculating $\Lambda^{\text{EN}}(\bm{x}+\bm{B}^{\mathsf{T}}\bm{z})$, i.e., we have
\begin{equation}\label{eq:feas}
	p_i\leq (\bm{\pi}^{\mathsf{T}}\bm{p})_i+x_i+z_j
\end{equation}
for every $i\in[n]$, $j\in[g]$ such that $B_{ji}=1$. Let $j\in[g]$. If $z^{\text{UB}}_j\geq z_j$, then for every $i\in[n]$ with $B_{ji}=1$, we have
\[
p_i\leq  (\bm{\pi}^{\mathsf{T}}\bm{p})_i+x_i+z_j= (\bm{\pi}^{\mathsf{T}}\bm{p})_i+x_i+z^\prime_j
\]
by \eqref{eq:feas}. If $z^{\text{UB}}_j<z_j$, then for every $i\in[n]$ with $B_{ji}=1$, we have
\[
(\bm{\pi}^{\mathsf{T}}\bm{p})_i+x_i+z_j^\prime= (\bm{\pi}^{\mathsf{T}}\bm{p})_i+x_i+z^{\text{UB}}_j \geq (\bm{\pi}^{\mathsf{T}}\bm{p})_i+x_i+\bar{p}_i \geq p_i
\]
by the definition of $\bm{z}^{\text{UB}}$. Hence,
\[
p_i\leq (\bm{\pi}^{\mathsf{T}}\bm{p})_i+x_i+z^\prime_j
\]
for every $i\in[n]$, $j\in[g]$ such that $B_{ji}=1$, i.e., $\bm{z}^\prime$ is feasible for the problem of calculating $\Lambda^{\text{EN}}(\bm{x}+\bm{B}^{\mathsf{T}}\bm{z}^\prime)$. It follows that $\Lambda^{\text{EN}}(\bm{x}+\bm{B}^{\mathsf{T}}\bm{z})\leq \Lambda^{\text{EN}}(\bm{x}+\bm{B}^{\mathsf{T}}\bm{z}^\prime)$, completing the proof of the claim.

With the above claim, we get
\[
\mathbb{P}\{\Lambda^{\text{EN}}(\bm{X}+\bm{B}^{\mathsf{T}}\bm{z})< \alpha\}=\mathbb{P}\{\Lambda^{\text{EN}}(\bm{X}+\bm{B}^{\mathsf{T}}\bm{z}^\prime)<\alpha\}\leq\lambda.
\]
Since $\bm{z}^\prime\geq \bm{z}^{\text{LB}}$, we also have $\mathbb{P}\{\bm{X}+\bm{B}^{\mathsf{T}}\bm{z}^\prime\geq 0\}=1$ so that $\bm{z}^\prime\in R^{\text{EN}}_{\alpha,\lambda}(\bm{X})$ and the proof is complete.
\Halmos
\endproof

\proof{Proof of \Cref{thm:ENscalar}.}
	(i) Since $R^{\text{EN}}_{\alpha,\lambda}(\bm{X})=R^{\text{EN}}_{\alpha,\lambda}(\bm{X})\cap\mathcal{Z}+\R^g_+$, we have
	\[
	\mathsf{s}(\bm{w},R^{\textup{EN}}_{\alpha,\lambda}(\bm{X}))=\mathsf{s}(\bm{w},R^{\textup{EN}}_{\alpha,\lambda}(\bm{X})\cap\mathcal{Z})+\mathsf{s}(\bm{w},\R^g_+)=\mathsf{s}(\bm{w},R^{\textup{EN}}_{\alpha,\lambda}(\bm{X})\cap\mathcal{Z}).
	\]
	Since $R^{\textup{EN}}_{\alpha,\lambda}(\bm{X})\cap\mathcal{Z}$ is a compact set and $\bm{z}\mapsto \bm{w}^{\mathsf{T}}\bm{z}$ is a continuous function, there exists $\bar{\bm{z}}\in R^{\text{EN}}_{\alpha,\lambda}(\bm{X})\cap\mathcal{Z}$ such that $\bm{w}^{\mathsf{T}}\bar{\bm{z}}=\mathsf{s}(\bm{w},R^{\textup{EN}}_{\alpha,\lambda}(\bm{X})\cap\mathcal{Z})$.
	
	(ii) Let $\ell\in\N$ and define $\bm{z}^{\ell}\coloneqq (\bar{\bm{z}}+\frac{1}{\ell}\bm{1}_g)\wedge \bm{z}^{\text{UB}}$. Clearly, $\bm{z}^\ell \in\mathcal{Z}$ and $\bar{\bm{z}}\leq \bm{z}^\ell$ since $\bar{\bm{z}}\in\mathcal{Z}$. Due to the monotonicity of $\Lambda^{\text{EN}}$ and the feasibility of $\bar{\bm{z}}$, we have
	\[
	\mathbb{P}\{\bm{X}+\bm{B}^{\mathsf{T}}\bm{z}^\ell\geq 0\}\geq \mathbb{P}\{\bm{X}+\bm{B}^{\mathsf{T}}\bar{\bm{z}}\geq 0\}=1.
	\]
	and
	\begin{equation}\label{eq:nonstrict}
		\mathbb{P}\{\Lambda^{\text{EN}}(\bm{X}+\bm{B}^{\mathsf{T}}\bm{z}^\ell)<\alpha\}\leq \mathbb{P}\{\Lambda^{\text{EN}}(\bm{X}+\bm{B}^{\mathsf{T}}\bar{\bm{z}})<\alpha\}\leq\lambda.
	\end{equation}
	We claim that
	\[
	\mathbb{P}\{\Lambda^{\text{EN}}(\bm{X}+\bm{B}^{\mathsf{T}}\bm{z}^\ell)<\alpha\}< \lambda.
	\]
	If $\mathbb{P}\{\Lambda^{\text{EN}}(\bm{X}+\bm{B}^{\mathsf{T}}\bar{\bm{z}})<\alpha\}<\lambda$, then the claim holds trivially by \eqref{eq:nonstrict}. Let us assume that
	\begin{equation}\label{eq:lambda}
		\mathbb{P}\{\Lambda^{\text{EN}}(\bm{X}+\bm{B}^{\mathsf{T}}\bar{\bm{z}})<\alpha\}=\lambda.
	\end{equation}
	Then, the claim can be rewritten as
	\begin{equation}\label{eq:claim}
		\mathbb{P}\{\Lambda^{\text{EN}}(\bm{X}+\bm{B}^{\mathsf{T}}\bm{z}^\ell)<\alpha\}< \mathbb{P}\{\Lambda^{\text{EN}}(\bm{X}+\bm{B}^{\mathsf{T}}\bar{\bm{z}})<\alpha\}.
	\end{equation}
	First, note that $\bar{\bm{z}}\neq \bm{z}^{\text{UB}}$ as we have 
	\[
	\mathbb{P}\{\Lambda^{\text{EN}}(\bm{X}+\bm{B}^{\mathsf{T}}\bm{z}^{\text{UB}})<\alpha\}\leq \mathbb{P}\{\Lambda^{\text{EN}}(\bm{B}^{\mathsf{T}}\bm{z}^{\text{UB}})<
	\alpha\}= \mathbb{P}\{\bm{1}_d^{\mathsf{T}}\bar{\bm{p}}<
	\alpha\}=0<\lambda
	\]
	due to \Cref{rem:zub} and the assumption that $\alpha<\bm{1}_{d}^{\mathsf{T}}\bar{\bm{p}}$. In particular, we have $\bm{z}^\ell=\bar{\bm{z}}+\frac{1}{\ell}\bm{1}_g$.
	
	Let us fix $\bm{z}\in\bm{z}^{\text{LB}}+\R^g_+$ and define
	\[
	S(\bm{z})\coloneqq\cb{\bm{x}\in \R^d_+\mid \Lambda^{\text{EN}}(\bm{x}+\bm{B}^{\mathsf{T}}\bm{z})<\alpha},\quad \bar{S}(\bm{z})\coloneqq\cb{\bm{x}\in \R^d_+\mid \Lambda^{\text{EN}}(\bm{x}+\bm{B}^{\mathsf{T}}\bm{z})\leq \alpha}.
	\]
	Using the dual formulation \eqref{dual} in the proof of \Cref{lem:enmon}, for each $\bm{x}\in \R^d_+$, we have
	\[
	\Lambda^{\text{EN}}(\bm{x})=\min\{\bm{\lambda}^{\mathsf{T}}\bar{\bm{p}}+\bm{\mu}^{\mathsf{T}}\bm{x}\mid (\bm{\lambda},\bm{\mu})\in \mathcal{D}^{\text{EN}}\},
	\]
	where $\mathcal{D}^{\text{EN}}\coloneqq \{(\bm{\lambda},\bm{\mu})\in\R^{2d}_+\mid \bm{\lambda}+(\bm{I}_d-\bm{\pi})^{\mathsf{T}}\bm{\mu}\geq \bm{1}_d\}$. Let $(\bm{\lambda}^1,\bm{\mu}^1),\ldots,(\bm{\lambda}^v,\bm{\mu}^v)$ denote the vertices of $\mathcal{D}^{\text{EN}}$, where $v\in\N$. Then,
	\[
	\Lambda^{\text{EN}}(\bm{x})=\min_{k\in[v]} \of{(\bm{\lambda}^k)^{\mathsf{T}}\bar{\bm{p}}+(\bm{\mu}^k)^{\mathsf{T}}\bm{x}},\quad \bm{x}\in\R^d_+.
	\]
	It follows that $\R^d_+\setminus S(\bm{z})$ is a polyhedral closed convex set whose interior relative to $\R^d_+$ equals $\R^d_+\setminus \bar{S}(\bm{z})$. In particular,
	\begin{equation}\label{eq:clbar}
		\cl_{\R^d_+} S(\bm{z})=\R^d_+\setminus \Int_{\R^d_+}(\R^d_+\setminus S(\bm{z}))=\R^d_+\setminus(\R^d_+\setminus\bar{S}(\bm{z}))=\bar{S}(\bm{z}),
	\end{equation}
	where $\cl_{\R^d_+}$ and $\Int_{\R^d_+}$ denote closure and interior in the usual topology relative to $\R^d_+$, respectively.
	
	Note that $S(\bm{z}^\ell)\subseteq S(\bar{\bm{z}})$. Hence, \eqref{eq:claim} is equivalent to $\mathbb{P}\{\bm{X} \in S(\bar{\bm{z}})\setminus S(\bm{z}^{\ell})\}>0$. We prove the stronger statement that $\mathbb{P}\{\bm{X} \in S(\bar{\bm{z}})\setminus \bar{S}(\bm{z}^{\ell})\}>0$, or equivalently, $\Leb_d(S(\bar{\bm{z}})\setminus \bar{S}(\bm{z}^\ell))>0$. Since $\Lambda^{\text{EN}}$ is continuous on $\R^d_+$, the set $S(\bar{\bm{z}})\setminus \bar{S}(\bm{z}^\ell)$ is a relatively open subset of $\R^d_+$. Hence, $\Leb_d(S(\bar{\bm{z}})\setminus \bar{S}(\bm{z}^\ell))>0$ follows once we show that $S(\bar{\bm{z}})\setminus \bar{S}(\bm{z}^\ell)\neq\emptyset$.
	
	To get a contradiction, suppose that $S(\bar{\bm{z}})\setminus \bar{S}(\bm{z}^\ell)=\emptyset$, so that $S(\bm{z}^\ell)\subseteq S(\bar{\bm{z}})\subseteq \bar{S}(\bm{z}^\ell)$. Hence, by \eqref{eq:clbar},
	\begin{equation}\label{eq:inc}
		\bar{S}(\bar{\bm{z}}) = \cl_{\R^d_+} S(\bar{\bm{z}})= \bar{S}(\bm{z}^\ell).
	\end{equation}
	Note that \eqref{eq:lambda} ensures that $\Leb_d(S(\bar{\bm{z}}))>0$. In particular, $S(\bar{\bm{z}})\neq\emptyset$. Let us fix $\bm{x}^0\in S(\bar{\bm{z}})$. Then, \eqref{eq:inc} implies that
	\[
	\Lambda^{\text{EN}}(\bm{x}^0+\bm{B}^\mathsf{T}\bm{z}^\ell)=\Lambda^{\text{EN}}\of{\bm{x}^0+\frac{1}{\ell}\bm{1}_d+\bm{B}^\mathsf{T}\bar{\bm{z}}}\leq \alpha,
	\]
	i.e, $\bm{x}^0+\frac{1}{\ell}\bm{1}_d\in \bar{S}(\bar{\bm{z}})$. Then, by \eqref{eq:inc}, $\bm{x}^0+\frac{1}{\ell}\bm{1}_d\in \bar{S}(\bar{\bm{z}})$. Repeating the same argument inductively yields that $\bm{x}^0+\frac{k}{\ell}\bm{1}_d\in \bar{S}(\bar{\bm{z}})$ for every $k\in\N$. Hence,
	\begin{equation}\label{eq:contr}
		\limsup_{k\rightarrow\infty} \Lambda^{\EN}
		\of{\bm{x}^0+\bm{B}^{\mathsf{T}}\bar{\bm{z}}+\frac{k}{\ell}\bm{1}_d}\leq \alpha.
	\end{equation}
	However, we clearly have $\bm{x}^0+\bm{B}^{\mathsf{T}}\bar{\bm{z}}+\frac{k}{\ell}\bm{1}_d \geq \bm{B}^{\mathsf{T}}\bm{z}^{\text{UB}}$ for large $k\in\N$. Hence, by 
	\Cref{rem:zub} and the monotonicity of $\Lambda^{\text{EN}}$, we have
	\[
	\Lambda^{\EN}
	\of{\bm{x}^0+\bm{B}^{\mathsf{T}}\bar{\bm{z}}+\frac{k}{\ell}\bm{1}_d}\geq \Lambda^{\text{EN}}(\bm{B}^{\mathsf{T}}\bm{z}^{\text{UB}})=\bm{1}^{\mathsf{T}}_d\bar{\bm{p}}>\alpha
	\]
	for large $k\in\N$, which is a contradiction to \eqref{eq:contr}. Hence, we get $S(\bar{\bm{z}})\setminus \bar{S}(\bm{z}^\ell)\neq\emptyset$. This completes the proof since $(\bm{z}^\ell)_{\ell\in\N}$ converges to $\bar{\bm{z}}$.
\Halmos
\endproof

\proof{Proof of \Cref{thm:ENscalar2}.}
(i) Since $\bm{v}\in\mathcal{Z}$, by \Cref{lem:dopt}, we have
\[
\mathsf{d}(\bm{v},R^{\textup{EN}}_{\alpha,\lambda}(\bm{X}))=\mathsf{d}(\bm{v},R^{\textup{EN}}_{\alpha,\lambda}(\bm{X})\cap\mathcal{Z}).
\]
Since $R^{\textup{EN}}_{\alpha,\lambda}(\bm{X})\cap\mathcal{Z}$ is a compact set and $\bm{z}\mapsto |\bm{v}-\bm{z}|_2$ is a continuous function, there exists $\bar{\bm{z}}\in R^{\text{EN}}_{\alpha,\lambda}(\bm{X})\cap\mathcal{Z}$ such that $|\bm{v}-\bar{\bm{z}}|_2=\mathsf{d}(\bm{v},R^{\textup{EN}}_{\alpha,\lambda}(\bm{X})\cap\mathcal{Z})$.

(ii) Note that the proof of \Cref{thm:ENscalar2}(ii) only uses the feasibility properties of $\bar{\bm{z}}$ but not its optimality for the weighted-sum scalarization. Since the weighted-sum and norm-minimizing scalarizations have the same feasible region, the same proof works here without the need for any changes.
\Halmos
\endproof

\proof{Proof of \Cref{cor:EN}.}
The corollary is an immediate consequence of \Cref{thm:scalar}, \Cref{thm:hausdorff}, \Cref{thm:ENscalar}, and \Cref{thm:ENscalar2}.
\Halmos
\endproof 

\section{Proof and Corollaries of \Cref{thm:milvp}}\label{app:milvp}

In this section, in addition to the proof of \Cref{thm:milvp}, we present an MILP formulation for the weighted-sum scalarization and an MIQP formulation for the norm-minimizing scalarization of the sensitive systemic value-at-risk for the Eisenberg-Noe model.

\proof{Proof of \Cref{thm:milvp}.}
Let us fix $\bm{x}^1,\ldots,\bm{x}^N\in\R^d$ and denote by $\tilde{R}^{\text{EN},N}_{\alpha,\lambda}(\bm{x}^1,\ldots,\bm{x}^N)$ the set on the right of \eqref{eq:milvp}. Let $\bm{z}\in\tilde{R}^{\textup{EN},N}_{\alpha,\lambda}(\bm{x}^1,\ldots,\bm{x}^N)$ with corresponding $\bm{p}^n\in[\bm{0}_d,\bar{\bm{p}}]$ and $y^n\in\{0,1\}$, for each $n\in[N]$, that satisfy the constraints of $\tilde{R}^{\text{EN},N}_{\alpha,\lambda}(\bm{x}^1,\ldots,\bm{x}^N)$. Then, for each $n \in[N]$, $\bm{p}^n$ is a feasible solution for the problem of calculating $\Lambda^{\text{EN}}(\bm{x}^n + \bm{B}^{\mathsf{T}}\bm{z})$ so that
\[
\Lambda^{\text{EN}}(\bm{x}^n + \bm{B}^{\mathsf{T}}\bm{z}) \geq \bm{1}_d^{\mathsf{T}}\bm{p}^n\geq \alpha y^n.
\]
Thus,
\[
\frac{1}{N}\sum_{n=1}^{N} 1_{[\alpha,+\infty)}(\Lambda^{\text{EN}}(\bm{x}^n+\bm{B}^{\mathsf{T}}\bm{z})) \geq \frac{1}{N}\sum_{n=1}^{N} 1_{[\alpha,+\infty)}(\alpha y^n) = \frac{1}{N}\sum_{n=1}^{N} y^n \geq 1-\lambda .
\]
Hence, $\bm{z}\in R^{\text{EN},N}_{\alpha,\lambda}(\bm{x}^1,\ldots,\bm{x}^N)$

Conversely, let $\bm{z}\in R^{\text{EN},N}_{\alpha,\lambda}(\bm{x}^1,\ldots,\bm{x}^N)$. For each $n\in [N]$, since $\bm{x}^n+\bm{B}^{\mathsf{T}}\bm{z}\geq 0$, there exists an optimal solution $\bm{p}^n\in [\bm{0}_d,\bar{\bm{p}}]$ for the problem of calculating $\Lambda^{\text{EN}}(\bm{x}^n + \bm{B}^{\mathsf{T}}\bm{z})$; in particular, we have $\bm{p}^n\leq \bm{\pi}^{\mathsf{T}}\bm{p}^n+\bm{x}^n+\bm{B}^{\mathsf{T}}\bm{z}$. For each $n\in[N]$, let $y^n\coloneqq 1_{[\alpha,+\infty)}(\bm{1}_d^{\mathsf{T}}\bm{p}^n)$; in particular, we have $\bm{1}_d^{\mathsf{T}}\bm{p}^n\geq \alpha y^n$ immediately. Finally, by the optimality of $\bm{p}^n$ for each $n\in[N]$, we get
\[
\frac{1}{N}\sum_{n=1}^N y^n=\frac{1}{N}\sum_{n=1}^N1_{[\alpha,+\infty)}(\bm{1}_d^{\mathsf{T}}\bm{p}^n)=\frac{1}{N}\sum_{n=1}^N1_{[\alpha,+\infty)}(\Lambda^{\text{EN}}(\bm{x}^n+\bm{B}^{\mathsf{T}}\bm{z}))\geq 1-\lambda.
\]
Hence, $\bm{z}\in \tilde{R}^{\text{EN},N}_{\alpha,\lambda}(\bm{x}^1,\ldots,\bm{x}^N)$.
\Halmos
\endproof

\begin{corollary}\label{cor:ws}
	Let $N\in\N$ and $\bm{x}^1,\ldots,\bm{x}^N\in\R^d$. Suppose that $\alpha \leq \bm{1}_d^{\mathsf{T}}\bar{\bm{p}}$ and let $\bm{w}\in\R^g_+\setminus\{\bm{0}_g\}$. Consider the MILP problem
	\begin{align}
		&\textup{minimize}\quad \bm{w}^{\mathsf{T}}\bm{z}\notag \\
		&\textup{subject to}\quad \sum_{n=1}^N y^n\geq N(1-\lambda),\notag \\
		&\quad\quad\quad\quad\quad\; \bm{1}_d^{\mathsf{T}}\bm{p}^n\geq \alpha y^n,\quad n\in[N],\notag \\ 
		&\quad\quad\quad\quad\quad\;  \bm{x}^n+\bm{B}^{\mathsf{T}}\bm{z}\geq 0,\quad n\in[N],\notag \\
		&\quad\quad\quad\quad\quad\;  \bm{p}^n\leq \bm{\pi}^{\mathsf{T}}\bm{p}^n+\bm{x}^n+\bm{B}^{\mathsf{T}}\bm{z},\quad n\in[N], \notag \\
		&\quad\quad\quad\quad\quad\; \bm{p}^n\in [\bm{0}_d,\bar{\bm{p}}],\ y^n\in\{0,1\},\quad n\in[N],\notag \\
		&\quad\quad\quad\quad\quad\; \bm{z}\in\R^g.\notag 
	\end{align}
	Then, the optimal value of this problem equals $\mathsf{s}(\bm{w},R^{\textup{EN},N}_{\alpha,\lambda}(\bm{x}^1,\ldots,\bm{x}^N))$.
	Moreover, the problem admits an optimal solution.
\end{corollary}

\proof{Proof.}
The claim about the optimal value is an immediate consequence of \Cref{thm:milvp}. By \Cref{prop:nonempty}, we have $R^{\text{EN},N}_{\alpha,\lambda}(\bm{x}^1,\ldots,\bm{x}^N)\neq\emptyset$, which implies that the MILP has a feasible solution. The existence of an optimal solution is guaranteed since the feasible region is compact and the objective function is continuous.
\Halmos
\endproof


\begin{corollary} \label{cor:nm}
	Let $N\in\N$ and $\bm{x}^1,\ldots,\bm{x}^N\in\R^d$. Suppose that $\alpha \leq \bm{1}_d^{\mathsf{T}}\bar{\bm{p}}$ and let $\bm{v}\in\R^g$. Consider the MIQP problem
	\begin{align}
		&\textup{minimize}\quad |\bm{v}-\bm{z}|_2^2\notag \\
		&\textup{subject to}\quad \sum_{n=1}^N y^n\geq N(1-\lambda),\notag \\
		&\quad\quad\quad\quad\quad\; \bm{1}_d^{\mathsf{T}}\bm{p}^n\geq \alpha y^n,\quad n\in[N],\notag \\ 
		&\quad\quad\quad\quad\quad\;  \bm{x}^n+\bm{B}^{\mathsf{T}}\bm{z}\geq 0,\quad n\in[N],\notag \\
		&\quad\quad\quad\quad\quad\;  \bm{p}^n\leq \bm{\pi}^{\mathsf{T}}\bm{p}^n+\bm{x}^n+\bm{B}^{\mathsf{T}}\bm{z},\quad n\in[N], \notag \\
		&\quad\quad\quad\quad\quad\; \bm{p}^n\in [\bm{0}_d,\bar{\bm{p}}],\ y^n\in\{0,1\},\quad n\in[N],\notag \\
		&\quad\quad\quad\quad\quad\; \bm{z}\in\R^g.\notag 
	\end{align}
	Then, the optimal value of this problem equals $(\mathsf{d}(\bm{v},R^{\textup{EN},N}_{\alpha,\lambda}(\bm{x}^1,\ldots,\bm{x}^N)))^2$.
	Moreover, the problem admits an optimal solution.
\end{corollary}

%

\proof{Proof.}
	The proof follows the same lines as the proof of \Cref{cor:ws}, thus we omit it.
\Halmos
\endproof

\section{Correctness Proof of the Grid Search Algorithms in \Cref{alg section}}\label{app:alg}

\proof{Proof of \Cref{thm:correctness}}
The finiteness of both algorithms is obvious since, the grid point $\bm{z}$ under consideration is removed from the grid in any case at the end of each iteration.

For \Cref{alg:acceptability}, the property $\hat{R}^{\textup{EN},N}_{\alpha,\lambda}(\bm{x}^1,\ldots,\bm{x}^N)\subseteq R^{\textup{EN},N}_{\alpha,\lambda}(\bm{x}^1,\ldots,\bm{x}^N)$ is clear by the structure of the algorithm.

The finiteness of \Cref{alg:acceptabilitynm} is also obvious since the current , by \Cref{lem:dopt}, we have $\mathsf{d}(\bm{z},R^{\text{EN},N}_{\alpha,
	\lambda}(\bm{x}^1,\ldots,\bm{x}^N))=\mathsf{d}(\bm{z},R^{\text{EN},N}_{\alpha,
	\lambda}(\bm{x}^1,\ldots,\bm{x}^N)\cap [\bm{z}^{\text{ideal},N},\bm{z}^{\text{UB}}])$, hence an optimizer $\bar{\bm{z}}\in R^{\text{EN},N}_{\alpha,
	\lambda}(\bm{x}^1,\ldots,\bm{x}^N)\cap [\bm{z}^{\text{ideal},N},\bm{z}^{\text{UB}}]$ exists for the grid point $\bm{z}$ of a given iteration. Then, $\bar{\bm{z}}+\R^g_+\subseteq R^{\text{EN},N}_{\alpha,
	\lambda}(\bm{x}^1,\ldots,\bm{x}^N)$ so that all grid points labeled as acceptable are in $R^{\text{EN},N}_{\alpha,
	\lambda}(\bm{x}^1,\ldots,\bm{x}^N)$. Hence, $\hat{R}^{\textup{EN},N}_{\alpha,\lambda}(\bm{x}^1,\ldots,\bm{x}^N)\subseteq R^{\textup{EN},N}_{\alpha,\lambda}(\bm{x}^1,\ldots,\bm{x}^N)$ follows.

To show that $\mathsf{h}(\hat{R}^{\textup{EN},N}_{\alpha,\lambda}(\bm{x}^1,\ldots,\bm{x}^N),R^{\textup{EN},N}_{\alpha,\lambda}(\bm{x}^1,\ldots,\bm{x}^N)) \leq \epsilon$, it is enough to check that
\[
\mathsf{d}(\bm{v},\hat{R}^{\textup{EN},N}_{\alpha,\lambda}(\bm{x}^1,\ldots,\bm{x}^N))\leq \epsilon
\]
for every fixed $\bm{v}\in R^{\textup{EN},N}_{\alpha,\lambda}(\bm{x}^1,\ldots,\bm{x}^N)$.

First, let us consider the case $\bm{v}\in [\bm{z}^{\text{ideal},N},\bm{z}^{\text{UB}}]$. By assumption, there exists $\bm{z}\in\texttt{G}$ with $\bm{v}\leq \bm{z}\leq \bm{v}+\frac{\epsilon}{\sqrt{g}}\bm{1}_g$. In particular, $|\bm{v}-\bm{z}|_2\leq\epsilon$. Since $\bm{v}\in R^{\textup{EN},N}_{\alpha,\lambda}(\bm{x}^1,\ldots,\bm{x}^N)$, we also have $\bm{z}\in R^{\textup{EN},N}_{\alpha,\lambda}(\bm{x}^1,\ldots,\bm{x}^N)$. Hence, $\bm{z}$ is labeled as acceptable by both algorithms so that $\bm{z}\in \hat{R}^{\textup{EN},N}_{\alpha,\lambda}(\bm{x}^1,\ldots,\bm{x}^N)$. Then,
\[
\mathsf{d}(\bm{v},\hat{R}^{\textup{EN},N}_{\alpha,\lambda}(\bm{x}^1,\ldots,\bm{x}^N))\leq |\bm{v}-\bm{z}|_2\leq \epsilon.
\]

In general, when $\bm{v}\in R^{\textup{EN},N}_{\alpha,\lambda}(\bm{x}^1,\ldots,\bm{x}^N)$, we can find $\bm{v}^{\prime}\in R^{\textup{EN},N}_{\alpha,\lambda}(\bm{x}^1,\ldots,\bm{x}^N)\cap [\bm{z}^{\text{ideal},N},\bm{z}^{\text{UB}}]$ such that $\bm{v}=\bm{v}^\prime+\bm{y}$ for some $\bm{y}\in\R^g_+$. Then, by the first case, there exists $\bm{z}^\prime \in \hat{R}^{\textup{EN},N}_{\alpha,\lambda}(\bm{x}^1,\ldots,\bm{x}^N)$ such that $|\bm{v}^\prime-\bm{z}^\prime|_2\leq \epsilon$. Note that $\bm{z}\coloneqq \bm{z}^\prime+\bm{y}\in \hat{R}^{\textup{EN},N}_{\alpha,\lambda}(\bm{x}^1,\ldots,\bm{x}^N)$ as well. Hence,
\[
\mathsf{d}(\bm{v},\hat{R}^{\textup{EN},N}_{\alpha,\lambda}(\bm{x}^1,\ldots,\bm{x}^N))\leq |\bm{v}-\bm{z}|_2=|\bm{v}^\prime-\bm{z}^\prime|_2\leq \epsilon,
\]
which completes the proof.
\Halmos
\endproof

\end{document}